\documentclass[aps,pra,twocolumn,showpacs,superscriptaddress,10p,longbibliography]{revtex4-1}
\usepackage{graphicx}
\usepackage{amssymb}
\usepackage{amsmath}
\usepackage{bm}
\usepackage{xcolor}
\usepackage{subfigure}
\usepackage{float}
\usepackage[colorlinks=true,linkcolor=red, citecolor = magenta]{hyperref}
\usepackage{gensymb}
\usepackage[sort&compress]{natbib}
\usepackage[shortlabels]{enumitem}
\usepackage{scalerel}
\usepackage{dsfont}
\newcommand{\expect}[1]{\langle{#1}\rangle}
\newcommand{\ket}[1]{|{#1}\rangle}
\newcommand{\bra}[1]{\langle{#1}|}
\newcommand{\braket}[2]{\langle{#1}|{#2}\rangle}
\newcommand{\abs}[1]{\left|{#1}\right|}

\begin{document}

\title{Topological delocalization in the completely disordered
  two-dimensional quantum walk}
  
\author{J\'anos K. Asb\'oth}
\affiliation{Department of Theoretical Physics and BME-MTA Exotic
  Quantum Phases Research Group, Budapest University of Technology and
  Economics, H-1111 Budapest, Hungary}
\affiliation{Institute for Solid State Physics and
  Optics, Wigner Research Centre, H-1525 Budapest P.O. Box 49,
  Hungary}
\email{janos.asboth@wigner.hu}

\author{Arindam Mallick}
\affiliation{Center for Theoretical Physics of Complex Systems,
  Institute for Basic Science (IBS), Daejeon 34126, Republic of Korea}
\affiliation{Institute for Solid State Physics and Optics, Wigner
  Research Centre, H-1525 Budapest P.O. Box 49, Hungary}
\email{marindam@ibs.re.kr}

\date{\today}

\begin{abstract}

We investigate numerically and theoretically the effect of spatial
disorder on two-dimensional split-step discrete-time quantum walks
with two internal ``coin'' states.
Spatial disorder can lead to Anderson localization, inhibiting the
spread of quantum walks, putting them at a disadvantage against
their diffusively spreading classical counterparts.
We find that spatial disorder of the most general type, i.e., 
position-dependent Haar random coin operators, does
not lead to Anderson localization, but to a diffusive spread instead.
This is a delocalization, which happens because disorder places the
quantum walk to a critical point between different anomalous
Floquet-Anderson insulating topological phases.
We base this explanation on the relationship of this general quantum
walk to a simpler case more studied in the literature, and for which
disorder-induced delocalization of a topological origin has been
observed.
We review topological delocalization for the simpler quantum walk,
using time-evolution of the wavefunctions and level spacing
statistics.
We apply scattering theory to two-dimensional quantum walks, and thus
calculate the topological invariants of disordered quantum walks,
substantiating the topological interpretation of the delocalization,
and finding signatures of the delocalization in the finite-size
scaling of transmission.
We show criticality of the Haar random quantum walk by calculating the
critical exponent $\eta$ in three different ways, and find $\eta\approx 0.52$
as in the integer quantum Hall effect.
Our results showcase how theoretical ideas and numerical tools from
solid-state physics can help us understand spatially random quantum
walks.

\end{abstract}
%\pacs{74.45.+c, 71.10.Pm, 73.23.-b, 74.72.-h}
\maketitle

\section{Introduction}

Discrete-Time Quantum Walks (or quantum walks for short) are the
quantum generalizations of random
walks \cite{kempe2003quantum,genske2013electric}.
They are promising components of quantum algorithms because they
spread faster than their classical
counterparts \cite{ambainis2020quadratic}.
These quantum walks can be described as periodic
sequences of unitary ``coin toss'' and ``shift'' operations applied to
a particle on a lattice.
If the particle is initially on one of the lattice sites, during the
quantum walk, its wavefunction spreads out over the lattice to
infinity in a ballistic way, i.e., with root-mean-square distance from
the origin proportional to time elapsed.
This is faster than the diffusive spread characteristic of classical
random walks.
%
%The speed of ballistic spread can be obtained by bringing tools from
%solid state physics: assigning an effective Hamiltonian to the walk
%and obtaining group velocities from its dispersion relation.  

Spatial disorder in the quantum walk parameters can be detrimental to
the fast spread of the walk.
Intuition from solid-state physics suggest that the combination of
disorder and coherence generically leads to Anderson localization
\cite{nagaoka1985theory}: all eigenstates assume an
exponentially localized envelope).
As a consequence, a particle starting from a single site cannot spread
off to infinity, its wavefunction remains within a bounded region, to
exponential accuracy.
Anderson localization can be tested numerically from the spectrum of
the Hamiltonian, alternatively from the conductance, or from the
spreading of the wavefunction \cite{sepehrinia2011numerical}.
Anderson localization indeed happens in quantum walks
\cite{joye2012dynamical,vakulchyk2017anderson}, and has already been
observed in
experiment \cite{schreiber2011decoherence,pankov2019anderson}.
We note this is very different than the dynamical effect of
fluctuating disorder: that, by inducing a loss of coherence, would
lead to diffusive spreading, i.e., a loss of the ``quantum
advantage''\cite{dur2002quantum,schreiber2011decoherence}.
This is also different from the so-called trapping effect in quantum
walks \cite{kollar2015strongly,machida2015localization,kollar2020complete}, which is a form of
localization without disorder, related to the presence of flat bands
in the quasienergy spectrum.

Another interesting feature of quantum walks is that they can have
topological phases, of the kind known from the physics of topological
insulators and superconductors \cite{hasan2010colloquium,kitagawa2010exploring,tarasinski2014scattering,cedzich2018topological}.
On one hand, this allows quantum walks to be used as simulators
for these solid state physics systems.
On the other hand, quantum walks also have topological phases beyond
those of topological insulators, which are prototypical of
periodically driven, Floquet systems \cite{rudner2013anomalous,asboth2015edge}.
Thus, they can be used as versatile toy models for periodically driven
systems in the nonperturbative limit of strong driving.

Recently, a striking disorder-induced \emph{delocalization} phenomenon
was observed for the simplest two-dimensional quantum
walk \cite{edge2015localization}, related to its topological phases.
%t was observed that topological
%phases of quantum walks impact the way in which they respond to
%disorder, leading to a
%
This quantum walk is the split-step walk on a square lattice, with two
internal (coin) states, and a real-valued coin operator.
Disorder, added to the walk via an onsite complex phase factor
(mimicking onsite potential disorder), leads to an anomalous
Floquet-Anderson insulator (AFAI) state \cite{titum2016anomalous}, with
Anderson localization of all bulk eigenstates yet with topologically
protected edge states present in the
spectrum.
%\cite{edge2015localization}.
%
If the coin operator is finetuned to a critical value at the
transition between two topological insulating phases, disorder does
not lead to Anderson localization, but rather to a diffusive spread
\cite{edge2015localization}.
This route to delocalization also explains the lack of Anderson
localization observed elsewhere for some two-dimensional
walks \cite{zeng2017discrete}.
Intriguingly, this delocalization can also be achieved if instead of
fine tuning the coin operator parameters, they are chosen to be
maximally disordered \cite{edge2015localization}.
This happens because coin disorder can tune the bulk topological
invariant and hence also put the system to a critical point---an
example is the case where the coin rotation angle is uniformly
random.
We note that this localization-delocalization transition has a
completely different physical origin than that induced by correlated
disorder \cite{mendes2019localization}.
It is also different from delocalization in one-dimensional quantum
walks \cite{obuse2011topological, zhao2015disordered,
  rakovszky2015localization} in that it does not require chiral
symmetry.

In this article we ask whether we see Anderson localization or
topological delocalization in the \emph{completely disordered} (Haar
random) general two-state quantum walk on the square lattice, where
the coin is picked from all $U(2)$ operators in a Haar uniform random
way.
We employ a numerical tool used in previous
work \cite{edge2015localization}, namely, time evolution of the
wavefunction, which is more efficient to compute for quantum walks
than for static Hamiltonians.
We also use other numerical tools, which have been applied to quantum
walks less often: (1) analysis of the level spacing distribution to
help identify Anderson localization and (2) calculation of the
transmission matrix both to calculate the topological invariants with
disorder and to find signatures in the finite-size scaling of Anderson
localization and diffusion.
We make use of the fact that when there is a complete phase disorder---as 
in the case when coin operators are Haar uniformly random---all
properties of the spectrum should be quasienergy independent, and thus
disorder averaging can be replaced/supplemented by quasienergy
averaging, which results in a substantial numerical advantage.
We find that the completely disordered (Haar random) two-dimensional
quantum walk is topologically delocalized, for a similar reason as
the more restricted quantum walk.

Our article is organized as follows.
In Sec.~\ref{sec:model} we introduce the split-step two-dimensional
quantum walk with two internal states and discuss the topological
properties in the clean, translational invariant setting.
In Sec.~\ref{sec:topo_scattering} we introduce the scattering matrix of
the quantum walk and use it to calculate the topological invariants in
the case with disorder.
In Sec.~\ref{sec:phase_disorder} we compute time evolution, the level
spacing statistics, and the finite-size scaling of the transmission, to
show that disorder in the phase parameter leads to Anderson
localization, except if the system is in a critical state.
In Sec.~\ref{sec:haar_disorder} we show using the numerical tools
mentioned above that turning the disorder to the ``most random'' case,
i.e., coin operator from Haar uniform random distribution, leads to
delocalization.
In Sec.~\ref{sec:haar_critical}
we calculate the critical exponent $\eta$ of the Haar random quantum walk in 
three different ways: autocorrelation of the position distribution, fractal analysis of position distribution, and 
time dependence of the position distribution peak at the origin.
We conclude in Sec.~\ref{sec:conclusion} and show some preliminary
investigation of the case of binary disorder in the coin angle
parameter $\theta$ in Appendix.

\section{Two-dimensional quantum walk with two internal states}
\label{sec:model}

The quantum walk we consider in this paper is the time evolution of a
particle with two internal (spin) states on a square lattice.  Its
wavefunction reads,
\begin{align}
\ket{\Psi(t)} &= \sum_{x} \sum_{y} \sum_{s=\pm 1} 
\Psi_{x,y,s}(t) \ket{x,y,s},
\end{align}
where $x,y\in\mathbb{Z}$ are lattice coordinates and we are interested
in the state at discrete times $t\in \mathbb{N}$.
The dynamics is given by a periodic sequence of internal rotations
(coin operator) and spin-dependent displacements (shift operator) on
the lattice.

The internal rotations are general $U(2)$ operations acting on the
internal degree of freedom, in a position-dependent way.  
We split off a position-dependent phase operator,
\begin{align}
  \hat{F} &= \sum_{x,y}\sum_{s=\pm 1} 
  e^{i \phi(x,y)}\ket{x, y, s}\bra{x,y, s},
  \label{eq:def_F}
\end{align}
and $\hat{R}_j$ denotes the remaining SU(2) rotation operator, 
\begin{align}
  \hat{R}_j &= \sum_{x,y} \ket{x, y}\bra{x,y} \otimes
  \hat{R}\left(\alpha_j(x,y), \beta_j(x,y),\theta_j(x,y)\right),
\end{align}
where $j$ is used to differentiate between different rotations, and
\begin{align}
  \label{eq:rotation_matrix}
  R(\alpha, \beta,\theta) &=
  \begin{pmatrix} e^{-i(\alpha+\beta)} \cos\theta & -e^{i(\alpha-\beta)} \sin\theta \\
    e^{-i(\alpha-\beta)} \sin\theta & e^{i(\alpha+\beta)} \cos\theta 
  \end{pmatrix}.
\end{align}
Note that this can be written succinctly in Euler angle representation
as $\hat{R} = e^{-i\beta \hat{\sigma}_z} e^{-i\theta \hat{\sigma}_y}
e^{-i\alpha \hat{\sigma}_z}$, where $\alpha$, $\theta$, and $\beta$,
respectively, are the first, second and third Euler angles.
Setting $\alpha=\beta=0$ reduces this coin operator to that used in
several quantum walk papers \cite{kitagawa2010exploring,kitagawa2012topological,edge2015localization}. %

The shifts are spin-dependent translations on the lattice,
\begin{subequations}
\label{eq:shifts_def}
  \begin{align}
  \hat{S}_{x} &= \sum_{x,y}\sum_{s=\pm 1} 
  \ket{x+s, y}\bra{x,y} \otimes\ket{s}\bra{s};\\
  \hat{S}_{y} &= \sum_{x,y}\sum_{s=\pm 1} 
  \ket{x, y+s}\bra{x,y} \otimes\ket{s}\bra{s}.
  \label{eq:Sy_def}
\end{align}
\end{subequations}

The wavefunction of the walk after $t\in\mathbb{N}$ timesteps reads
\begin{align}
  \ket{\Psi(t)} &= \hat{U}^t \ket{\Psi(0)},
\end{align}
where the timestep operator, $\hat{U}$ represents the effect of one
period of the quantum walk,
\begin{align}
  \label{eq:def_timestep}
  \hat{U} &= \hat{F} \hat{S}_{y} \hat{R}_2
  \hat{S}_{x} \hat{R}_1.
\end{align}
As initial state we took $\ket{\Psi(0} = \ket{x=0,y=0,s=+1}$---due to
the disorder, the internal state of the initial condition plays no
role in the time evolution, so any choice of initial state will give
qualitatively the same results
The parameters of the timestep operator
are the position-dependent angle variables: a phase $\phi(x,y)$, and
the rotation parameters $\theta_{1,2}(x,y)$, $\alpha_{1,2}(x,y)$,
$\beta_{1,2}(x,y)$.
In the formulas that follow we will often suppress the explicit
position and spin dependence for better readability.

\subsection{Sublattice symmetry of the quantum walk}
\label{subsec:sublattice}

The two-dimensional quantum walk we consider here has a sublattice
structure.
There are four sublattices on the square lattice of $x,y\in\mathbb{Z}$,
according to whether $x$ and $y$ are even ($e$) or odd ($o$), with corresponding projectors defined
as, e.g., 
\begin{align}
  \label{eq:sublattice_projector_def}
  \hat{\Pi}_{e,o} = \sum_{x =\text{ even}} \sum_{y= \text{ odd}}
    \ket{x,y}\bra{x,y},
\end{align}
with $\hat{\Pi}_{o,e}$, $\hat{\Pi}_{o,o}$, and $\hat{\Pi}_{e,e}$
defined similarly.
Since the quantum walk we consider has one shift
along both $x$ and $y$ in every timestep, $\hat{U}$ switches
sublattices, according to $(e,e) \leftrightarrow (o,o)$ and $(e,o)
\leftrightarrow (o,e)$.
Therefore, a quantum walk started from one of the 4 sublattices never
interferes with a walk started from any other sublattice---assuming 
that boundary conditions also respect the sublattice
structure, e.g., are periodic with both $L_x$ and
$L_y$ even.
(We will later also use absorbing boundary conditions, which always
respect the sublattice structure independent of system size.)

The sublattice structure has important consequences for the spectrum
of $\hat{U}$, which can be appreciated by first considering the
spectrum of $\hat{U}^2$.
The operator $\hat{U}^2$ is block diagonal in the sublattice
basis, i.e.,
\begin{align}
  \hat{U}^2 = \sum_{j=e,o} \sum_{l=e,o} \hat{\Pi}_{j,l} \hat{U}^2
  \hat{\Pi}_{j,l}.
\end{align}
Take an eigenstate $\ket{\Psi}$ of $\hat{U}^2$ on the
$(e,e)$ sublattice,
\begin{align}
  \hat{U}^2 \ket\Psi &= e^{-2i\varepsilon} \ket\Psi,
  \end{align}
with $0\le \varepsilon < \pi$.
First, note that $\hat{U}\ket\Psi$ is on the $(o,o)$ sublattice, and
is an eigenstate of $\hat{U}^2$ with the same eigenvalue
$e^{-2i\varepsilon}$.
Second, we can use $\ket\Psi$ to generate two eigenstates of
$\hat{U}$, since 
\begin{align}
  \label{eq:eigstates_from_U2}
  \hat{U} &
  \left( \ket\Psi \pm e^{i\varepsilon} \hat{U} \ket\Psi \right) =
  \pm e^{-i\varepsilon}\left( \ket\Psi
  \pm e^{i\varepsilon} \hat{U} \ket\Psi \right).
\end{align}
These relations also hold for an eigenstate $\ket{\Psi'}$ on the $(e,o)$
sublattice, with $\hat{U}\ket{\Psi'}$ on the $(o,e)$ sublattice.

The results of the previous paragraph can be
rephrased \cite{asboth2015edge} as \emph{sublattice symmetry} of
$\hat U$, represented by a unitary and Hermitian sublattice operator,
\begin{align}
  \hat{\Gamma} \hat{U} \hat{\Gamma} &= -\hat{U}, \,\,\text{with}& 
  \hat{\Gamma} &= \hat{\Pi}_{e,e} + \hat{\Pi}_{e,o}
  - \hat{\Pi}_{o,o} - \hat{\Pi}_{o,e}. 
\end{align}
Every eigenstate $\ket{\Phi}$ of the
walk has a sublattice symmetric partner,
\begin{align}
  \label{eq:sublattice_partners}
  \hat{U}\ket{\Phi}&=e^{-i\varepsilon} \ket{\Phi} \quad \Rightarrow \quad
  \hat{U}\hat{\Gamma}\ket{\Phi}&=e^{-i(\varepsilon +\pi)} \hat{\Gamma} \ket{\Phi},
\end{align}
which is related to Eq.~\eqref{eq:eigstates_from_U2} by $\ket{\Phi}=(\ket\Psi
+ e^{i\varepsilon}\hat{U}\ket\Psi)/\sqrt{2}$.

The sublattice symmetry we defined above is \emph{not} the same as the
sublattice symmetry familiar from solid state physics, which is also
known as chiral symmetry. Chiral symmetry states $\hat\Gamma
\hat{H}\hat\Gamma = -\hat{H}$, linking eigenstates of a Hamiltonian at
energy $\varepsilon$ to eigenstates at $-\varepsilon$.
The quantum walk we consider here has chiral symmetry only if the 
parameters are finetuned.
For chiral symmetry, we need $\phi=0$, and
$\alpha_2(x,y)=\beta_1(x,y)$ and $\alpha_1(x,y) = \beta_2(x,y)$,
moreover, either $\theta_1(x,y)=\theta_2(x,y) +n \pi$ (with chiral
symmetry by $\hat{\sigma}_x$), or $\theta_1(x,y)=-\theta_2(x,y)+n \pi$
(with chiral symmetry by $\hat{\sigma}_y$).

\subsection{Parameters $\alpha$ and $\beta$ represent a vector potential}

The parameters $\alpha_{1,2}(x,y)$ and $\beta_{1,2}(x,y)$ of the coin
rotations can be understood to represent a vector
potential \cite{yalccinkaya2015two,arnault2016quantum,sajid2018creating,cedzich2019quantum, Mallick:2019hvd}.
This is best seen by writing the time-step operator in a different
timeframe \cite{asboth2013bulk}, which amounts to a similarity
transformation on the original time-step operator
of Eq.~\eqref{eq:def_timestep}:
\begin{align}
  \hat{U}_2 &=  e^{-i \alpha_1 \hat{\sigma}_z} \hat{U} e^{i \alpha_1 \hat{\sigma}_z}  = e^{-i\phi} \hat{S}_y'  e^{-i\theta_2\hat{\sigma}_y}
  \hat{S}_x'e^{-i\theta_1\hat{\sigma}_y};\\
 \hat{S}_y' &= e^{-i\alpha_1\hat{\sigma}_z}\hat{S}_y e^{-i\beta_2
   \hat{\sigma}_z};\quad  \hat{S}_x' =
 e^{-i\alpha_2\hat{\sigma}_z}\hat{S}_x e^{-i\beta_1 \hat{\sigma}_z}.
 \label{eq:shift_gauge}
\end{align}
First, if $\alpha_{1,2}$ and $\beta_{1,2}$ are independent of
position, then their effect on the quantum walk is just a gauge
transformation.
To prove this, we rewrite Eq.~\eqref{eq:shift_gauge} as
\begin{align} \hat{S}_x' &=
         e^{-i(\beta_1+\alpha_2)\hat{x}} \hat{S}_x
         e^{i(\beta_1+\alpha_2) \hat{x}},
\end{align}
where $\hat{x}$ = $\sum_x x \ket{x}\bra{x}$ is the position operator
corresponding to position coordinate $x$. Thus $\hat{S}'_x$ is a
unitary transformed version of $\hat{S}_x$, and the transforming
operator commutes with rotations, phase operations, and $\hat{S}_y$ as
well. Similar statements hold for $\hat{S}'_y$. Thus, in this case,
the timestep operator $\hat{U}_2$ of a quantum walk is unitary
equivalent to the timestep operator of the same walk with
$\alpha_{1,2}=\beta_{1,2}=0$.

For $\alpha_{1,2}(x,y)$ and $\beta_{1,2}(x,y)$ depending on position,
we find a more complicated situation.  For this more general case, it
is helpful to think of the quantum walk as a nearest-neighbor hopping
model on the square lattice with time-dependent hopping
amplitudes \cite{asboth2015edge}. The angles $\alpha_{1,2}(x,y)$ and
$\beta_{1,2}(x,y)$ can then be included in this hopping model by the
use of Peierls phases: For a link from site $(x,y)$ to $(x\pm 1, y)$,
the corresponding Peierls phase is $\pm(\beta_1(x,y) + \alpha_2(x\pm
1,y))$, and from $(x,y)$ to $(x,y\pm 1)$, it is $\pm(\beta_2(x,y) +
\alpha_1(x,y\pm 1))$. A caveat: These Peierls phases depend on the
sublattice.

\subsection{The quantum walk is topological}

The quantum walk we consider has topological phases and associated
chiral edge states similar to the simpler two-dimensional quantum
walk.
That simpler quantum walk (i.e., the vector-potential-like parameters
$\alpha=\beta=0$), without spatial disorder has been recognized to
host robust edge states, even though the Chern numbers of both of its
quasienergy bands vanish \cite{kitagawa2012topological}.
The bulk topological invariant explaining the presence of the edge
states is a winding number, the
RLBL-invariant \cite{rudner2013anomalous,asboth2015edge,asboth2017spectral,sajid2018creating}
we will call $\nu$.
Since spatially constant vector-potential-like parameters
$\alpha,\beta$ can be gauged away, we can conclude that our quantum
walk has a winding number that depends only on the values of
$\theta_1$ and $\theta_2$.
The invariant follows a harlequin pattern \cite{asboth2015edge}, which
can be put in a concise (although somewhat obfuscated) formula,
\begin{align}
  \label{eq:clean_topo_number}
  \nu &= \text{sgn} \left[ \sin(\theta_1 - \theta_2)
    \sin(\theta_1+\theta_2) \right],
\end{align}
with $\nu=0$ signaling the critical, gapless cases, where the topological
invariant is not well defined.
Since we would like to study disordered quantum walks, we
need to find a way to calculate this topological invariant with all
parameters of the quantum walk spatially dependent.

\section{Topological invariant of the disordered quantum walk using scattering theory}
\label{sec:topo_scattering}

We will compute the bulk topological invariant for disordered quantum
walks by detecting the edge states, in a scattering setup, borrowing
ideas from numerics on Floquet
systems \cite{fulga2016scattering,rodriguez2019topological,liu2020anomalous}.
Scattering theory has already been applied to one-dimensional quantum
walks to obtain topological invariants even in the disordered
case \cite{tarasinski2014scattering,rakovszky2015localization}, and
such quantized signatures have even been measured
experimentally \cite{barkhofen2017measuring}.
We will first sketch the concept and then give the concrete recipe
for numerical implementation.  

\subsection{Conceptual setup and implementation}
\label{subsec:scattering_setup}

To calculate the topological invariant, we need to (1) define a
scattering setup for the quantum walk with semi-infinite leads
attached, (2) implement periodic and open boundary conditions in the
transverse direction, i.e., implement a ``cut'' connecting the two
leads, (3) define the scattering matrix, and then (4) give a recipe
for its calculation.

\begin{figure}[!h]
 \subfigure{\includegraphics[width =\columnwidth]{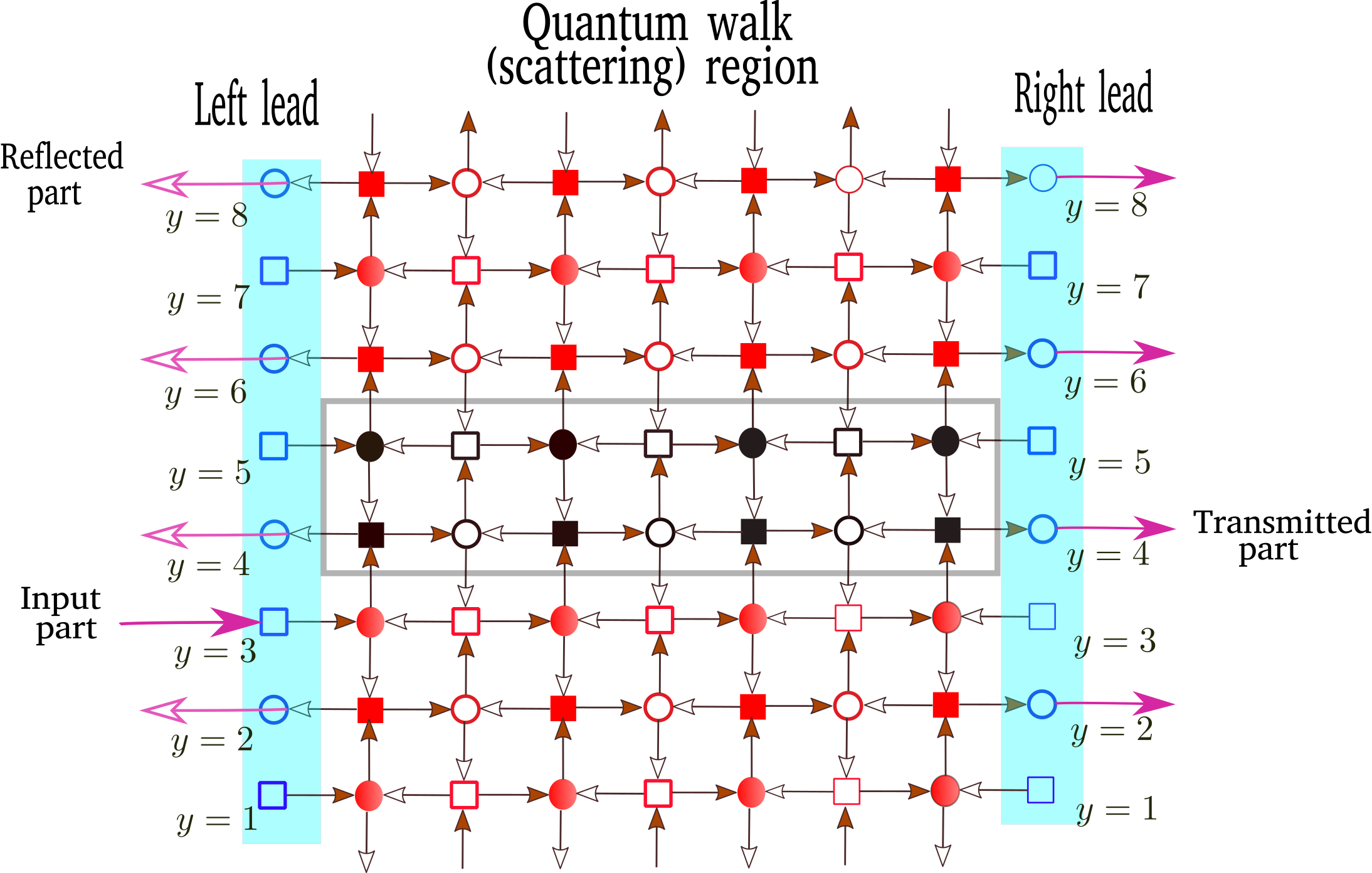}}
 \caption{Scattering setup: A two-dimensional split-step quantum walk
   takes place in the scattering region (middle, with different
   symbols denoting different sublattices), which has left and right
   leads (shaded background) attached. In the leads, rotations and
   $y$-shifts are omitted. For input from a single row as marked,
   shaded (empty) arrows denote propagation direction of $s=1$
   ($s=-1$) component. The walk can leave the system in the $s=-1$
   state in the left, or $s=1$ state in the right lead, in the marked
   rows. A cut is realized by setting the coin parameters in two rows
   in the scattering region (marked by a rectangle and black symbols)
   according to Eq.~\eqref{eq:def_cuts}. }
 \label{fig:scatter_pic}
\end{figure}

(1) We start with a scattering setup. We take a rectangle, $1 \le x
\le L_x$ and $1 \le y \le L_y$ (with $L_y$ even), as the ``system'',
and semi-infinite extensions along $x$
as the leads.
For later use, we define the projector $\hat{\Pi}_\text{sys}$ to the
system as
\begin{align}
  \hat\Pi_\text{sys} &= \sum_{x=1}^{L_x}\sum_{y=1}^{L_y}\sum_{s=\pm
  1} \ket{x,y,s}\bra{x,y,s}.
\end{align}
The left lead is at $x\le 0$ and $1\le y \le L_y$, while the right lead
is at $x > L_x$ and $1\le y \le L_y $.
In the leads, the quantum walk is simplified, with all coin parameters
set to 0, and the $y$-shift omitted.
Thus in the leads, the quantum walk just propagates the particle
right if its spin is up $(+1)$ and left if its spin is down $(-1)$.
We use periodic boundary conditions along $y$, i.e., in the
$\hat{S}_y$ shift, Eq.~\eqref{eq:Sy_def}, $y+s$ should be replaced by
$(y+s-1) \mod L_y +1$.

(2) To realize edge states, we need to define a cut region connecting
the two leads, a quantum walk equivalent of open boundary conditions.
We do this in the simplest possible way, while not breaking sublattice
symmetry \cite{asboth2015edge}: the cut is a row of sites,
$y=y_\text{cut}$ (or more rows of sites can be used to the same
effect; we used two rows in the simulation, see black sites in
rectangle in Fig.~\ref{fig:scatter_pic}), where the angle parameters
are set so that a walker impinging on the cut is certainly reflected
back from it.
We have two different choices here,
\begin{subequations}
  \label{eq:def_cuts}
\begin{align}
  &\text{cut $A$:}\quad \theta_1(x,y_\text{cut})=0,&\quad
  \theta_2(x,y_\text{cut})=\frac{\pi}{2};\\
  &\text{cut $B$:}\quad \theta_1(x,y_\text{cut})=\frac{\pi}{2},&\quad
  \theta_2(y=y_\text{cut})=0.
\end{align}
\end{subequations}
To be specific we set all other coin parameters in the cut to
$\alpha_j=\beta_j= 0$, although giving them any other value would not
have any important effect.
These two choices of coin parameters both ensure reflection---constituting a ``bulk'' with all flat
bands \cite{kitagawa2012topological}---but have different topological
invariant: $\nu_A=-1$ for the first case and $\nu_B = +1$ for the
second case.
Thus, if the bulk of the scattering region has a topological invariant
$\nu$ that is different from $\nu_\text{cut}$, with ``cut'' being A or
B, we expect chiral edge states in the bulk
propagating in opposite directions directly above and below the cut,
 $\abs{\nu-\nu_\text{cut}}$ of them in both directions. 

(3) To define the scattering matrix we need to construct scattering
eigenstates of the quantum walk and isolate their
reflected/transmitted parts.
This is done in the same way as for the one-dimensional quantum
walk \cite{tarasinski2014scattering}, and so we only give the
definition of the process here, and refer the readers to
Ref.~\cite{tarasinski2014scattering} for details.
At quasienergy $\varepsilon$ (i.e., eigenvalue $e^{-i\varepsilon}$),
there are $L_y$ scattering states, eigenstates of the quantum walk,
originating from a particle incident
from the left at position $y=n$ (with $1 \le n \le L_y$), which read
\begin{align}
  \ket{\Psi_{n,\varepsilon}} &= \sum_{t=-\infty}^\infty
    e^{i\varepsilon t} \hat{U}^t  \ket{0,n,+1},
\end{align}
where $\hat{U}$ includes the leads (which are at this point
semi-infinite).

The reflection (transmission) matrix element
$\mathfrak{r}_{mn}$ ($\mathfrak{t}_{mn}$) is the probability amplitude
of the part of the scattering state that is in the left (right) lead,
propagating towards the left (right), at $y=m$. These can be obtained
as
\begin{align}
 \mathfrak{r}_{mn}(\varepsilon) &= \braket{0,m,-1}{\Psi_{n,\varepsilon}};\\
 \mathfrak{t}_{mn}(\varepsilon) &=  \braket{L_x+1,m,+1}{\Psi_{n,\varepsilon}} .
\end{align}
The eigenvalues of the transmission matrix,
$\mathfrak{t}(\varepsilon)^\dagger \mathfrak{t}(\varepsilon)$, are the
transmission eigenvalues.
In the Landauer-B\"uttiker formalism \cite{nazarov2009quantum}, quantum
transport takes place via independent channels, and there are $L_y$ of
these in this system, since there are $L_y$ right-propagating modes in
the lead at each quasienergy.
The degree to which each channel is open for transport is quantified
by the corresponding transmission eigenvalue \cite{nazarov2009quantum}.
The total transmission at a given quasienergy is therefore the trace
of the transmission matrix,
\begin{align}
  T(\varepsilon) &= \sum_{mn} \abs{\mathfrak{t}_{mn}(\varepsilon)}^2.
  \label{eq:total_transmission_def}
\end{align}

(4) Having sketched the conceptual definition of the scattering
matrix, we need to circumvent the need for semi-infinite leads, as in
the one-dimensional case \cite{tarasinski2014scattering}.
First, note that the leads are nonreflective, and thus the part of the
wavefunction exiting the system can safely be projected out at the end
of every timestep.
For a quantum walk on an $L_x\times L_y$ rectangle, we can thus
realize the (relevant part of the) $x$-leads by a single column of
sites at $x=0, y=1,\ldots,L_y$.
We now take periodic boundary conditions along $x$ (as well as $y$),
i.e., in the $\hat{S}_x$ shift, Eq.~\eqref{eq:shifts_def}, replace $x+s$
by $x+s \mod (L_x+1)$.
However, at the end of every timestep, we first read out
the contents of the extra column $x=0$---with $s=+1$ being the
transmitted part and $s=-1$ the reflected part of the
wavefunction---and then erase it.

Summarizing all this, the recipe for the transmission matrix reads
\begin{align}
  \mathfrak{t}_{mn}(\varepsilon) &= 
           \lim_{t_{\text{max}}\to\infty} 
  \sum_{t=0}^{t_\text{max}} e^{i\varepsilon t} \mathfrak{t}_{mn}(t);\\
  \mathfrak{t}_{mn}(t+1) &= \bra{0,m,+1}
  \,\hat{U} \,(\hat{\Pi}_\text{sys} \hat{U})^t \,\ket{0,n,+1},
\end{align}
where $\hat{U}$ contains the leads of a single column of sites at $x=0$,
as described in the previous paragraph.  From this, the total
transmission at any quasienergy $\varepsilon$ can be calculated using
Eq.~\eqref{eq:total_transmission_def}.

\subsection{Topological invariants from transmission}

We can infer the bulk topological invariant from the presence and
number of topologically protected edge states at any quasienergy
$\varepsilon$ via the calculation of the transmission with and without
cut.
First, by calculating the transmission without cut, we can check if
the system is insulating, i.e.,
\begin{align}
    \text{no cut}:&\quad
  T(\varepsilon) \underset{L\to \infty}{\longrightarrow} 0,
\end{align}
where $L$ gives the scale of the system size, i.e., $L_x/L$ and
$L_y/L$ constant.
This requirement is a prerequisite for topologically protected edge
states.
Then, we calculate the transmission with either type of cut, which
will only be due to the edge states above or below the cut (whichever
carries current from left to right), 
\begin{align}
    \text{with cut A/B}:&\quad
  T_{A/B}(\varepsilon) \underset{L\to \infty}{\longrightarrow} \abs{\nu_{A/B}-\nu},
\end{align}
where we use $T_A$ to denote total transmission with cut $A$, and
$T_B$ for cut $B$.%

We can thus infer the bulk topological invariant,
\begin{align}
  \label{eq:nu_from_T}
  \nu &= \text{sgn}(T_A - T_B) \frac{T_A+T_B}{2},
\end{align}
where the dependence on quasienergy $\varepsilon$ was suppressed for
readability, and the sign function is defined to be $\text{sgn}(x)=+1$ if
$0<x$, $-1$ if $x<0$, and $0$ if $x=0$.
Note that because of the sublattice symmetry of the quantum walk, for
an input from a mode with even $n$, the transmission will only be to a
mode with odd/even $m$ (and have nonvanishing amplitudes for even/odd
$t$) for odd/even $L_x$.

\subsection{Topological invariants of quantum walks with maximal phase disorder}
\label{subsec:top_inv_disorder}

\begin{figure}[!h]
  \includegraphics[width = \columnwidth]
    {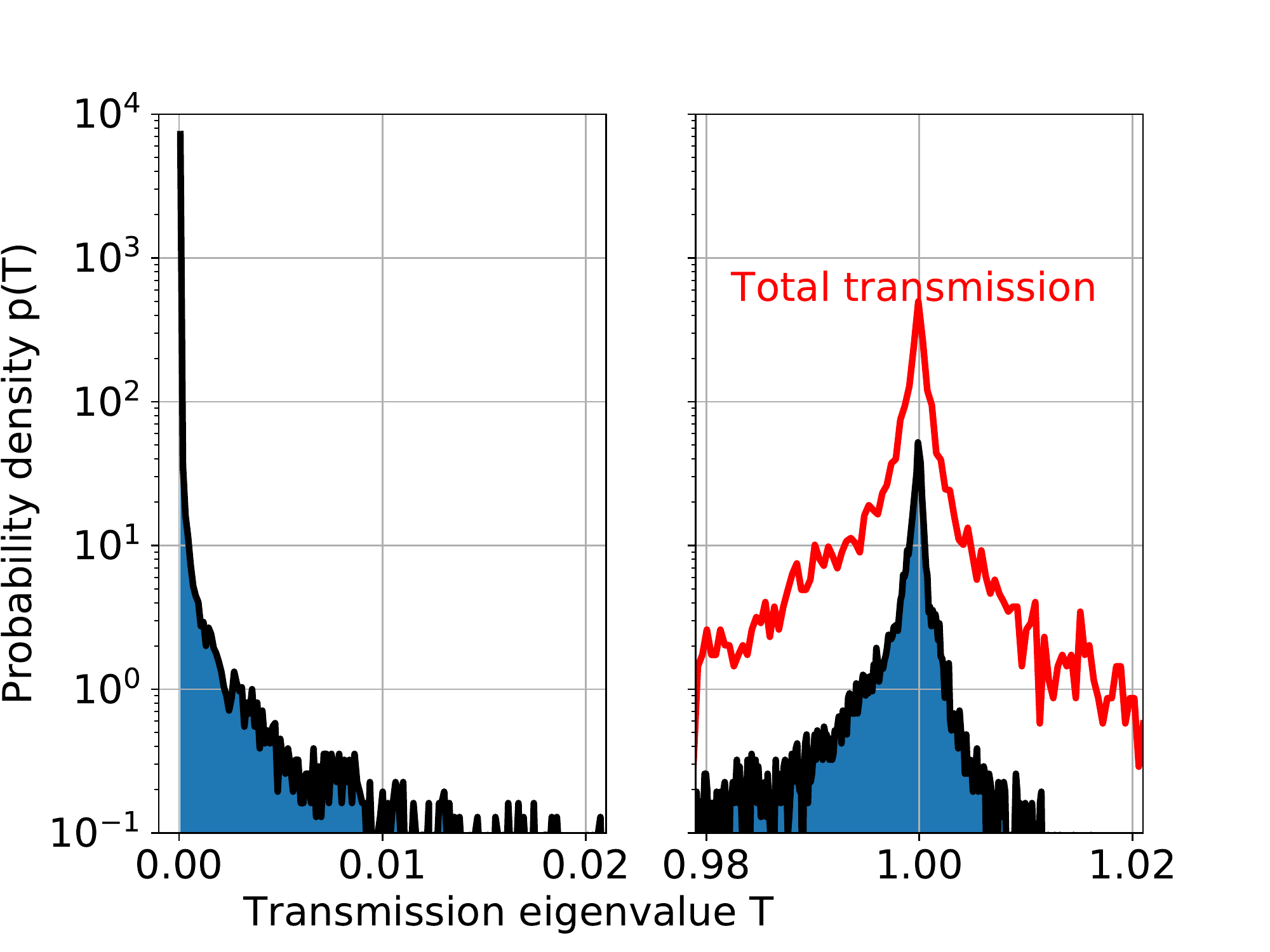}%
    \caption{Distributions of transmission eigenvalues and total
      transmission, for an Anderson localized quantum walk with
      $\theta_1=0.2\pi$, $\theta_2=0.4\pi$, complete phase disorder,
      and a cut $B$ that hosts edge states.  System size was
      $L_x=29$,$L_y=30$, and number of timesteps simulated
      $t_\text{max}=8192$.  At each quasienergy, we would expect 28
      closed and 2 open channels, with transmission eigenvalues close
      to $0$ and $1$, respectively.  The distributions of transmission
      eigenvalues fit this expectation, with 93\% below 0.01, and 6\%
      above 0.98---the distributions are shown. A small number of
      eigenvalues above 1 is a finite-time effect, their number
      decreases if $t_\text{max}$ is increased. Also shown is the
      distribution of the total transmissions (per sublattice, in red
      unfilled curve), sharply peaked around $1$.}
\label{fig:T_distributions}
\end{figure}

Maximal phase disorder---when $\phi$ in Eq.~\eqref{eq:def_F} is
chosen for each site randomly and uniformly from the interval
$[-\pi,\pi)$---randomizes the quasienergy of each
eigenstate, and therefore all quantities should be on average
quasienergy independent.
This goes for the spectrum of transmission eigenvalues, the total
transmission, even the shapes of the edge states' wavefunctions.
We illustrate this in Fig.~\ref{fig:T_distributions} by the example of
a quantum walk with $\theta_1=0.2\pi$, $\theta_2=0.4\pi$, complete
phase disorder, and a cut of type $B$.
We chose $L_x=29$, $L_y=30$, and $t_\text{max}=8192$, which after
Fourier transformation gave us 8192 quasienergy values and 30
transmission eigenvalues at each of these.
The distribution of the 8192 total transmission values is sharply
peaked around 2 (1 per sublattice), which shows that the total
transmission is quasienergy independent.
Moreover, the distribution of the more than 245000 transmission
eigenvalues reveals 93\% of them is around 0 and 6\% is around 1, so that
at every quasienergy 28 eigenvalues are almost 0 and 2 are almost 1.
This confirms that the the distribution of transmission eigenvalues is
quasienergy independent, moreover, that we have almost no
transmission across the bulk (due to Anderson localization, discussed
in Sec.~\ref{sec:phase_disorder}) and almost perfect transmission
mediated by the edge states.

With maximal phase disorder we can thus average quantities over
quasienergy, and this amounts to some disorder averaging.
This simplifies the calculation of the average value of the total
transmission, Eq.~\eqref{eq:total_transmission_def}: its
quasienergy-averaged value reads
\begin{align}
  \label{eq:average_total_transmission}
  \overline{T} &=\lim_{t_{\text{max}}\to\infty}
    \sum_{m=1}^{L_y}  \sum_{n=1}^{L_y}  \sum_{t=0}^{t_\text{max}}
       {\abs{\mathfrak{t}_{mn}(t)}^2}.
\end{align}

\begin{figure}[!h]
 ~~~~ \subfigure[]{\includegraphics[width = 0.5\columnwidth]
    {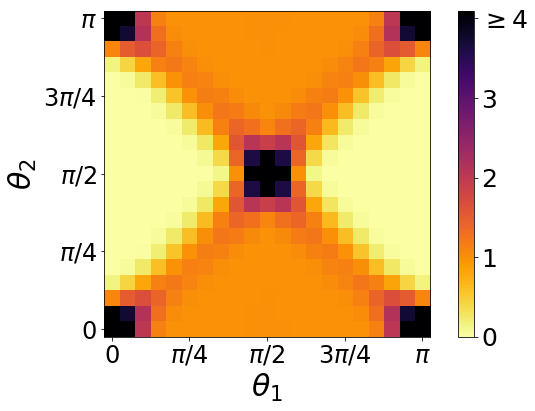}}%
\subfigure[]{\includegraphics[width = 0.5\columnwidth]
  {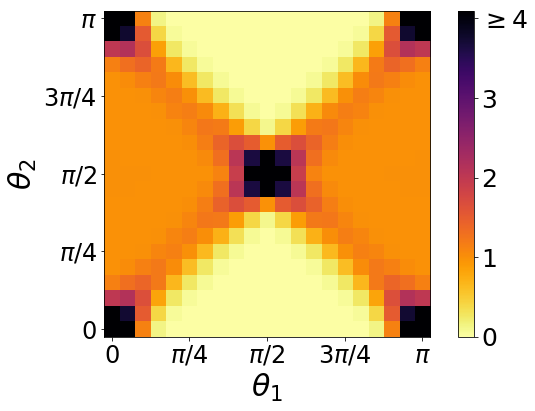}}\\
\subfigure[]{\includegraphics[width = 0.5\columnwidth]
  {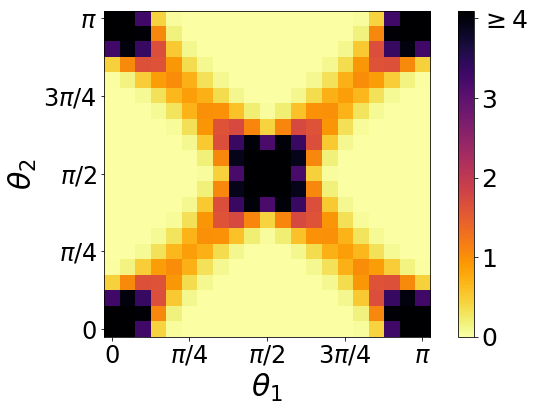}}%
\subfigure[]{\includegraphics[width = 0.42\columnwidth]
  {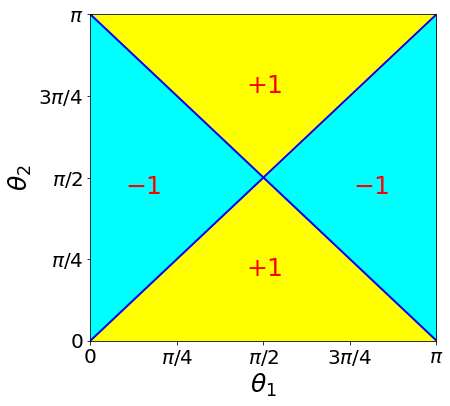}}
\caption{Half of the total quasienergy-averaged transmission
  $\overline{T}$, for two-dimensional quantum walks with
  position-independent parameters $\theta_j, \alpha_j, \beta_j$, and
  with maximally disordered $\phi$. Contribution to $\overline{T}$
  from all even input channels is shown; odd input channels have
  contribution that is indistinguishable. Results are independent of
  the magnetic parameters (set to $\alpha_1 = 0.25\pi$, $\alpha_2 =
  0.4\pi$, $\beta_1 = 0.1\pi$, $\beta_2 = 0.3\pi$).  Dependence of the
  transmission on $\theta_1$ and $\theta_2$ is shown, with (a) a cut
  of type $A$, (b) a cut of type $B$, and (c) no cut. Without cuts,
  Anderson localized phases with low transmission are separated by
  lines of large transmission. With cuts, we have quantized
  transmission of 1 (total transmission of 2) in topological phases
  where edge states contribute to the transmission.  (d) shows the
  values of the invariant for the translation invariant case,
  Eq.~\eqref{eq:clean_topo_number}, which is in good agreement from
  the combination of (a)-(c) according to Eq.~\eqref{eq:nu_from_T}. }
\label{fig:topo_trans}
\end{figure}
We illustrate how the method of obtaining topological invariants from
total transmission with and without cuts works by numerical results,
for the two-dimensional quantum walk with position-independent
$\theta$, $\alpha$ and $\beta$ parameters, but completely disordered
$\phi$.

We expect from previous work \cite{edge2015localization} that the
topological invariant should be given by
Eq.~\eqref{eq:clean_topo_number}.
We calculated transmission for $t_\text{max}=2000$ timesteps, on a
system size of $L_x = 39$ and $L_y = 60$, considering only even input
channels (odd channels have identical contribution).
The results, shown in Fig.~\ref{fig:topo_trans}, largely confirm our
expectations.
We see anomalous Floquet-Anderson insulator phases characterized by
low transmission in the case without cuts, and to a good approximation
quantized transmission in the case with cuts (total transmission of 1
for even input channels is shown, same results for odd input channels
were obtained), topological invariants matching up with
Eqs.~\eqref{eq:nu_from_T} and \eqref{eq:clean_topo_number}.
These insulating phases are separated by lines of critical states,
where transmission is high both with and without cuts.
For the cases where quasienergy-averaged total transmission was 1,
this was due to a single open channel (transmission eigenvalue
$\approx$ 1), with other channels almost completely closed
(transmission eigenvalue $\approx$ 0), as illustrated
by a concrete example in
Fig.~\ref{fig:T_distributions}.
We defer analysis of the finite-size scaling of the
total transmission near the topological phase transition to the next
section.

\section{Disorder in phase and magnetic parameters}
\label{sec:phase_disorder}

Before turning to the two-dimensional quantum walk with all parameters
random, we revisit the problem of a quantum walk with fixed $\theta$
parameters and fully random $\alpha, \beta$ or $\phi$.
When disorder is only in the phase $\phi$, this reduces to the case of
the split-step quantum walk, studied in Ref.~\cite{edge2015localization}, since position independent angle
parameters $\alpha$ and $\beta$ can always be gauged away. 
There, it was found that in the simple split-step walk, with angle
parameters $\theta_1, \theta_2$ fixed to generic values, phase
disorder leads to Anderson localization.
On the other hand, when $\theta_1, \theta_2$ is tuned to a topological
phase transition, phase disorder leads to a diffusive spread of the
wavefunction---a consequence of the critical nature of the system.

The numerical results in this section complement those of
Ref.~\cite{edge2015localization} by (1) a more in-depth analysis
of the way the wavefunction of an initially localized particle
spreads, (2) different numerical tools, the analysis of level
repulsion statistics and of the finite-size scaling of transmission.
We will also investigate the effect of having disorder not in $\phi$,
but in the vector-potential-like parameters $\alpha$ and $\beta$.
This is very similar to having onsite phase disorder but in a
spin-dependent way \cite{edge2015localization}.
Before showing our numerical results, we discuss a technical detail
that makes the simulation more efficient: a rotated basis.

\subsection{Rotated basis}

 \begin{figure}[!h]
 \includegraphics[width = 7.8cm]{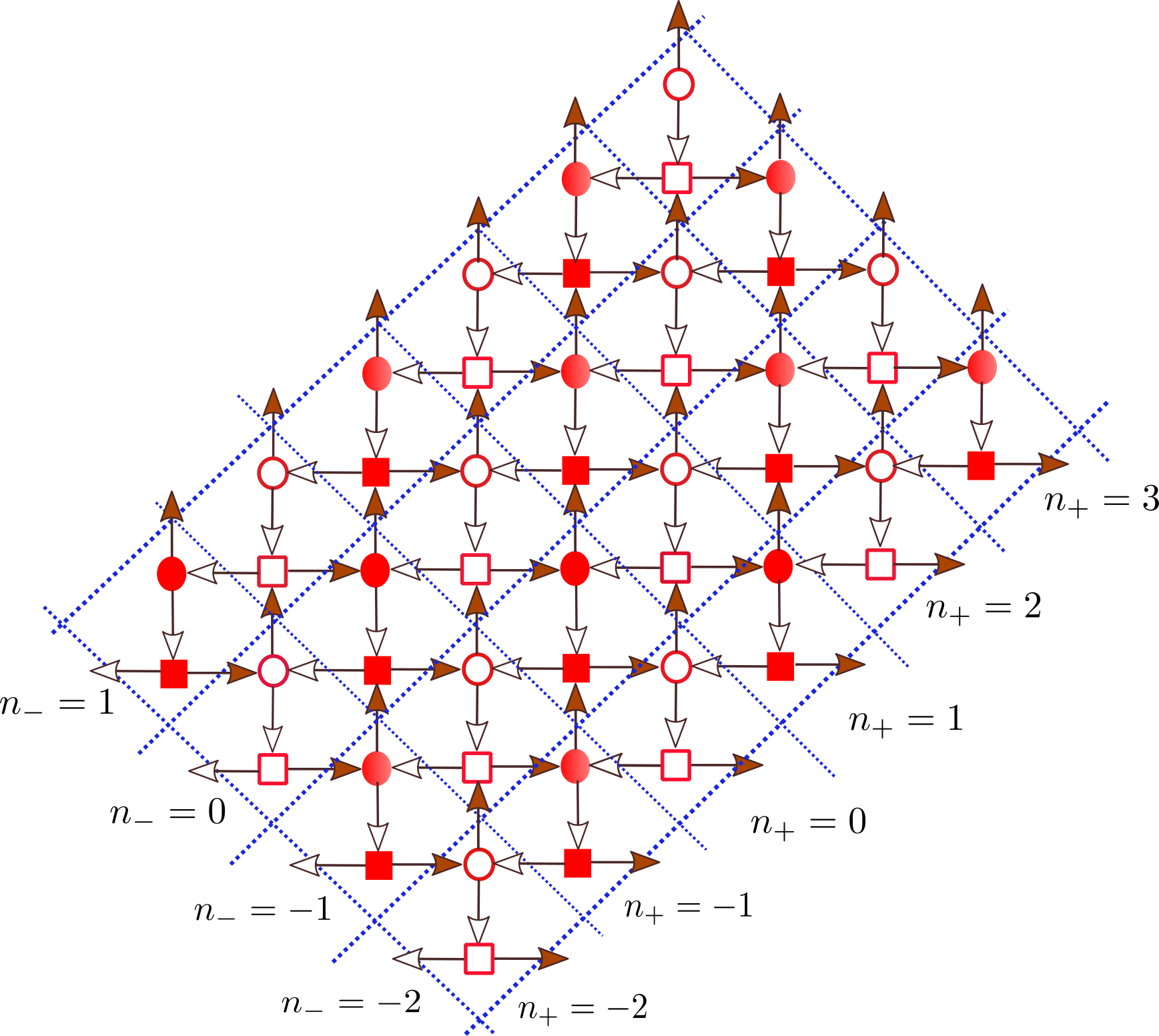}
 \caption{The rotated basis for the quantum walk, with squared-shape
   unit cells containing two sites each from different sublattices
   (different symbols), and unit cell indices $n_+$ and $n_-$
   shown. For a walk started from $x=0,y=0$, i.e., the $\square$ site in
   unit cell $n_+=n_-=0$, the walk only progresses along the solid and
   empty arrows (in the $s=1$ and $s=-1$ state, respectively).}
 \label{fig:basis_tran}
\end{figure}

For the time evolution of the quantum walk with fixed $\theta_{1,2}$,
we used a rotated basis, with square-shaped unit cells containing two
sites each from different sublattices, as shown in
Fig.~\ref{fig:basis_tran}.
Because of the sublattice symmetry, for a walk started from one
sublattice, at any time during its time evolution, its wavefunction
will only have support on at most one site per unit cell.
Thus it is enough to keep track of the integer valued unit cell
indices $n_+$ and $n_-$, and discard the sublattice information.
On the $(e,e)$ and $(o,o)$ sublattices, these unit cell indices, and
the corresponding coordinates $x_\pm$, are related to the integer valued site
coordinates $x,y$ by 
\begin{align}
  n_\pm &= \frac{y\pm x}{2};&
  x_\pm &= \frac{y\pm x}{\sqrt{2}}.
\end{align}
When rewriting the timestep operator of the quantum walk in this
rotated basis, we need to start with the shifts defined in
Eq.~\eqref{eq:shifts_def}.
For the quantum walk started from the $(e,e)$ or $(o,o)$ sublattice
(a square in Fig.~\ref{fig:basis_tran}), we need
\begin{subequations}
  \begin{align}
  \hat{S}_{1\square} &=  \sum_{n_+,n_-} \Big(
  \ket{n_+, n_- - 1, +1} \bra{n_+,n_-,+1} \nonumber \\
  \quad& \quad \quad + \ket{n_+-1, n_-, -1} \bra{n_+,n_-,-1}
  \Big);\\
  \hat{S}_{2\square} &=  \sum_{n_+,n_-} \Big(
  \ket{n_++1, n_-+1, +1} \bra{n_+,n_-,+1} \nonumber \\
  \quad & \quad \quad + \ket{n_+, n_-, -1} \bra{n_+,n_-,-1}
  \Big),
  \end{align}
  \label{eq:S_diag_def}
\end{subequations}
and obtain the timestep operator as
\begin{align}
  \label{eq:quantum_walk_ee}
  \hat{U}_\square &= \hat{F} \hat{S}_{2\square} \hat{R}_2
  \hat{S}_{1\square} \hat{R}_1,
\end{align}
where the parameters of the operators $\hat{F}$, $\hat{R}_j$ have to
be chosen to match the position of the walker.

For the sake of completeness, a quantum walk started from the $(o,e)$
or $(e,o)$ sublattices (circle in Fig.~\ref{fig:basis_tran}) has
modified shifts,
\begin{subequations}
  \begin{align}
  \hat{S}_{1\circ} &=  \sum_{n_+,n_-} \Big(
  \ket{n_++1, n_-, +1} \bra{n_+,n_-,+1} \nonumber \\
  \quad& \quad \quad + \ket{n_+, n_-+1, -1} \bra{n_+,n_-,-1}
  \Big);\\
  \hat{S}_{2\circ} &=  \sum_{n_+,n_-} \Big(
  \ket{n_+, n_-, +1} \bra{n_+,n_-,+1} \nonumber \\
  \quad & \quad \quad + \ket{n_+-1, n_--1, -1} \bra{n_+,n_-,-1}
  \Big),
  \end{align}
  \label{eq:Scirc_diag_def}
\end{subequations}
and timestep operator, 
\begin{align}
  \label{eq:quantum_walk_eo}
  \hat{U}_\circ &= \hat{F} \hat{S}_{2\circ} \hat{R}_2
  \hat{S}_{1\circ} \hat{R}_1.
\end{align}

With the rotated basis, a factor of two is gained in efficiency for
the representation of the wavefunction, i.e., an array of size
$L_+\times L_-$ is enough for a simulated area of $L_+ \times L_-$
unit cells with $2 L_+ L_-$ sites.
Besides, we will show later that during time evolution, the
disorder-averaged position distribution is generically anisotropic,
and elongated either along the diagonal or along the anti-diagonal
direction (in the original, $x$-$y$ basis), depending on the values of
$\theta_1$ and $\theta_2$.
Thus, to minimize finite-size effects, it is more efficient to use
rectangular rather than square shaped simulation area, elongated along
the direction in which the position distribution is elongated, i.e.,
rotated by $45^\circ$.
This is straightforward to implement in the rotated basis.

\subsection{Time evolution of wavefunction}

A direct test of Anderson localization is simulation of the time
evolution of the wavefunction for a particle started from a single
site.
In case of Anderson localization, the root-mean-squared displacement
(distance from the origin, square root of the variance of position)
remains bounded, and the long-time limiting form of the
disorder-averaged probability distribution falls off
exponentially with distance from the origin.
In contrast, for a critical system, diffusive spread of the
wavefunction is expected, with position variance increasing linearly
with time, and the disorder-averaged probability distribution 
approaching a Gaussian shape.
Both signatures for both cases have been observed for the split-step
quantum walk with phase disorder \cite{edge2015localization}, although
for a limited range of parameters, with $\theta_1+\theta_2=\pi/2$.
As we show below, departing from this limitation on the rotation
angles $\theta$ changes the dynamics qualitatively, introducing
anisotropy.

For the numerical simulation, we found it advantageous to work with
absorbing boundary conditions.
Absorbing boundary conditions, as already used for the scattering
calculation in Sec.~\ref{subsec:scattering_setup}, simply means
setting the value of the wavefunction at the boundary to 0 at the end
of every timestep.
This allows us to track the finite-size error, or the ``leaving
probability'' directly by the norm of the wavefunction,
\begin{align}
  \label{eq:pleave_def}
  p_\text{leave} = 1 - \sum\limits_{n_+ = 1}^{L_+} \sum\limits_{n_- =
    1}^{L_-} \sum\limits_{s = \pm 1}|\psi_{n_+, n_-, s}
  (t_\text{max})|^2.
\end{align}
Note that the absorbing boundary conditions as defined here are a
``quick-and-dirty'' way to emulate a finite segment of an infinite
plane: in fact, a small part of the wavefunction is reflected back
from the absorbing boundaries.
Such reflections are numerical artifacts, which we can estimate
by monitoring the error $p_\text{leave}$.

\begin{figure}[!h]
\includegraphics[width = 6cm]{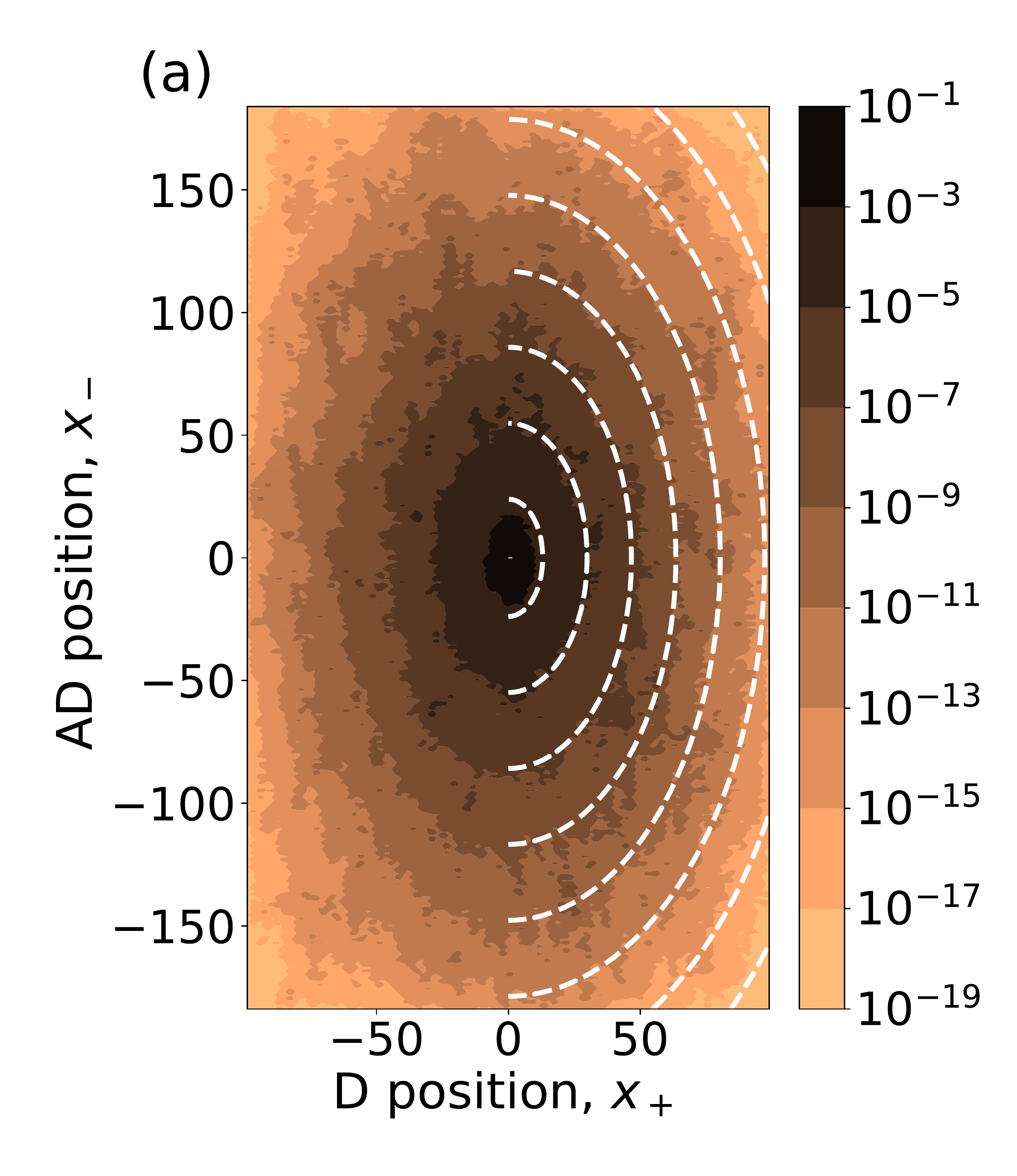}
\includegraphics[width = \columnwidth]{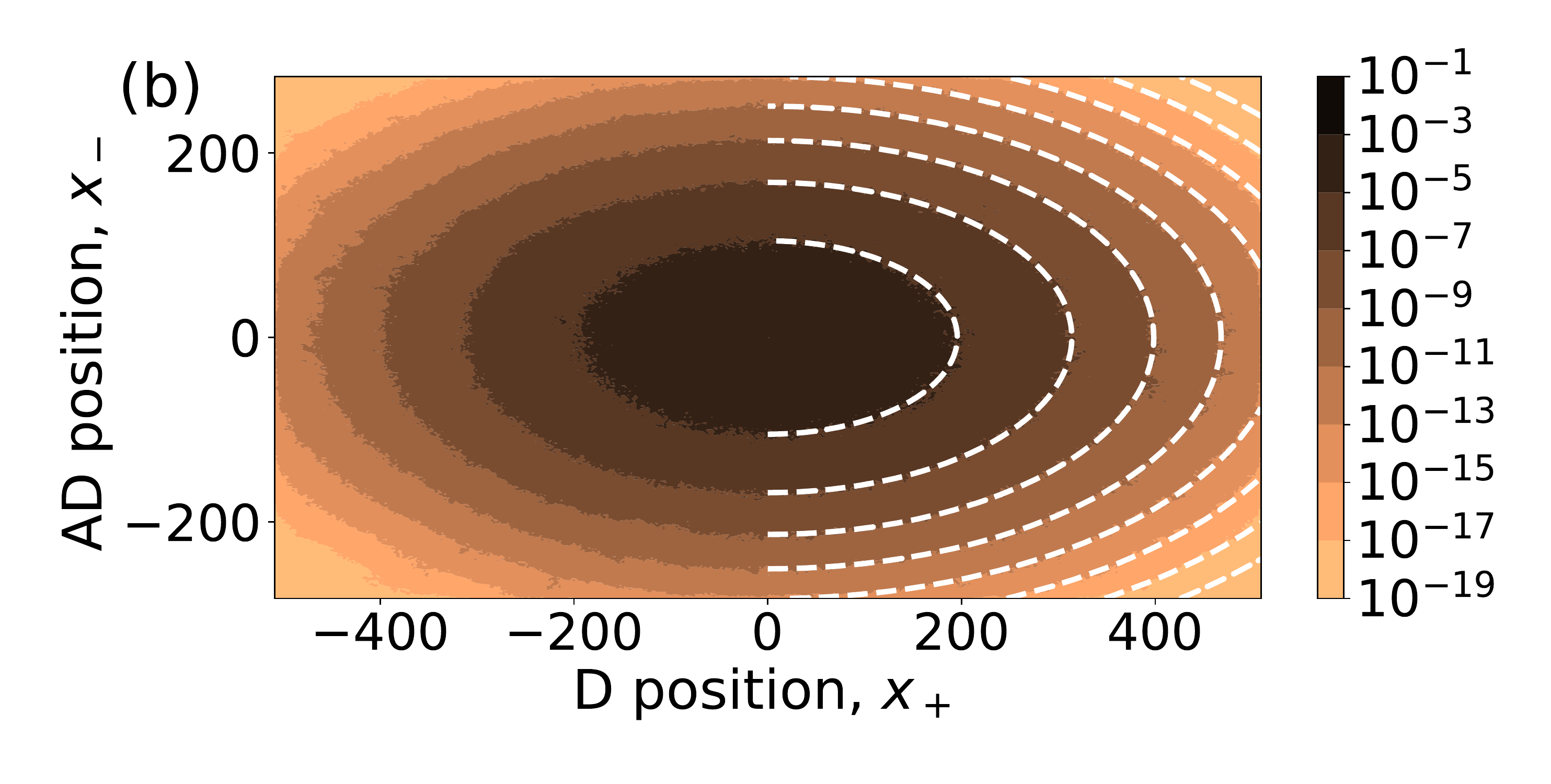}
\caption{Disorder-averaged (envelope of the) position distribution,
  for two-dimensional quantum walks with position-independent
  $\theta_j, \alpha_j, \beta_j$, and maximally disordered $\phi$,
  together with fitted curves (shown in half of the semi-log plot). In
  (a), a generic case, with $\theta_1=0.2\pi$, and $\theta_2=0.4\pi$,
  we find Anderson localization, with good fit from
  Eq.~\eqref{eq:ansatz_localized} for the average over 100 disorder
  realizations after 10000 timesteps.  In (b), with
  $\theta_1=\theta_2=0.2\pi$, we have a critical case with diffusive
  spread, with a good fit from Eq.~\eqref{eq:ansatz_diffusive} for the
  average over 10 disorder realizations after 3000 timesteps. }
\label{fig:fig1_evolve}
 \end{figure}

We have noticed that the disorder-averaged shape of the position
distribution as the wavefunction of the quantum walk spreads is
anisotropic, with contour lines having elliptical shapes, ellipses
elongated either along the diagonal ($x+y$) or along the antidiagonal
($x-y$) direction.
Typical examples are shown in Fig.~\ref{fig:fig1_evolve}.
The direction of elongation depends on the parameters
$\theta_j$, following a harlequin pattern, which can be put in a
concise (although somewhat obfuscated) formula: the dominant direction
is
\begin{align}
  x \pm y, \quad \text{with }\,\, \pm &= \text{sgn} \left[ \cos(\theta_1 -
    \theta_2) \cos(\theta_1+\theta_2) \right],
  \label{eq:dominant_axis}
\end{align}
with an isotropic position distribution if $\cos(\theta_1 - \theta_2)
\cos(\theta_1+\theta_2)=0$.
We do not have a complete understanding of why the elliptical contour
lines are not rotated by any angle other than $\pm 45^\circ$.
There are, however, two special cases that are straightforward to
check by considering two consecutive timesteps (as in Ref.~\cite{asboth2012symmetries} for a one-dimensional quantum walk).
If either $\theta_1$ or $\theta_2$ is set to $n \pi$, the quantum walk
is one-dimensional, spreading only along the diagonal.
If either $\theta_1$ or $\theta_2$ is set to $(n+1/2) \pi$, it spreads
only along the anti-diagonal (hence if one of them is $n\pi$, the other
$m\pi+\pi/2$, with $n,m\in\mathbb{Z}$, no spreading at all).
This is consistent with Eq.~\eqref{eq:dominant_axis}.

We now turn to the rate at which the position distribution spreads and
its shape in the long-time limit, to find signatures of Anderson
localization vs.~diffusion.
For generic cases, i.e., when $\theta_1 \neq \pm\theta_2+n\pi$, with
$n\in\mathbb{N}$, with maximal phase disorder, we find signatures of
Anderson localization.
The envelope of the disorder-averaged position
distribution in the long-time limit can be fitted very well with
\begin{align}
  p(x_+,x_-,t) &\propto
  \exp\left[- \sqrt{ \left(\frac{x_+}{\zeta_+(t)}\right)^2
    + \left(\frac{x_-}{\zeta_-(t)}\right)^2 }\, \right],
  \label{eq:ansatz_localized}
\end{align}
where $\zeta_+$ and $\zeta_-$ increase slowly with time (we expect
them to saturate in the long-time limit); these denote the
localization lengths along the diagonal and anti-diagonal directions,
with the variance of the position given by $\expect{x^2+y^2} \approx
3(\zeta_+^2+\zeta_-^2)$.
By ``envelope'', we mean that on every second site the wavefunction is
0 because of sublattice symmetry. 
Eq.~\eqref{eq:ansatz_localized} is an exponentially localized form,
with contours that have elliptic shapes, tilted by $45^\circ$.
We show an example in Fig.~\ref{fig:fig1_evolve}{\color{red}(a)}, where
$\theta_1=0.2\pi$, $\theta_2=0.4\pi$, and the fitted values of the
localization lengths are $\zeta_+=3.7$ and $\zeta_-=6.7$ (with maximal
disorder in the magnetic parameters as well, $\zeta_+=3.5$ and
$\zeta_-=6.8$).

For the critical cases, i.e., when $\theta_1 = \pm\theta_2+n\pi$, the
wavefunction spreads out in a diffusive way.
We find that a good fit for the envelope is
\begin{align}
  p(x_+,x_-,t) \propto
  \exp\left[- \frac{x_+^2}{4D_+ t}  - \frac{x_-^2}{4D_- t} \right],
  \label{eq:ansatz_diffusive}
\end{align}
with $D_+$ and $D_-$ denoting the diffusion coefficients along the
diagonal and antidiagonal directions,
with the variance of the position given by
$\expect{x^2+y^2} \approx 2(D_+ + D_- )t$.
We show an example in Fig.~\ref{fig:fig1_evolve}{\color{red}(b)}, where
$\theta_1=\theta_2=0.2\pi$, and the fitted values of the diffusion
coefficients are $D_+=1.1$ and $D_-=0.31$ (unaffected by maximal
disorder in the magnetic parameters).

\begin{figure}[!h]
\includegraphics[width=\columnwidth]{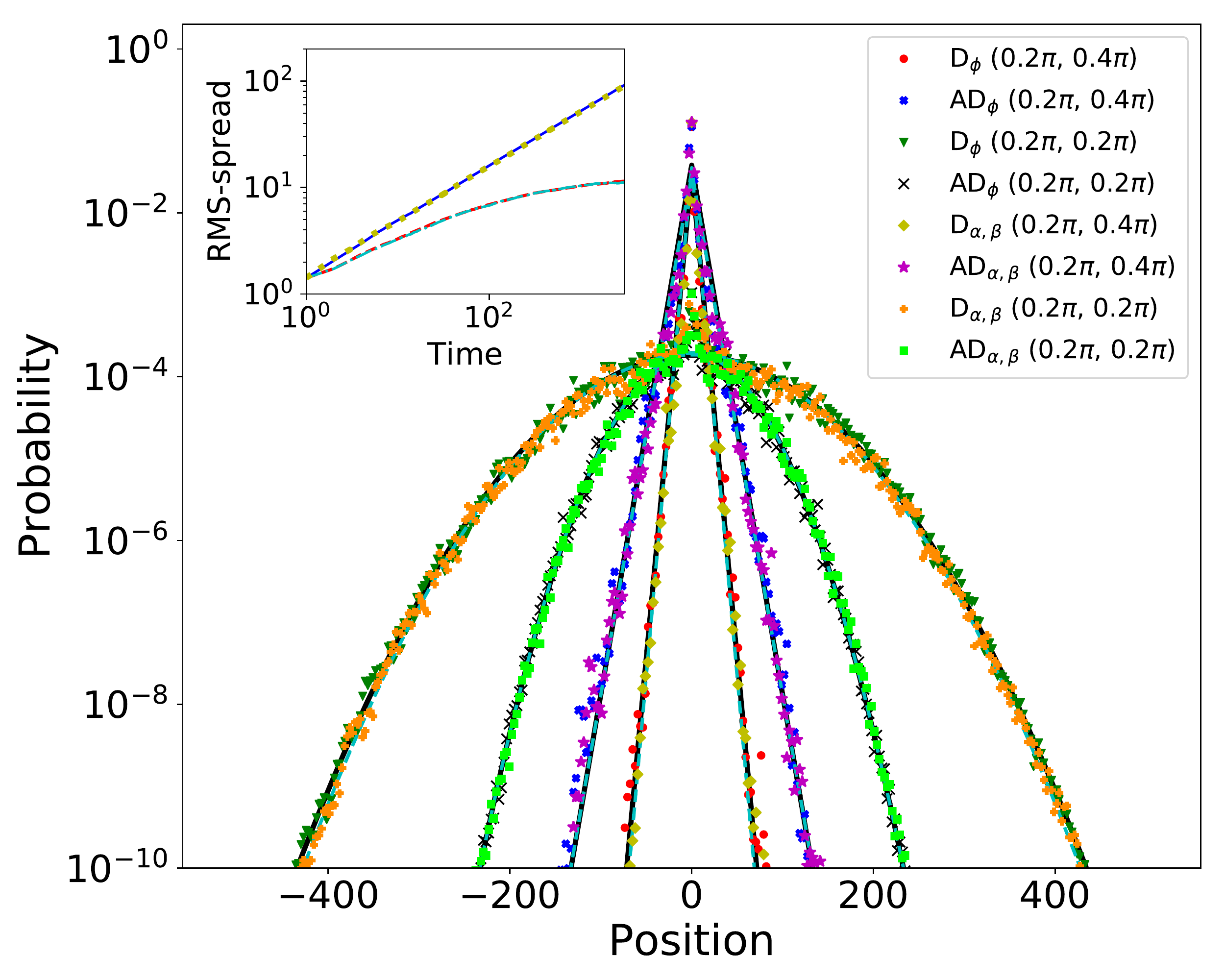}
\caption{Diagonal (D, along $x_-=0$, only even $x_+$) and antidiagonal
  (AD, along $x_+=0$, only even $x_-$) cuts of the position
  distributions of Fig.~\ref{fig:fig1_evolve} (semi-log plot). Phase
  disorder (suffix $\phi$) and disorder in magnetic parameter (suffix
  $\alpha,\beta$) give almost indistinguishable results. Continuous
  lines show the fitted theory curves---they are almost completely
  covered by the numerical data points, indicating excellent fit,
  except for the close vicinity of the origin. The inset shows the time
  dependence of the square root of the variance (log-log plot), which scales
  diffusively for the critical case and appears to approach
  saturation for the generic case.}
\label{fig:diagonal_ad_cuts}
 \end{figure}

We show more details on diagonal and antidiagonal cuts of the
disorder-averaged position distributions in
Fig.~\ref{fig:diagonal_ad_cuts}.
We took these cuts from the simulation runs represented in
Fig.~\ref{fig:fig1_evolve}.
We note that the parameters $\zeta_\pm$ and $D_\pm$ in
Eqs.~\eqref{eq:ansatz_localized} and \eqref{eq:ansatz_diffusive} were
obtained by fitting the analytical curves to these cuts.
The fits, as already seen in Fig.~\ref{fig:fig1_evolve}, are quite
good, except for the diffusive case in the vicinity of the origin,
where we observe a spike (more on this in
Sec.~\ref{subsec:time_dependence_origin}).
We also plot in Fig.~\ref{fig:diagonal_ad_cuts} the numerical results
for cases with the maximal disorder taken in $\alpha$ and $\beta$,
rather than in $\phi$; the effects of these two types of disorder are
the same, consistent with previous results \cite{edge2015localization}.
In the inset we show the measured time dependence of the
root-mean-squared width.
For the critical case, $\theta_1=\theta_2=0.2\pi$, this is a linear
function on the log-log plot, with $\sqrt{\expect{x^2+y^2}} = 1.5
t^{0.5}$, i.e., diffusive scaling.
For the generic case, $\theta_1=0.2\pi$, $\theta_2=0.4\pi$, the
variance grows slower, consistent with the expectation that it would
eventually saturate.

\subsection{Level spacing statistics}

A frequently used tool to characterize Anderson
localization/criticality is the level spacing
statistics \cite{sepehrinia2011numerical}.
As often applied in Hamiltonian systems, first an energy $E$ is fixed,
and then from each disorder realization the gap around $E$ is taken,
i.e., $\delta_j = \varepsilon_{+,j}-\varepsilon_{-,j}$, with
$\varepsilon_{\pm,j}$ denoting the first level above/below $E$ in the
$j$th disorder realization.
The ensemble of normalized level spacings $s_j$ is then defined as
\begin{align}
  s_j &= \frac{\delta_j}{\expect{\delta}},&
\end{align}
with $\expect{\ldots}$ denoting the disorder average.

In an Anderson localized system, eigenstates separated by
large distances cannot be coupled by local perturbations, therefore
their energies (in our case, quasienergies) are essentially
independent.
Thus the normalized level spacing has exponential probability
distribution (in this context also called Poissonian),
\begin{align}
\label{eq:levelspacing_localized}
  p(s) &= p_\text{loc}(s) = e^{-s},
\end{align}
with $p(s)$ denoting the probability density.
In a critical system, on the other hand, the eigenstates have extended
wavefunctions, which can be coupled by local perturbations and
hybridize, and therefore we expect level repulsion to occur.
The normalized level spacings in this case are expected to follow a
Wigner--Dyson distribution, which for the case relevant for us
(so-called Gaussian Unitary Ensemble) reads,
\begin{align}
\label{eq:levelspacing_gue}
  p(s) &= p_\text{GUE}(s) =
  \frac{32}{\pi ^{2}} s^{2} e^{{-{\frac {4}{\pi }}s^{2}}}.
\end{align}

When we compute level spacing statistics for quantum walks, we need to
pay attention to the sublattice structure (see
Sec.~\ref{subsec:sublattice}).
The timestep operator $\hat{U}$ of a walk describes two independent
quantum walks, taking place on different sublattices ($x+y=$ even vs.~$x+y=$ odd).
Thus $\hat{U}$ has two sets of energy levels whose level spacing
distributions should be calculated separately (just as in the case
of Hamiltonians with unitary symmetries): There can be no level
repulsion between levels from different sets.
To account for this, we start from the spectrum of $\hat{U}^2$, to
obtain that of $\hat{U}$, as in Sec.~\ref{subsec:sublattice}.
For a quantum walk on $2N$ sites (with $N$ even, and with proper
boundary conditions), we need to diagonalize the $N \times N$ blocks
of $\hat{U}^2$ on the $(e,e)$ sublattice (denoted by $\square$ in
Fig.~ \ref{fig:basis_tran}), and on the $(e,o)$ sublattice (denoted by
$\circ$).
From the corresponding eigenvalues of $\hat{U}^2$, namely,
$e^{2i\varepsilon^\square_j}$ and $e^{2i\varepsilon^\circ_j}$, with
$1\le j < N$, and $0\le \varepsilon^{\square/\circ}_j < \pi$, we
obtain the $2N$ independent quasienergy level spacings of the spectrum
of $\hat{U}$ as
\begin{align}
  \label{eq:delta_circ_def}
  \delta^{\square}_j &= (\varepsilon^{\square}_{(j+1) \text{ mod } N} -
  \varepsilon^{\square}_{j}) \mod \pi,
\end{align}
with $j=1,\ldots,N$, and with analogous definitions for
$\delta_j^\circ$.

Having maximal phase disorder, i.e., $\phi$ of Eq.~\eqref{eq:def_F}
equidistributed in the interval $[-\pi,\pi)$, gives us a huge boost in
numerical efficiency for obtaining the level statistics.
As already discussed in Sec.~\ref{subsec:top_inv_disorder}, we can
treat all of the quasienergies on the same footing and for a quantum
walk on $2L_+ L_- = 2N$ sites ($N$ cells in the rotated basis) obtain
$2N$ values of normalized level spacing as
\begin{align}
  s_j &= \frac{N}\pi \delta_j^\square;&
  s_{j+N} &= \frac{N}\pi \delta_j^\circ,
\end{align}
with the level spacings $\delta^\square_j$ and $\delta^\circ_j$
for $j=1,\ldots,N$ defined in Eq.~\eqref{eq:delta_circ_def}.
Thus, we obtain a level spacing ensemble of size $2N$ by only
diagonalizing two unitary matrices---a boost in numerical efficiency.
On the downside, to obtain all the level spacings we fully diagonalize
these large unitaries of size $N\times N$, where $N=L_+ L_-$, with
a numerical cost of $O(N^3)$.
This constrains us to system sizes of the order of $N\approx 20000$,
which turns out to be large enough for the cases we considered here.
We could go beyond that by repeatedly sampling the spectrum of
$\hat{U}^2$ using iterative algorithms that obtain only part of the
spectrum, such as the Arnoldi method \cite{lehoucq1998arpack} (which
works applies ideas used in the Lanczos algorithm to more general
matrices).

\begin{figure}[!h]
  \includegraphics[width = 9.3cm]{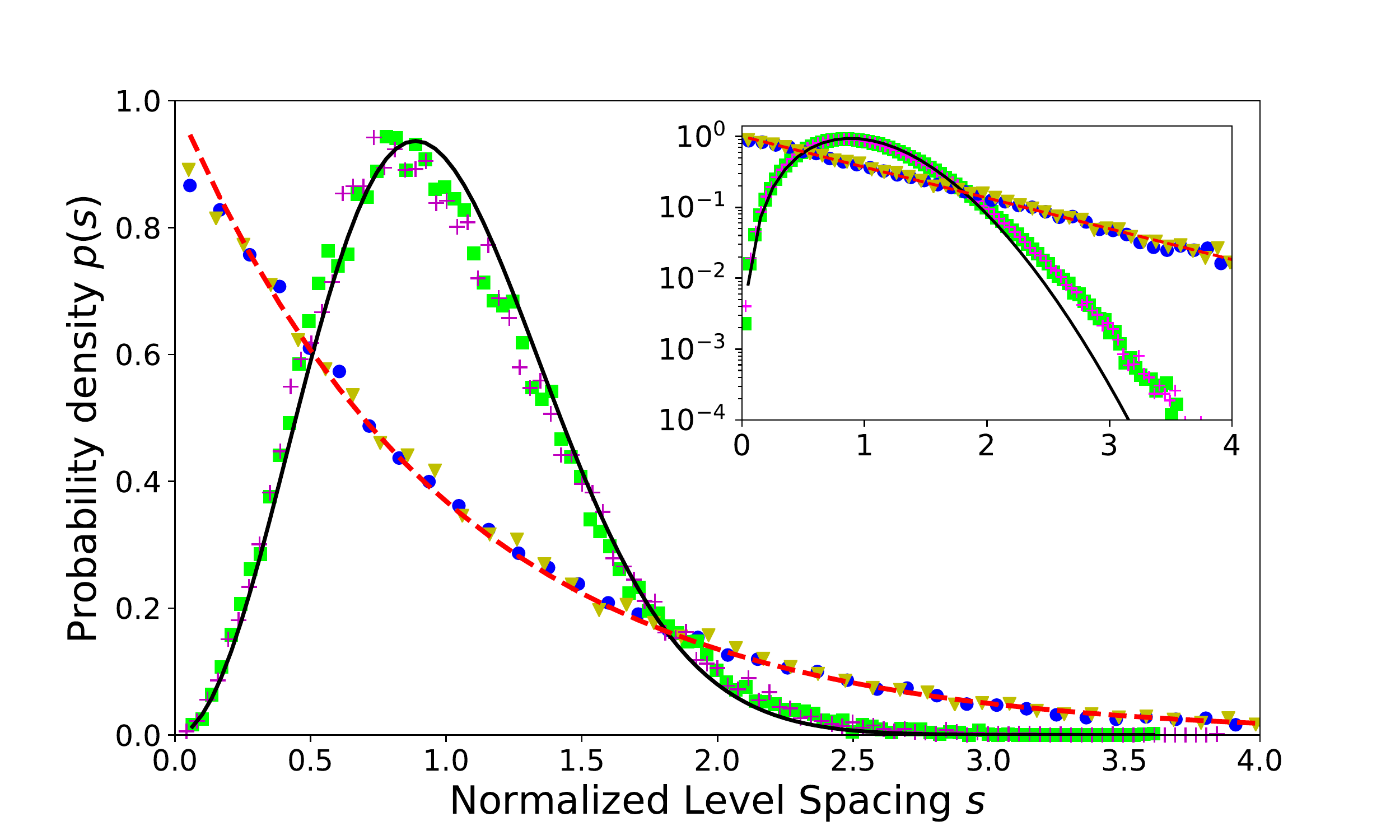}
\caption{Level-spacing distributions for the quantum
  walks of Fig.~\ref{fig:diagonal_ad_cuts}. Angle parameters
  $(\theta_1,\theta_2)$ were fixed, at ($0.2\pi,0.4\pi$)---localized---or
  ($0.2\pi, 0.2\pi$)---diffusive. For both cases, we considered
  maximal phase disorder (blue dots, green squares, respectively) or
  magnetic phase disorder (yellow triangles, purple crosses,
  respectively). For each parameter set a single disorder realization
  was used.  Calculations were done in the rotated basis with periodic
  boundary conditions, with system size optimized for numerical
  efficiency: $L_+=112, L_-=196$ for the localized cases, $L_+=204,
  L_-=110$ for the diffusive cases. Lines show the theoretical
  distributions, which fit well with the numerical data [dashed:
  Eq.~\eqref{eq:levelspacing_localized} and continuous:
  \eqref{eq:levelspacing_gue}, with no free parameters]. For better
  visibility we terminate the horizontal axis at $s=4$; in the case
  with $\theta_1=0.2\pi$ and $\theta_2=0.4\pi$, we observed even
  larger normalized level spacings, almost up to
  $s=10$. Semilogarithmic inset shows that in the diffusive case, for
  large $s$ the distribution is systematically above the Wigner-Dyson
  formula, and falls off roughly exponentially. For the inset, to have
  enough statistics in the critical case, we diagonalized 100 disorder realizations of a
  system with $L_+=136$, $L_-=74$. }
 \label{fig:levelspacings_1}
\end{figure}

We show numerically obtained level spacing distributions, and the
theoretical expectations for them, in Fig.~\ref{fig:levelspacings_1}.
Here we used the same examples as in Fig.~\ref{fig:diagonal_ad_cuts}:
a generic quantum walk with $\theta_1 = 0.2 \pi$, $\theta_2 = 0.4\pi$
and a critical quantum walk with $\theta_1 = \theta_2 = 0.2\pi$---in
both cases, results collected from a single sublattice are shown.
For both cases we considered two types of maximal disorder: disorder
in $\phi$ and $\alpha,\beta$ fixed or the other way around.
For the critical case, we see reasonable agreement between the level spacing
distributions and the Wigner-Dyson distribution of
Eq.~\eqref{eq:levelspacing_gue}, with slight deviations.
Notably, we see more instances of large level spacings than
Wigner-Dyson would predict, their number appears to fall off roughly
exponentially for $s>2.5$ rather than according to a Gaussian tail,
as shown in a logarithmic inset.
Such a departure of the level spacing statistics from Wigner-Dyson has
also been seen at critical points of the two-dimensional Anderson
transition \cite{obuse2005critical}.
For the generic, Anderson localized case, we observe a very good
agreement with the exponential distribution of
Eq.~\eqref{eq:levelspacing_localized}, with some level repulsion
showing up at the $s\ll 1$ end---states that are almost degenerate---which
we attribute to finite size effects.
Using the rotated basis for these calculations, allowed us to
adapt the shape of the simulated area to the expected shape of the
wavefunctions, and thus minimize finite-size effects.

\subsection{Scaling of transmission}
\label{subsec:scaling_transmission}

The calculation of the transmission matrix of a disordered
two-dimensional quantum walk gives us yet another numerical tool to
differentiate between Anderson localization and diffusive spread.
Scaling up the system size while keeping the system shape constant,
total transmission across an insulator should decrease exponentially,
while for the diffusive case (e.g., a metal), we expect a total
transmission that is roughly constant \cite{nazarov2009quantum}.

For the finite-size scaling of the transmission, we calculated
quasienergy- and disorder-averaged total transmissions for three
different system sizes with the same shape of $L_y/L_x=3/2$ (together
with the extra column of sites for the leads), for different values of
$\theta_1$ and $\theta_2$.
We kept $\theta_1+\theta_2=0.6\pi$, and used values of $\theta_1$ from
$\pi/10$ to $\pi/2$, thus tuning the system across a topological phase
transition which takes place at $\theta_1=\theta_2=0.3 \pi$.
For each value of $\theta_1,\theta_2$, we calculated transmission for
three different system sizes, $L_x=19, L_y=30$; $L_x=39, L_y=60$; and
$L_x=59, L_y=90$.
We used Eq.~\eqref{eq:average_total_transmission}, for the numerics,
with the number of timesteps $t_\text{max}$ chosen such that the
incident walker from any input lead will have left the system with
probability above $99\%$, namely, $t_\text{max} = 1000$ for $L_x=19,
L_y=30$; $t_\text{max} = 2000$ for $L_x=39, L_y=60$; and $t_\text{max}
= 3000$ for $L_x=59, L_y=90$.
This criterion for choosing $t_\text{max}$ can be written using
Eq.~\eqref{eq:pleave_def} as $p_\text{leave} > 0.99$, with the
conditional wavefunction
$ \ket{\Psi_n(t+1)} = \hat{U}(\hat{\Pi}_\text{sys}\hat{U})^t\ket{0,n,+1}$.

\begin{figure}[!h]
\includegraphics[width = \columnwidth]{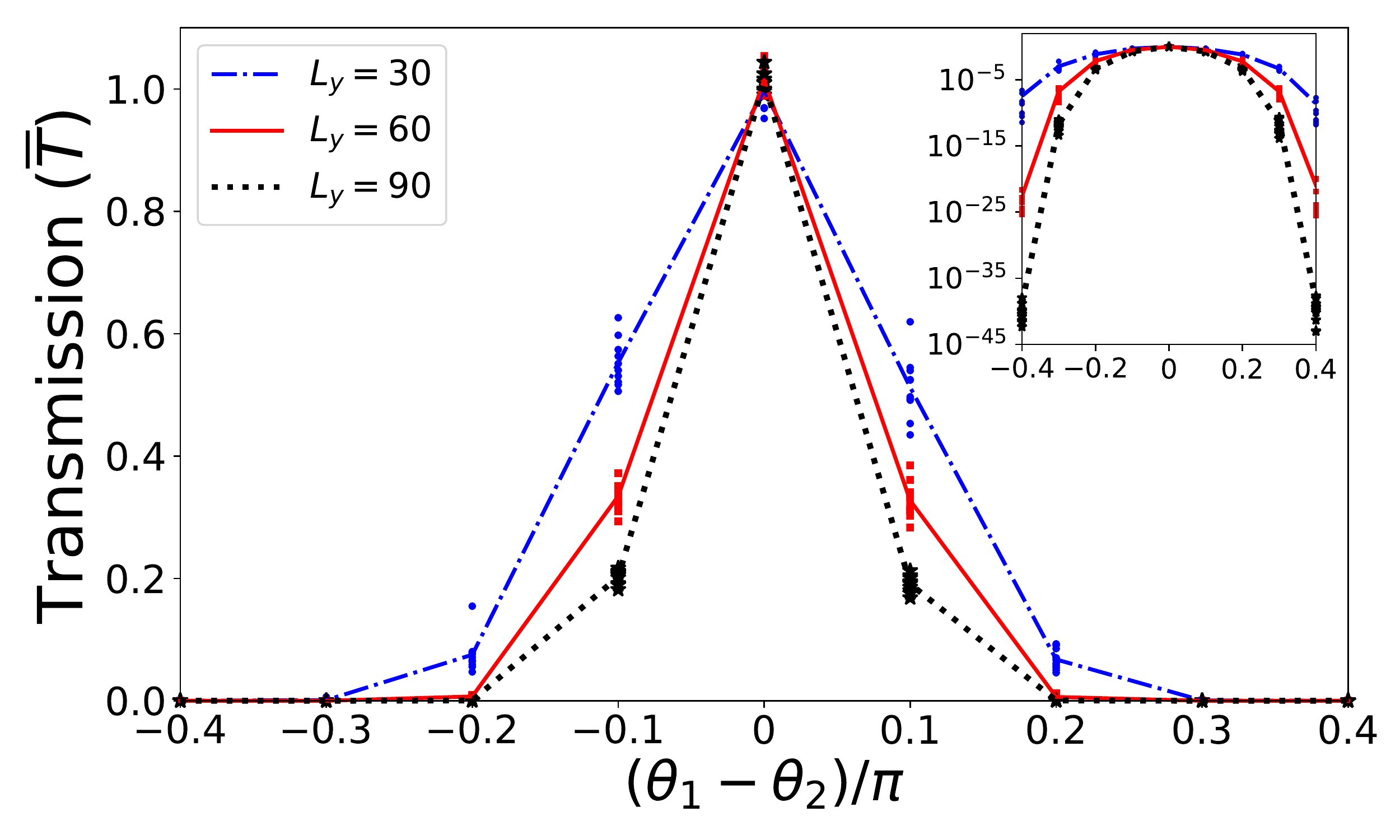}
\caption{Finite-size scaling of the
  quasienergy-averaged total transmission for quantum walks with
  maximal phase disorder but fixed $\theta_1$ and $\theta_2$. Results
  for three different system sizes are shown, with fixed aspect ratio
  $L_y/L_x=3/2$. Ten disorder realizations were used for each setting,
  with individual results shown by markers and averages by lines.  The
  inset shows the transmission data on a semilogarithmic scale, confirming
  exponential decrease of total transmission for the cases with
  $\theta_1\neq\theta_2$ as the system size is scaled up.  }
 \label{fig:transmission_phasetransition}
\end{figure}

Our numerical results for the quasienergy-averaged total transmission,
shown in Fig.~\ref{fig:transmission_phasetransition}, confirm that
tuning $\theta_1$ and $\theta_2$ indeed drives the quantum walk with
complete phase disorder across a quantum phase transition between
different localized phases.
We observe that for $\theta_1\neq\theta_2$, the total transmission
decreases exponentially (see semilogarithmic plot in the inset).
For $\theta_1=\theta_2=0.3\pi$, on the other hand, the transmission is
unchanged as the system size is scaled up, a numerical signature of
diffusive transport.

\section{Haar random quantum walk is diffusive}
\label{sec:haar_disorder}

We finally turn to the question raised in the title: Is the completely
disordered two-dimensional quantum walk Anderson localized or does it
spread out to infinity?
Thus, we take a two-dimensional quantum walk as in
Eq.~\eqref{eq:def_timestep}, with both rotation operators $\hat{R}_j$
chosen randomly from $U(2)$ according to the Haar
measure \cite{mezzadri2006generate}, i.e., Haar random
operators.

We can realize Haar random coins with the operators of
Eqs.~\eqref{eq:def_F} and \eqref{eq:rotation_matrix} by taking their
parameters from properly defined distributions \cite{ozols2009generate}.
The parameters $\alpha_j(x,y)$, $\beta_j(x,y)$, with $j=1,2$, and
$\phi(x,y)$ have to be uncorrelated random variables, uniform in the
interval $[-\pi,\pi)$.
The parameters $\theta_j(x,y)$ need to be generated as
\begin{align}
  \label{eq:haar_random}
 \theta_j(x,y) &= \arcsin\Big(\sqrt{\zeta_j(x,y)}\Big)
\end{align}
with $\zeta_j(x,y)$ uncorrelated uniform random in the interval $[0,1]$.

\begin{figure}[!h]
\includegraphics[width = 8.8cm]{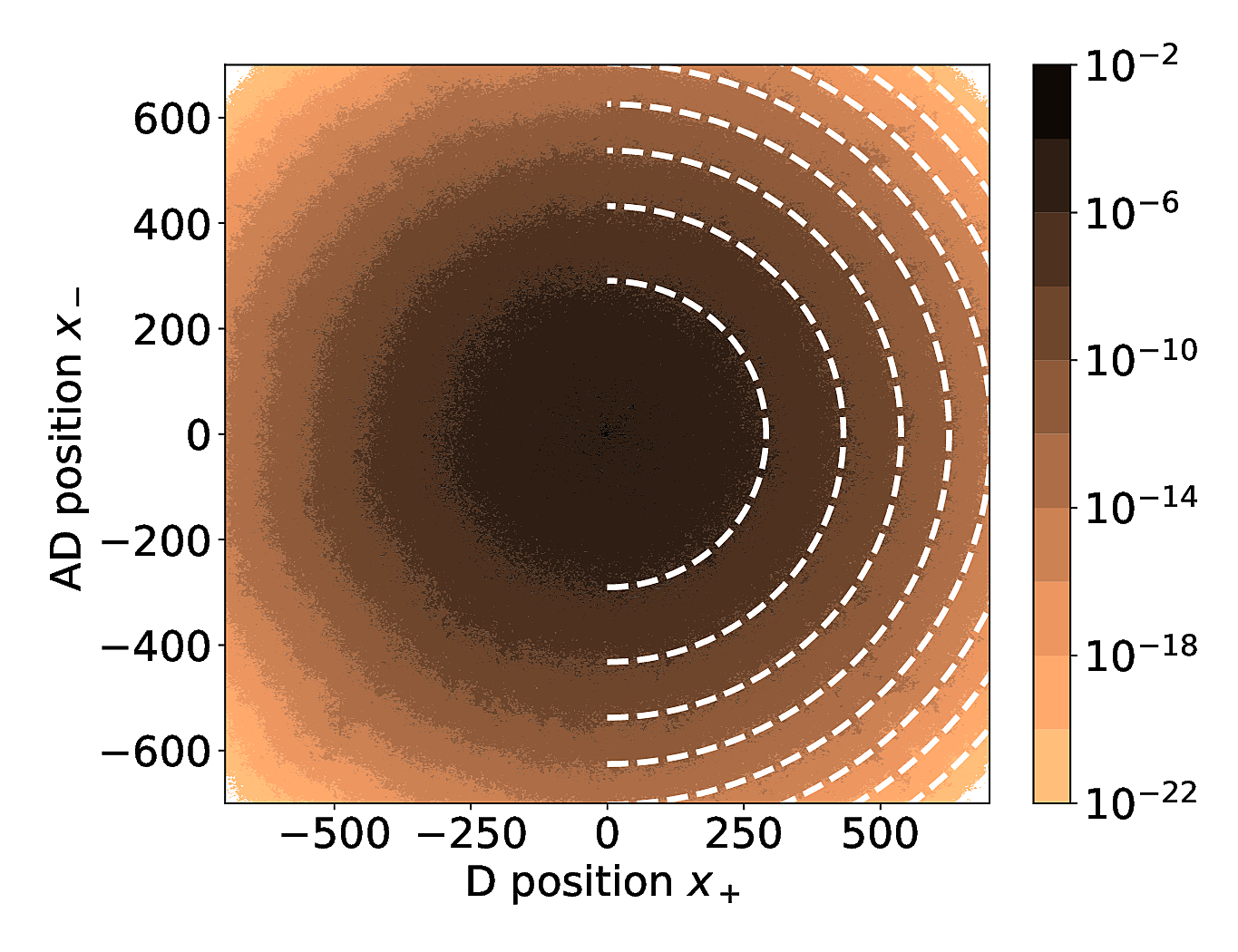}
\caption{Position distribution of the Haar random two-dimensional
  quantum walk after 10000 timesteps, with fitted theoretical
  distribution, Eq.~\eqref{eq:ansatz_diffusive} (semi-log plot). No
  disorder averaging used, result from a single disorder realization
  is shown.  Absorbing boundary conditions were used, the leaving
  probability from the boundary is $p_\text{leave} < 10^{-9}$.}\label{fig:haar_evolve}
\end{figure}

\begin{figure}[!h]
\includegraphics[width = \columnwidth]{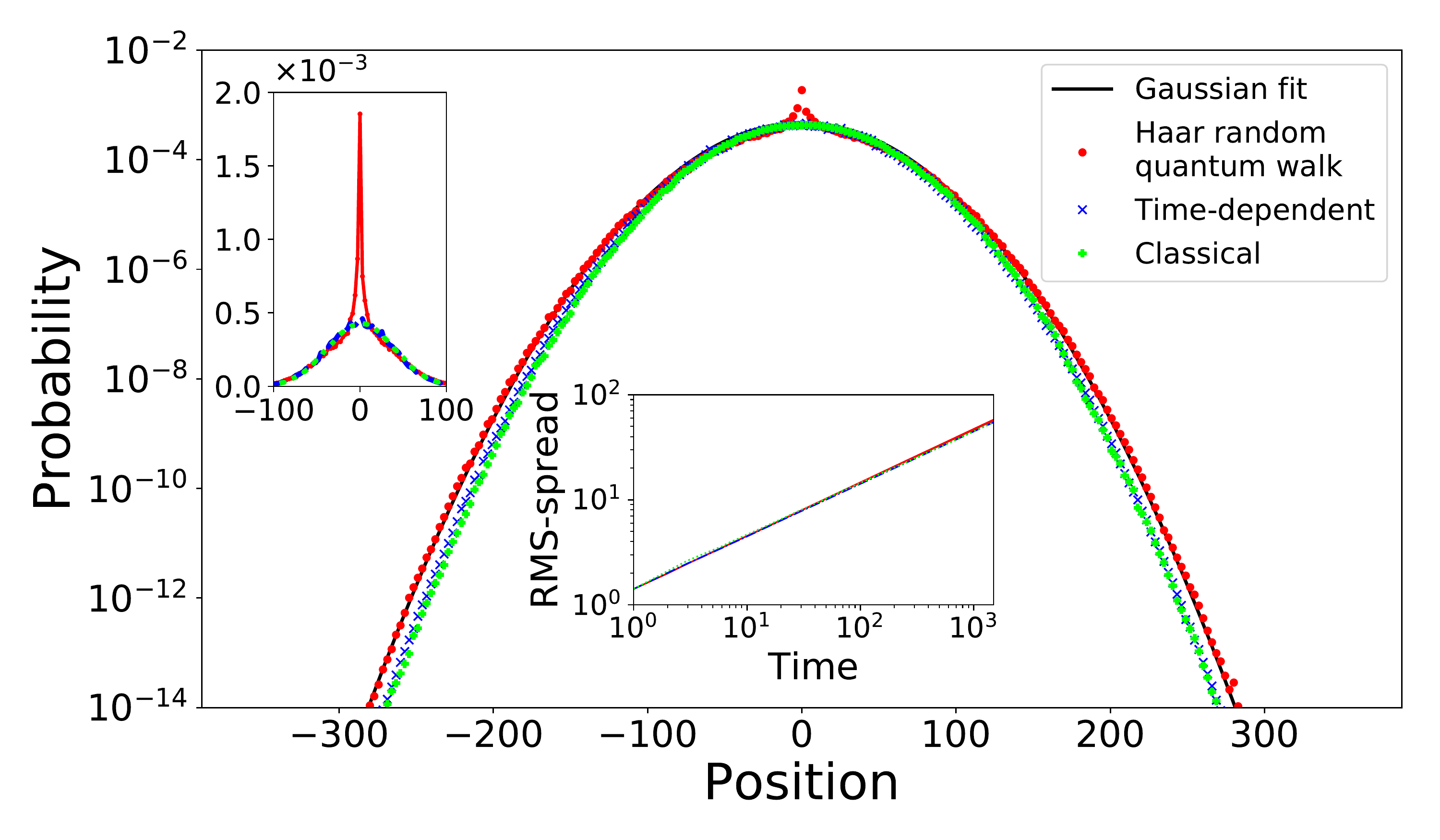}
\caption{Cross-sectional cut of the position distribution of
  Fig.~\ref{fig:haar_evolve}, with the fitted Gaussian curve (semi-log
  plot). Results for the $D$ and $AD$ cross-sectional cuts were
  indistinguishable; we show their average here. The fit is excellent,
  everywhere, except for the small local peak near the origin---shown
  in an inset.  A second inset shows the time dependence of the
  root-mean-square-width of the full position distribution (log-log
  plot), which matches well with a diffusive scaling
  $\sqrt{\expect{r^2(t)}} = 1.4\cdot \sqrt{t}$.
  %: $y = 0.51 x + 0.14$, indicates the diffusive nature.
}
\label{fig:haar_cuts}
\end{figure}

In Fig.~\ref{fig:haar_evolve} the position distribution is shown after
10000 timesteps on a system of size $1001\times 1001$.
Even without disorder averaging, the distribution is quite smooth and
roughly isotropic.
It corresponds to a Gaussian, can be fitted quite well with the
diffusive ansatz of Eq.~\eqref{eq:ansatz_diffusive}, with diffusion
coefficients $D_+=D_-=0.54$.
In Fig.~\ref{fig:haar_cuts}, we show the fit with the diffusive curve
of the cross-sectional cut of the disorder-averaged position
distribution (from 200 random realizations, after 1500 timesteps) of
the Haar random quantum walk.
It is only near the origin that the Gaussian fit is not very good; as
shown in the inset, we find here a pronounced peak, just as with the
phase disordered quantum walks of Fig.~\ref{fig:diagonal_ad_cuts}.
We analyze this peak in more detail later in Sec.~\ref{subsec:time_dependence_origin}.
The simulations of Fig.~\ref{fig:haar_cuts} were run on systems of
$501 \times 501$ unit cells in the rotated basis, with absorbing
boundary condition, error $p_\text{leave}<10^{-15}$.

We also have numerical evidence---shown in Fig.~\ref{fig:haar_cuts}---that
coherence plays almost no role in the way the Haar random
quantum walk spreads.
We show on the plot the position distributions of two classicalized
variants of the quantum walk, after the same number of timesteps, on a
same system size.
The first classicalized variant is a time-dependent Haar random
quantum walk, obtained by generating new disorder realizations of the
rotation matrices for every timestep (average of 200 random
realizations of the walk is shown).
The second classicalized variant was obtained with a single disorder
realization but with all coherence omitted from the quantum walk.
Here, we replaced the unitary timestep operator $\hat U$ of
Eq.~\eqref{eq:def_timestep}--more precisely, of
Eq.~\eqref{eq:quantum_walk_ee}---with the corresponding stochastic
operator, i.e., replaced all complex phase factors by 1 and the
parameters $\pm\sin\theta_j(x,y)$ and $\cos\theta_j(x,y)$ by
$\sin^2\theta_j(x,y)$ and $\cos^2\theta_j(x,y)$, respectively.
For this second classicalized variant, the positive-valued function
$\Psi(x,y)$ is interpreted as probability instead of probability
amplitude.
Here no disorder averaging was needed, as a single random realization
already gave results with negligible statistical fluctuations.

\begin{figure}[!h]
\includegraphics[width = \columnwidth]{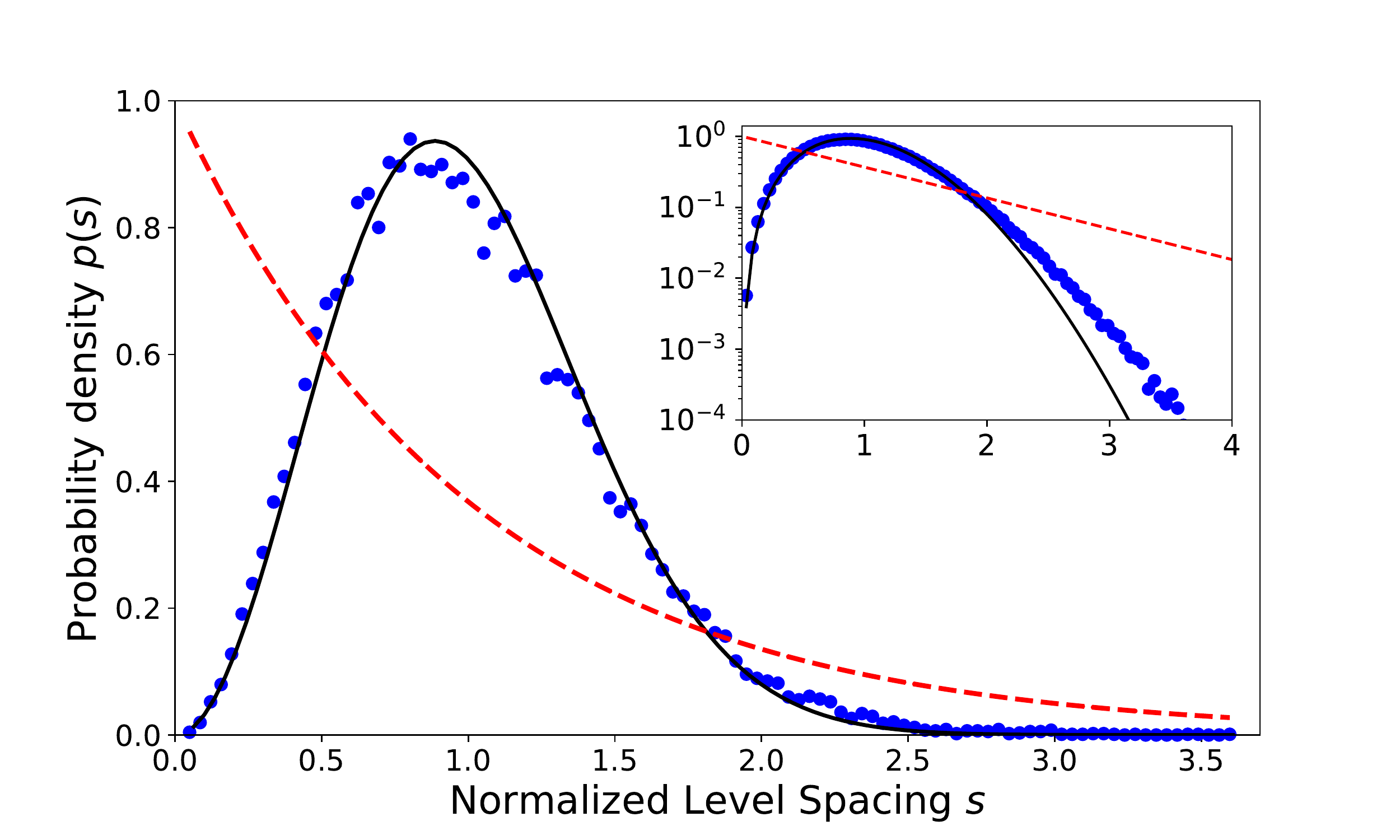}
\caption{Level spacing distribution for the Haar random
  two-dimensional quantum walk on a $160 \times 160$ square lattice,
  on the rotated basis, with periodic boundary conditions. The
  distribution shows signs of criticality; it is very close to the
  Wigner-Dyson distribution, Eq.~\eqref{eq:levelspacing_gue}, and not
  to the exponential distribution of
  Eq.~\eqref{eq:levelspacing_localized} typical for Anderson
  localization. Semilogarithmic inset shows that for large $s$ the
  distribution is systematically above the Wigner-Dyson formula and
  falls off roughly exponentially---in the same way as for the
  diffusive case in Fig.~\ref{fig:levelspacings_1}. For the inset, to have
  enough statistics, we diagonalized 100 disorder realizations on a
  system with $L_+=L_-=100$. }
\label{fig:haar_levelspacing}
\end{figure}

\begin{figure}[!h]
\includegraphics[width = \columnwidth]{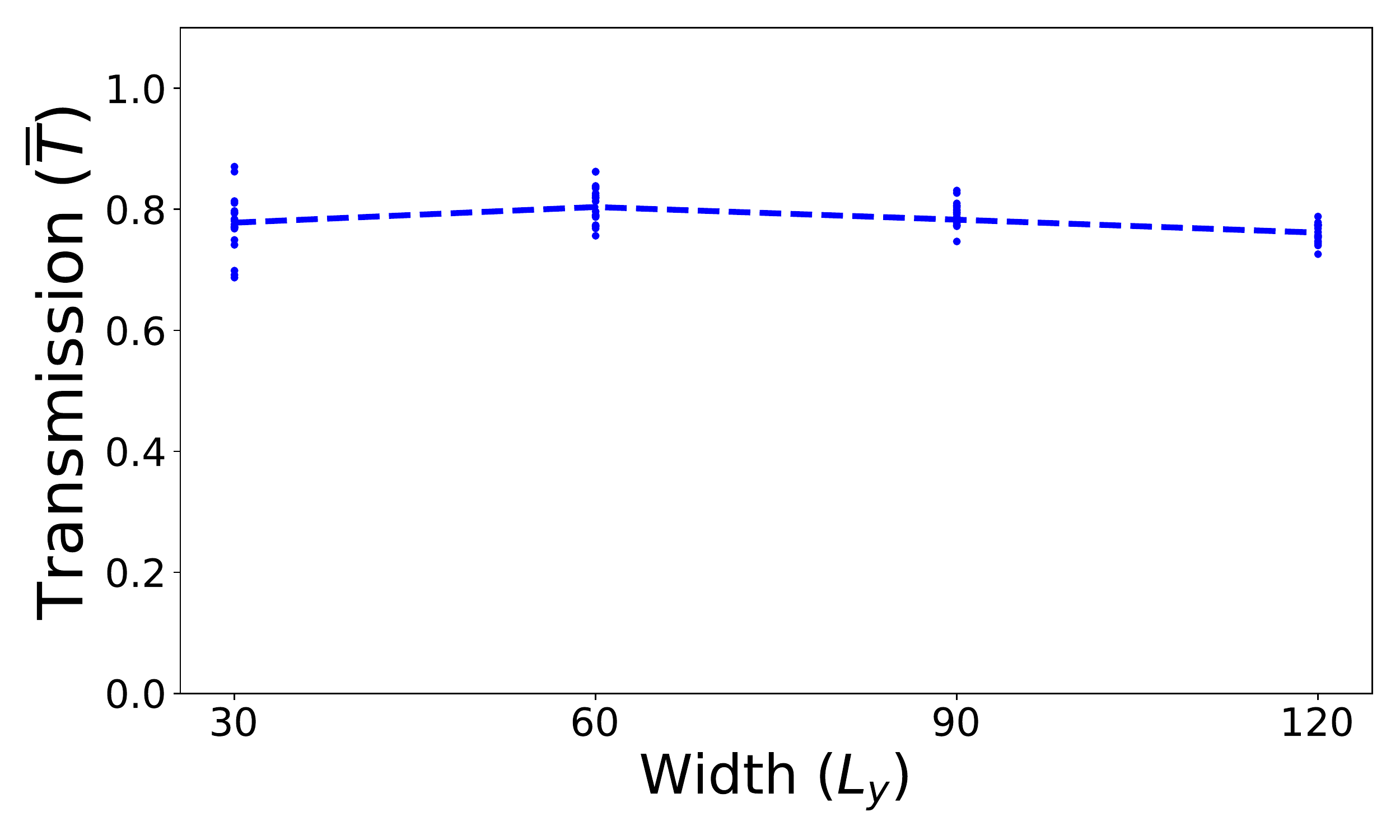}
\caption{Finite-size scaling of the
  quasienergy-averaged total transmission for Haar random
  two-dimensional quantum walk. As system size is increased by a
  factor of 4, keeping aspect ratio at $L_y/L_x=3/2$, total
  transmission is roughly a constant, indicating diffusive
  transmission. Forty disorder realizations were used for each setting;
  individual results are shown by markers and averages by lines.}
 \label{fig:Haar_transmission}
\end{figure}

Figure \ref{fig:haar_levelspacing} shows the level spacing
distribution, which matches quite well the Wigner-Dyson distribution of
Eq.~\eqref{eq:levelspacing_gue}.
As in the case of the critical quantum walk with fixed
$\theta_1=\theta_2$ (Fig.~\ref{fig:levelspacings_1}), we again
see small deviations, notably the probability of large values of the
level spacing $s$ seems to fall off exponentially with $s$, as shown
in the logarithmic inset (as in the two-dimensional Anderson transition
\cite{obuse2005critical}).
Good agreement with Wigner-Dyson indicates the presence of extended
states---or, in other words, the absence of Anderson localization.
The distribution was obtained for a single disorder realization, on a
$160\times 160$ lattice, on the rotated basis, with periodic boundary
conditions.

The quasienergy-averaged total transmission, shown in
Fig.~\ref{fig:Haar_transmission}, is roughly unchanged as the system
size is increased by a factor of 4, again indicating diffusive
transmission.
We used Eq.~\eqref{eq:average_total_transmission} for the transmission
calculations, with parameters similar to those
of Fig.~\ref{fig:transmission_phasetransition}, i.e., four different
system sizes, with integration times $t_\text{max}$ matched for error
$<0.01$, namely, 
$L_x=19, L_y=30, t_\text{max} = 1000$;
$L_x=39, L_y=60, t_\text{max} = 2000$;
$L_x=59, L_y=90, t_\text{max} = 3000$;
$L_x=79, L_y=120, t_\text{max} = 4000$.

\begin{figure}[!h]
\subfigure{~~\includegraphics[width = 8.5cm]{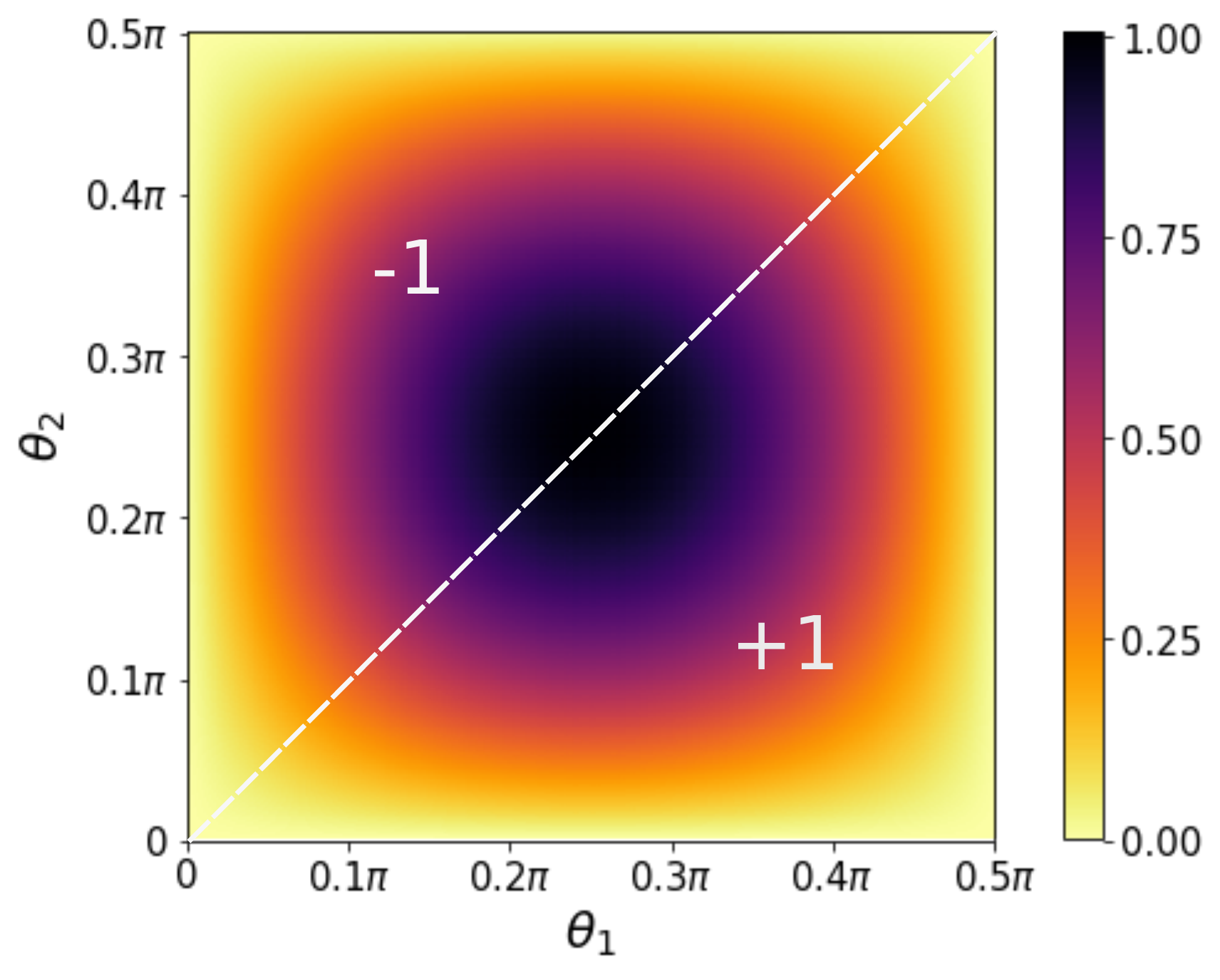}}
\caption{Probability distribution of the quantum walk parameters
  $(\theta_1, \theta_2)$, when they are selected according to
  Eq.~\eqref{eq:haar_random} to realize the Haar uniform distribution,
  with the corresponding value of the topological invariant ($\pm 1$)
  shown.  Because of the symmetry of the distribution, pairs of points
  $(\theta_1,\theta_2)$ and $(\theta_2,\theta_1)$ have the same
  probability density. These pairs have opposite topological invariant---see 
  also Fig.~\ref{fig:topo_trans}---but lead to the same
  localization lengths in the maximally disordered phase limit.}
\label{fig:haar_theta_probs}
\end{figure}

The diffusive spread of the Haar random two-dimensional quantum walk
is due to disorder-induced topological criticality.
Parameters $\alpha$ and $\beta$ have no impact on the topological
phase, which is determined solely by $\theta_1$ and $\theta_2$.
Although we are not sampling $\theta_1$ and $\theta_2$ uniformly, as
in Ref.~\cite{edge2015localization}, we are using a distribution,
using Eq.~\eqref{eq:haar_random}, shown in
Fig.~\ref{fig:haar_theta_probs}, which has the same probability
density for $(\theta_1,\theta_2)=(\theta_A,\theta_B)$ as
$(\theta_1,\theta_2)=(\theta_B,\theta_A)$, for any
$\theta_A,\theta_B$.  Thus, the qualitative argument used from
Ref.~\cite{edge2015localization}, related to network models of
topological phase transitions, still applies.

\section{Haar random quantum walk is critical}
\label{sec:haar_critical}

In the previous section, we have shown that the Haar random quantum
walk spreads diffusively, which we attributed to a topological
delocalization. Briefly summarized, we picked the coin operators in an
unbiased random way, uniformly according to the Haar distribution. Had 
we deviated from this unbiased randomness by distorting the
distribution in a generic way, we would have obtained a highly
disordered two-dimensional quantum walk, with a bulk topological
invariant (winding number) that is +1 or -1, depending on the
deviation. Thus the Haar random quantum walk is a critical case
between quantum walks that have localized bulk and topologically
protected edge states---just like the critical quantum walk of
Sec.~\ref{sec:phase_disorder}. 

We expect based on the arguments above that the Haar random quantum
walk should not only evade Anderson localization, but also display the
same type of critical behavior as the integer quantum Hall effect. In
the integer quantum Hall effect too, tuning to a transition between
two phases with different Chern numbers results in a divergence of the
bulk localization length, which is required because edge states on
opposing edges need to hybridize in order for the Chern number to
change. This rough analogy between these two topological transitions
has been precisely phrased for anomalous Floquet-Anderson insulators
using field theoretic tools \cite{woo2019quantum}.

We substantiate the topological delocalization picture by extracting a
critical exponent used in the integer quantum Hall effect from the
numerical results on the Haar random and the critical quantum walks.
This is the exponent $\eta$, which characterizes critical
wavefunction \cite{huckestein1995scaling} by (1) describing the
autocorrelation of the position distribution, (2) controlling the
fractal scaling of the second moment of the position distribution, and
(3) controlling the time decay of the probability of return to the
origin.  We use each of these three properties to extract numerical values for
$\eta$.  We have found that these approaches roughly agree and give
$\eta\approx0.52$, which is consistent with the value of $\eta$ for
the integer quantum Hall effect ($\eta=0.524 \pm 0.003$ in
Ref.~\cite{evers2008multifractality}).

\subsection{Autocorrelation of position distribution}

The first method to obtain the critical exponent $\eta$ is via the
autocorrelation function of the position distribution of the
eigenstates. For a random realization of the disorder, we obtain
eigenstates of $\hat{U}^2$ using sparse matrix diagonalization (in the
rotated basis, with periodic boundary conditions, obtained using the
Arnoldi method with shift-invert, with quasienergies
$\varepsilon\approx 0$).  Because of the sublattice symmetry, as
discussed in Sec.~\ref{subsec:sublattice}, each eigenvalue $e^{-i
  \varepsilon}$ of $\hat{U}^2$ is twice degenerate, with eigenstates
$\ket{\Psi_1}$ and $\ket{\Psi_2}$. 
We can obtain the position distribution corresponding to eigenstates
of $\hat{U}$ with eigenvalue $\pm e^{-i\varepsilon/2}$ by summing over
the two degenerate eigenstates,
\begin{align}
  p(n_+, n_-) &= \frac{1}{2}\sum_{j=1,2} \sum_{s=\pm 1} {\abs{\Psi_j(n_+, n_-,s)}}^2.
\end{align}
The autocorrelation function $R(r)$ is obtained from the position
distribution for various values of $r$, as
\begin{align}
  R(r) &= \sum_{n_+ = 1}^{L_+} \sum_{n_- = 1}^{L_-} p(n_+, n_-)
  p(n_+ + r~\text{mod}~L_+, n_-).
\end{align}

For critical wavefunctions, this autocorrelation
function decays as a power law as a function of
distance \cite{huckestein1995scaling}, with exponent $-\eta$, i.e., 
\begin{align}
  \overline{R(r)} \propto r^{-\eta}.
\end{align}
Here, the overbar denotes averaging over all eigenstates
(different disorder realizations and various quasienergy values). 

\begin{figure}[!h]
  \includegraphics[width=0.5\columnwidth]{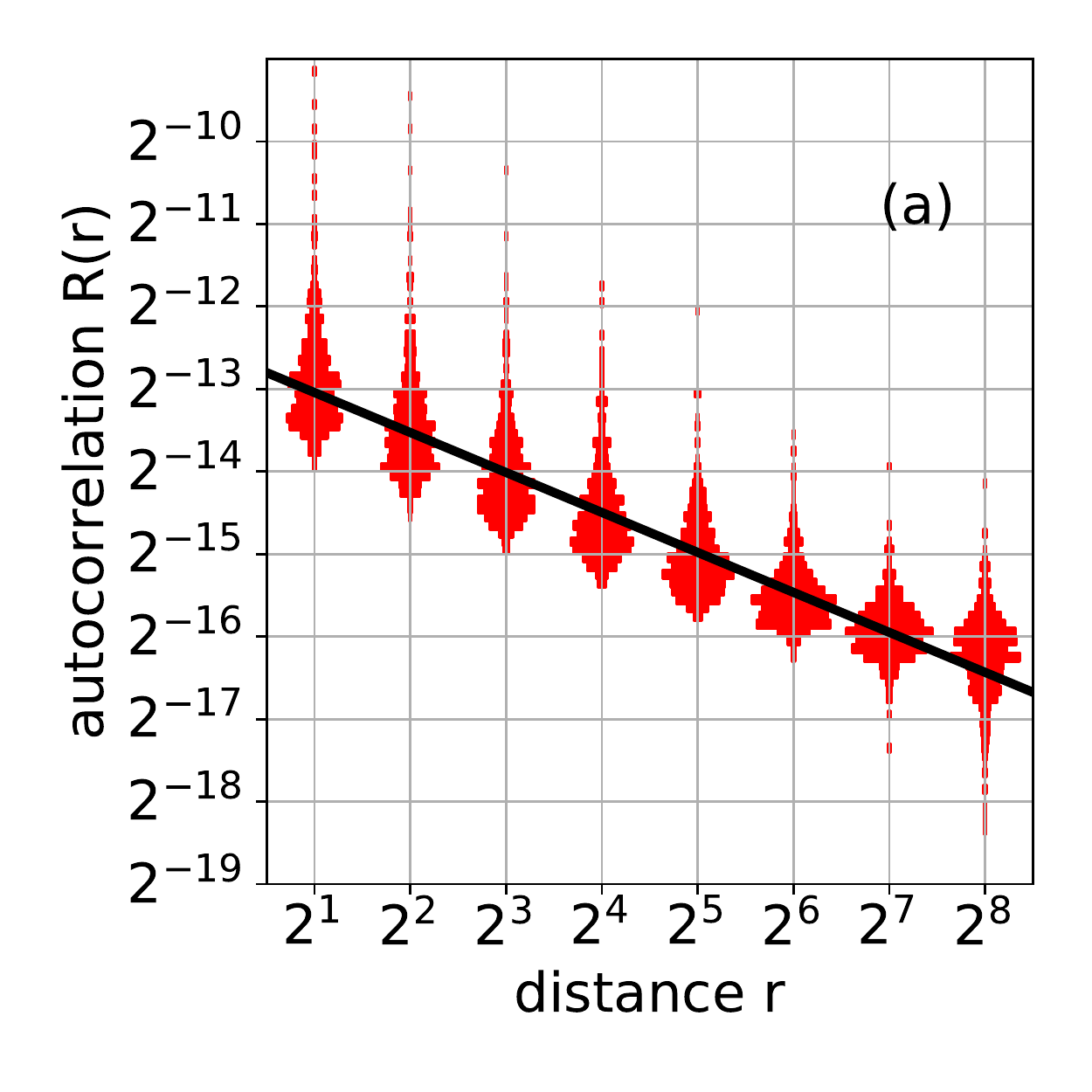}%
  \includegraphics[width =0.5\columnwidth]{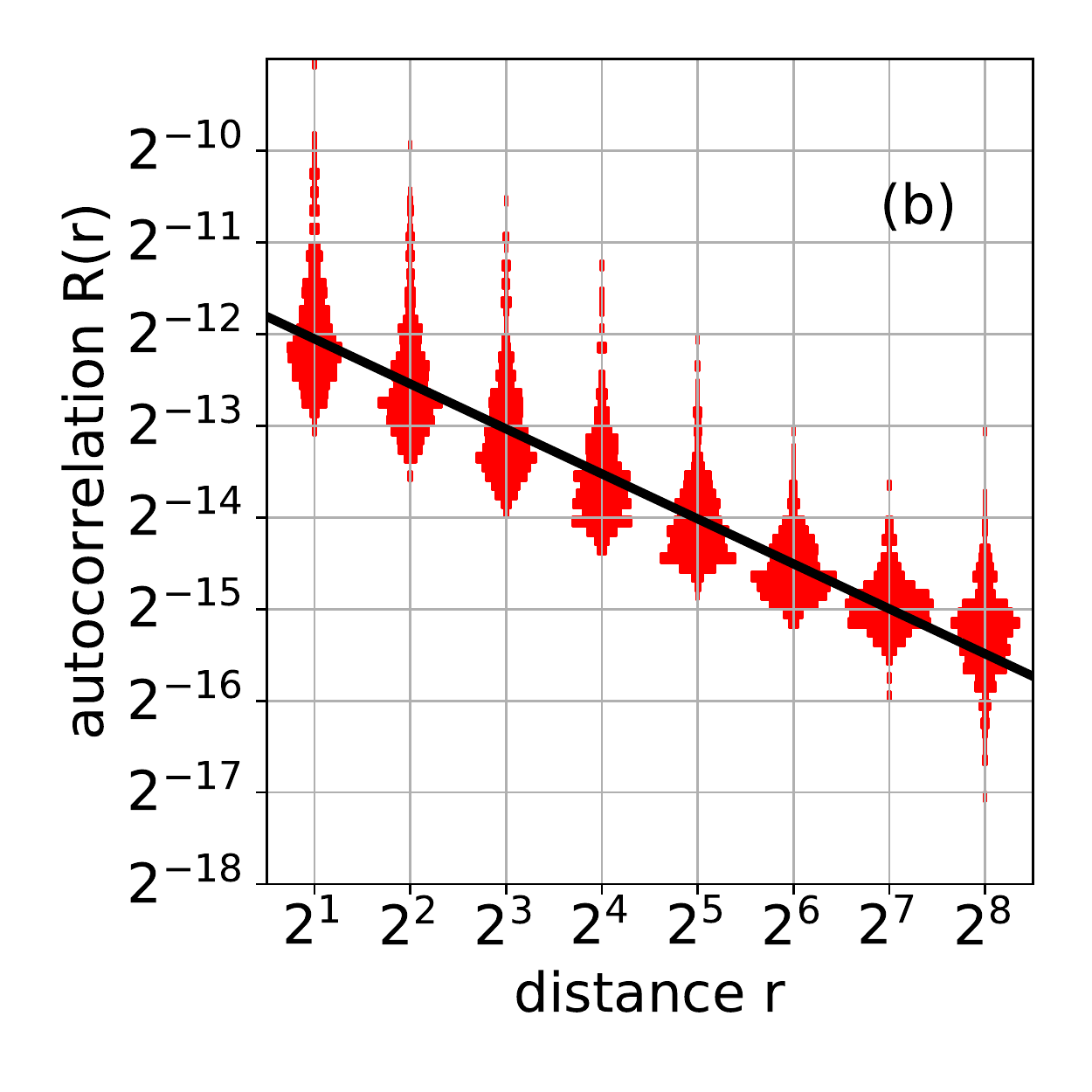}
\caption{Autocorrelation of position distributions of
  eigenstates of (a) Haar random quantum walk and (b) a critical
  quantum walk with $\theta_1=\theta_2=0.2 \pi$ with maximal phase
  disorder (disorder in magnetic parameter gave indistinguishable
  results). For both cases, 400 position distributions of eigenstates
  were calculated in the rotated basis, with periodic boundary
  conditions, on a lattice 512$\times$512 and 512$\times$256,
  respectively. Histograms of the logarithms of autocorrelation values
  for fixed $r$ (red shading) are consistent with power law decay;
  the fitted exponents are $\eta = 0.516\pm 0.02$ and $\eta = 0.510\pm
  0.02$  in the two cases.}
\label{fig:haar_autocorrelation}
\end{figure}

We show our numerical results on the autocorrelation $R(r)$ for both
the Haar random quantum walk and for a critical quantum walk with
fixed rotation angles $\theta_1=\theta_2=0.2 \pi$ and maximal phase
disorder, in Fig.~\ref{fig:haar_autocorrelation}. We calculated $R(r)$
for 8 values of $r$ from 400 position distributions (40 random
disorder realizations, for each realization 20 eigenstates giving 10
position distributions because of twofold degeneracy). We show
histograms of the logarithms of the autocorrelation values for all
values of $r$. The distributions clearly vary around a linear function
on the log-log plot, which confirms power law decay.  We estimated
$\eta$ by fitting a straight line to $\ln R$-$\ln r$. We omitted the
values $r> L_+/4$, since the distributions at $r=L_+/2$ appear to be
slightly above the expected value, influenced by the boundary
conditions (which we confirmed by calculations at smaller system
sizes).  We also omitted the two smallest values of $r$ from the fit,
to decrease the influence of lattice effects. We obtained
$\eta=0.516\pm 0.02$ and $\eta=0.510\pm 0.02$ for the Haar random
quantum walk and the critical quantum walk with fixed rotation
angles, respectively (with 99.7\% confidence level). These are
compatible with $\eta=0.524\pm 0.03$ calculated for the critical
states of the integer quantum Hall effect \cite{evers2008multifractality}.

\subsection{Fractal analysis of position distribution}

The second method to obtain the critical exponent $\eta$ uses a
fractal analysis of the position distribution of eigenstates at the
transition.  For each position distribution (obtained as detailed
above), we coarse grained using boxes (squares) of side $l$ (we chose
$l=2^j$ for integer $j$ values) and calculated averages of the
squares of the coarse grained distributions,
\begin{multline}
  \label{eq:coarse_graining}
  \expect{p^2(l)} = \frac{l^2}{L_+ L_-} 
  \sum_{m_+= 0}^{L_+ /l -1}\,
  \sum_{m_-= 0}^{L_- /l -1}
  [p_l (m_+, m_-)]^2; \\
  p_l (m_+, m_-) = 
  \sum_{n_+= m_+ l+1}^{m_+ l + l } \,
  \sum_{n_-= m_-  l+1}^{m_- l + l }
  p(n_+, n_-).
\end{multline}
For critical wavefunctions, this quantity increases with box size $l$
as a power law \cite{huckestein1995scaling},
\begin{align}
  \label{eq:scaling_fractal}
  \overline{\expect{p^2(l)}} \propto l^{4-\eta}.
\end{align}

\begin{figure}[!h]
\includegraphics[width = 0.5\columnwidth]{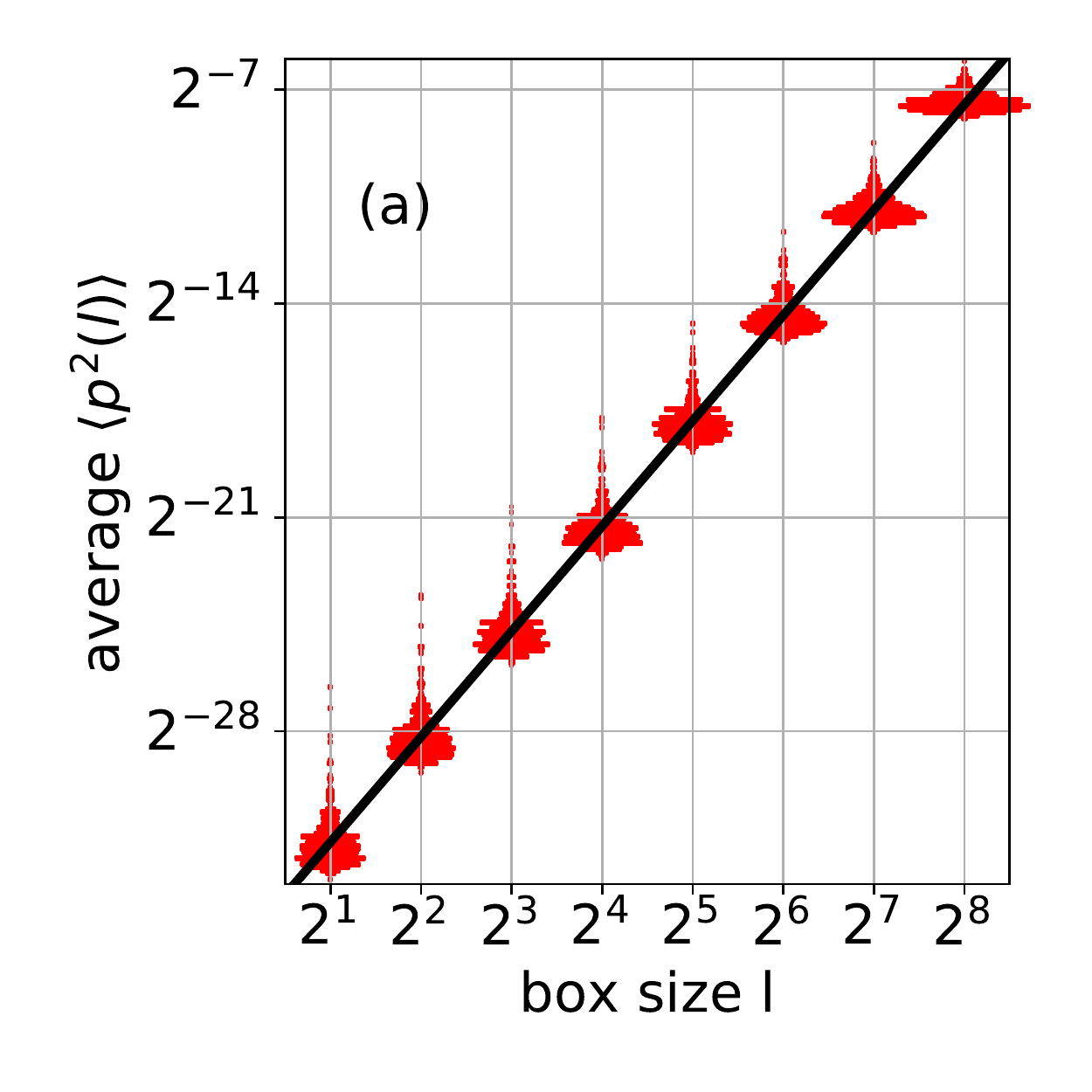}%
\includegraphics[width = 0.5\columnwidth]{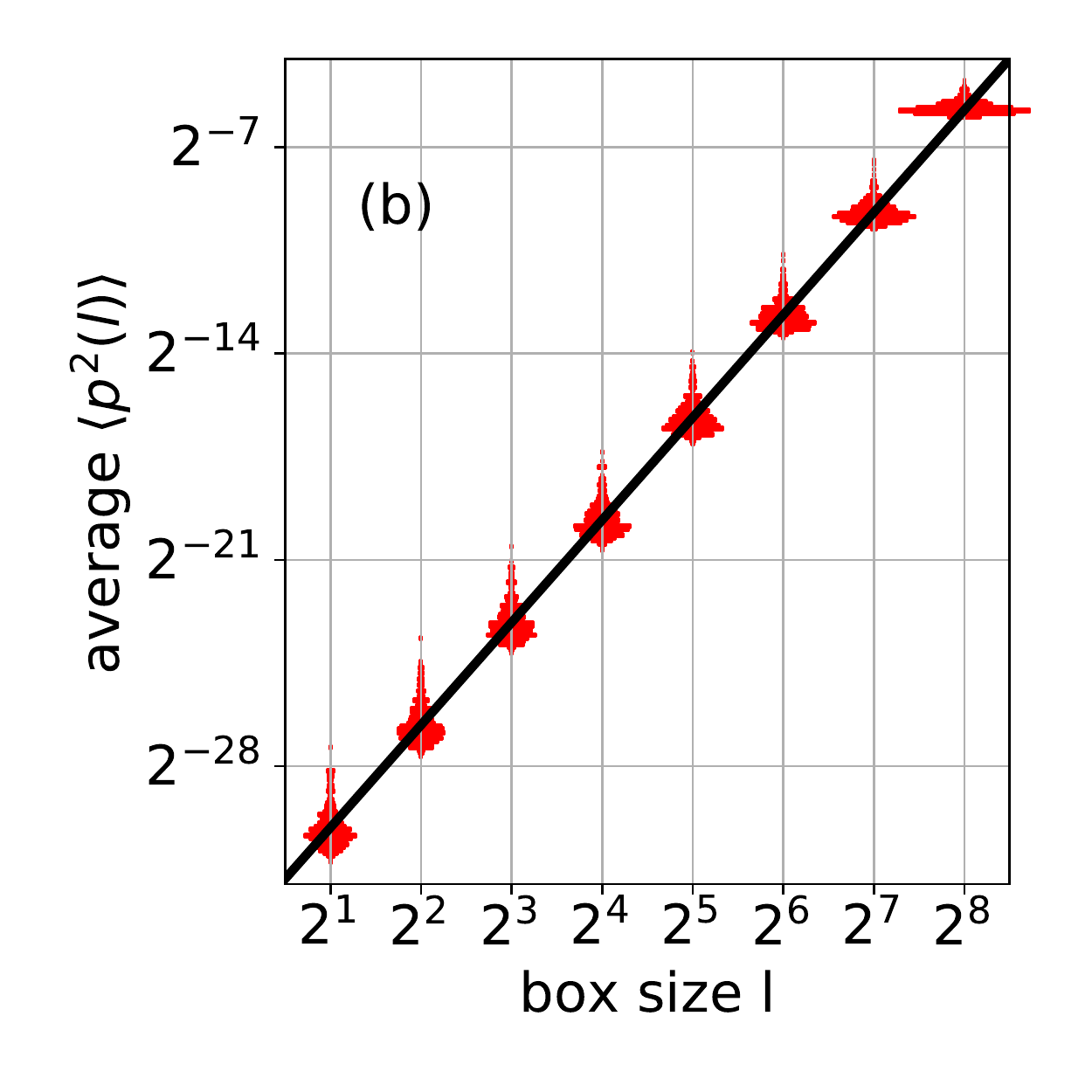}
\caption{Fractal analysis of the position distributions
  of (a) the Haar random quantum walk and (b) a critical quantum walk
  with fixed rotation angles $\theta_1$ = $\theta_2$ = $0.2 \pi$ with maximal
  phase disorder, using the same wavefunctions as for
  Fig.~\ref{fig:haar_autocorrelation}.  Averages are calculated after
  various degrees of coarse graining as per
  Eq.~\eqref{eq:coarse_graining}.  Histograms of the logarithms for
  fixed box size $l$ (red shading) are consistent with power law as
  per Eq.~\eqref{eq:scaling_fractal}; the fitted exponents give $\eta=0.55 \pm 0.02$ 
  for the Haar random and $0.520\pm 0.015$ for
  the critical quantum walk with fixed rotation angles.}
\label{fig:haar_crit_fractal}
\end{figure}

We show our numerical results on the fractality of $\expect{p^2}$ for
both the Haar random and for a critical quantum walk with fixed
rotation angles $\theta_1=\theta_2=0.2 \pi$ and maximal phase disorder
in Fig.~\ref{fig:haar_crit_fractal}. We used the same 400 position
distributions as for the autocorrelation calculation above. We show
the distributions of $\log(\expect{p^2(l)})$, which are clustered
around a straight line on a log-log plot, confirming the fractal
scaling.  The estimate of $\eta$ obtained by linear least-squares fit
is $\eta=0.55 \pm 0.02$, and $\eta=0.52 \pm 0.01$ for the Haar random
and the critical quantum walk with fixed rotation angles,
respectively, both reasonably compatible with
%$\eta=0.524\pm 0.03$
%calculated for
the value of $\eta$ in the integer quantum Hall effect.

\subsection{Time dependence of the position distribution peak at the origin}
\label{subsec:time_dependence_origin}

The third method to obtain the critical exponent is observing the time
dependence of the probability of return to the origin.  We have seen
in the previous section that on large length scale, the Haar random
quantum walk seems to spread diffusively, with the position
distribution well approximated by a two-dimensional Gaussian 
[Eq.~\eqref{eq:ansatz_diffusive}] whose width increases
linearly with time, and value near the origin decreases $\propto 1/t$.
However, close to the origin, there is a peak that can be observed.
The height of this peak is $p_0(t)$, the probability of being at the
origin,
\begin{align}
  p_0(t) = \sum_{s=\pm 1} \abs{\Psi(0,0,s,t)}^2.
  \end{align}
For a diffusively spreading probability distribution, we would expect
$p_0(t) \propto 1/t$.  For time evolution of critical wavepackets,
a power law with a different exponent is expected \cite{huckestein1995scaling},
\begin{align}
  \label{eq:scaling_return_prob}
  \overline{\expect{p_0(t)}} \propto t^{-1+\eta/2}.
\end{align}
Here we used the overbar to denote averaging over disorder
realizations, with time-evolved wavefunctions all started in the state
$\ket{0,0,+1}$. 

\begin{figure}[!h]
\includegraphics[width = 0.5\columnwidth]{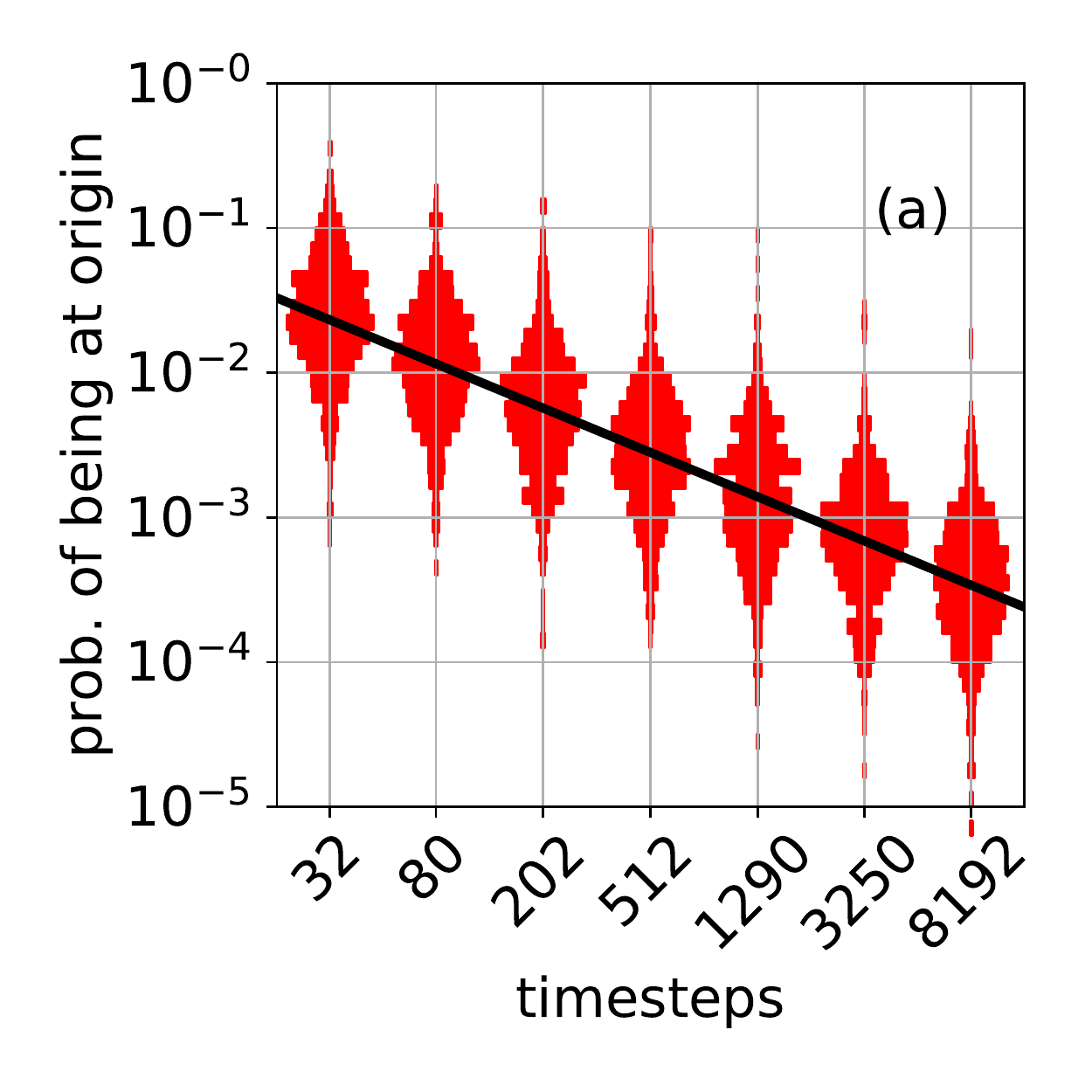}%
\includegraphics[width = 0.5\columnwidth]{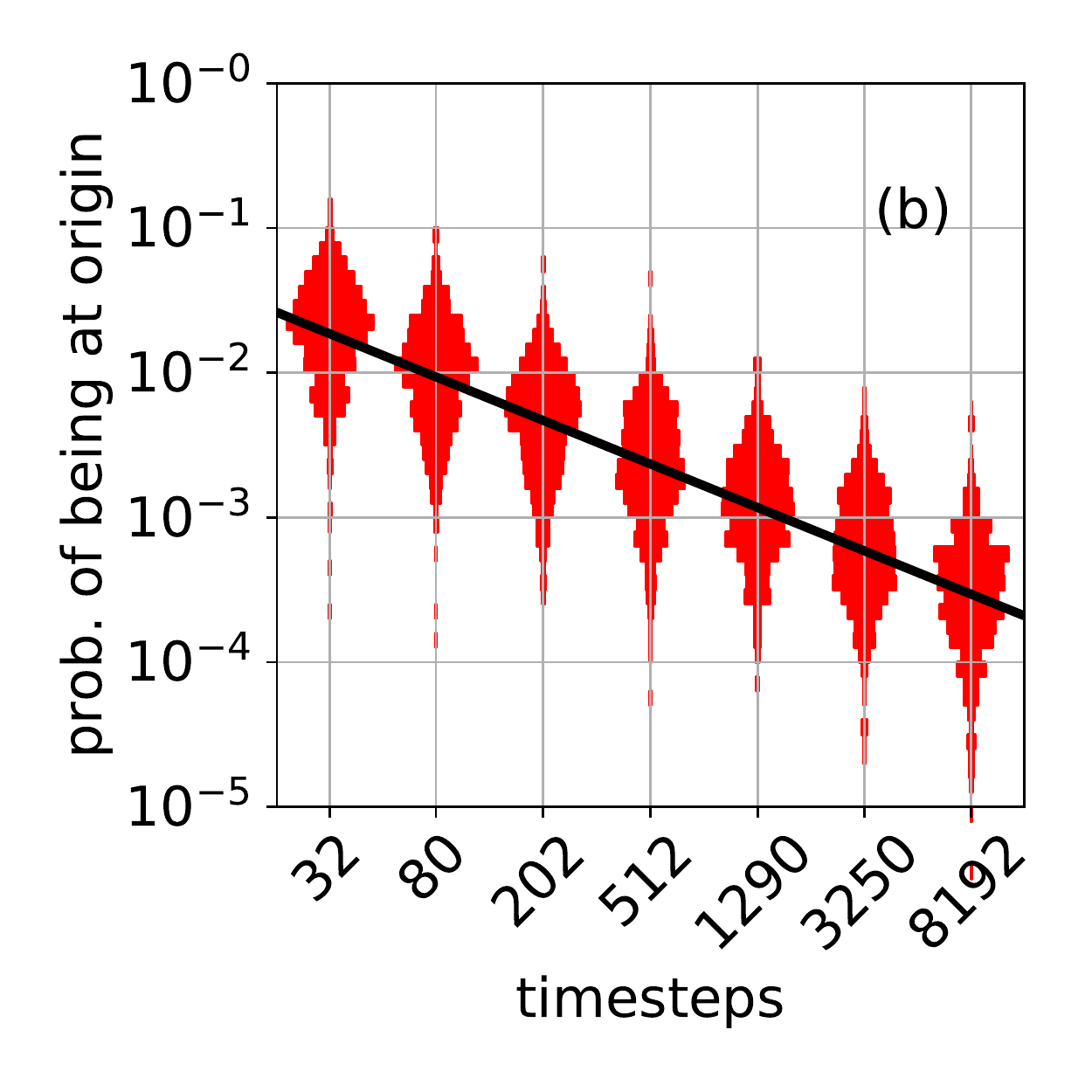}
\caption{Time dependence of probability of being at the
  origin for (a) the Haar random quantum walk and (b) a critical
  quantum walk with fixed rotation angles $\theta_1=\theta_2=0.2 \pi$
  with maximal phase disorder. Histograms of $\log(p_0(t))$ are
  consistent with power law as per Eq.~\eqref{eq:scaling_return_prob}.
  Fitted linear to the averages on the log-log plot of (a) Haar random
  quantum walk has slope $-0.76\pm 0.03$, which gives $\eta = 0.48\pm
  0.06$, (b) critical quantum walk with fixed rotation angles has
  slope $-0.75\pm 0.03$, which gives $\eta = 0.50\pm 0.06$.}
\label{fig:haar_crit_return_to_origin}
\end{figure}

We show our numerical results on the time dependence of the
probability of return to the origin (more precisely, of being at the
origin) for both the Haar random and for a critical quantum walk with
fixed rotation angles $\theta_1=\theta_2=0.2 \pi$ and maximal phase
disorder in Fig.~\ref{fig:haar_crit_return_to_origin}. 
For both cases we used 600 disorder realizations.  Distributions of $\log(p_0(t))$
are clustered around a straight line on a log-log plot, confirming the
power law decay. The estimate of $\eta$ obtained by linear
least-squares fit is $\eta=0.48 \pm 0.06$ for the Haar random quantum
walk and $\eta=0.50 \pm 0.06$ for the critical quantum walk with
fixed rotation angles. Both values are compatible with $\eta=0.524\pm
0.03$ of the integer quantum Hall effect.

\section{Discussion and conclusion}
\label{sec:conclusion}
We have studied Anderson localization and topological delocalization
in two-dimensional split-step quantum walks with two internal (coin)
states and complete phase disorder.
We first reviewed the known case of phase disorder on walks with real
coin operators, which results in an anomalous Floquet-Anderson
insulator.
Here we used the numerical tools of wavefunction spread, level spacing
distribution, and scaling of transmission.
We simulated wavefunction spread in a rotated basis, better adapted
to the observed nonisotropic shape of the position distributions
(elongated along the diagonal or antidiagonal direction in the $x$-$y$
basis).
Moreover, we calculated the topological invariants for these
disordered quantum walks using scattering theory, thus substantiating
the topological explanation of the delocalization given in
Ref.~\cite{edge2015localization}.
We then have shown how the numerical tools and analytical arguments
carry over to the case of completely disordered two-dimensional
quantum walks, i.e., with $U(2)$ coin operators taken randomly and
uniformly according to the Haar measure.
We have found that this maximal disorder does not lead to Anderson
localization but results in a diffusive spread of the quantum walk.
The absence of localization is explained by the observation that Haar
random disorder tunes the system to a critical state, between
different anomalous Floquet-Anderson phases.
We substantiated this explanation by calculating the critical exponent
$\eta$ using three different approaches and found good agreement
with the value for the quantum Hall effect.

Complete phase disorder was crucial in boosting the efficiency of
the numerical tools we used by ensuring that---statistically 
speaking---all quasienergies are equivalent.
This is quite different from the case of topological delocalization in
one-dimensional chiral symmetric quantum
walks \cite{obuse2011topological, zhao2015disordered,
  rakovszky2015localization}, which only happens at specific
quasienergy values ($\varepsilon=0, \pi, \pm \pi/2$).
We could thus interpret the time evolution of the spreading of the
wavefunction in a straightforward way and use quasienergy
averaging as a way of disorder averaging.
We have found that this holds if complete disorder is in the phase
$\phi$ or in the magnetic parameters $\alpha$ and $\beta$---as could
be expected from earlier work \cite{edge2015localization}.

It would be interesting to consider whether the topological
delocalization occurs in split-step quantum walks in higher
dimensions.
For the one-dimensional case, Haar random coins result in Anderson
localization, which has even been rigorously
proven \cite{ahlbrecht2011disordered}.
There in the absence of any symmetries (Cartan class A), no
topological invariant exists, and thus generic disorder destroys
topological phases rather than drive the system to criticality.
The same holds in every odd dimension \cite{hasan2010colloquium}, and
thus we expect Anderson localization for Haar
random split-step quantum walks in any odd dimensions.
However, in even dimensions we have the possibility of
disorder-induced topological delocalization, and it would be
interesting to find if this occurs, e.g., for four-dimensional
split-step quantum walks with two internal states.

We also wonder how sensitive our conclusions are to the number of
internal states of the quantum walk, which we set to two. 
Two internal states is the smallest number with which a discrete-time
quantum walk can be constructed, but it requires the split-step
construction for a quantum walks in two or more dimensions. 
Moving to higher number of internal states represents a challenge for
the description of topological phases because of the higher number of
coin parameters.
Moreover, with a larger internal coin space, it is quite possible that
the localization length would be significantly larger, making it
harder to observe numerical signatures of Anderson localization.

Our work also points to some open problems concerning a more complete
picture of the localization-delocalization transition in disordered
two-dimensional quantum walks.
First, we lack the precise condition for Anderson localization of
quantum walks for more general types of disorder---we have such a
precise condition for the one-dimensional
case \cite{rakovszky2015localization}).
We have made some first steps in this direction for the numerical
investigation of binary disorder; the results are in Appendix.
Second, we do not have an analytical understanding of why we obtain
diffusion, rather than sub- or superdiffusion in the critical case.
For both of these questions, mapping to network models of
topological transitions \cite{delplace2017phase}, or other theoretical
tools \cite{woo2019quantum}, could be used.

\acknowledgements

This work was partially supported by the Institute for Basic Science,
Project Code IBS-R024-D1.
A.M. thanks the Institute for Solid State Physics and Optics, Wigner
Research Centre for hospitality and financial support.
J.K.A.  acknowledges support from the National Research, Development
and Innovation Fund of Hungary within the Quantum Technology National
Excellence Program (Project No. 2017-1.2.1-NKP-2017-00001), FK124723
and K124351, and from the BME-Nanotechnology FIKP grant (BME FIKP-NAT).

\appendix 

\section*{Appendix: Localization under binary disorder in rotation angles}
\label{sec:binary_disorder}

\begin{figure}[h]
\subfigure[]{\includegraphics[width = 0.49\columnwidth]{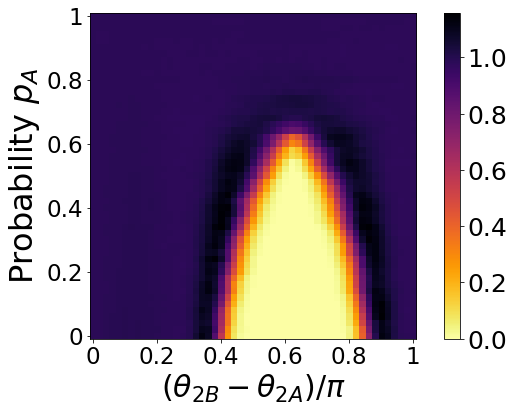}} %
\subfigure[]{\includegraphics[width = 0.49\columnwidth]{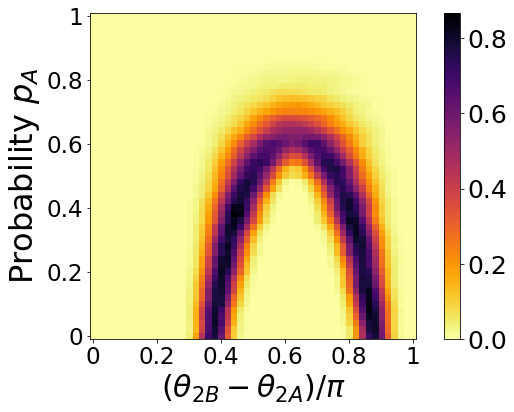}}
\caption{Total transmission $\overline{T}$ on one sublattice in a
  quantum walk with binary disorder---as described in
  Eqs.~\eqref{eq:binary_disorder_def}, \eqref{eq:binary_theta_A},
  \eqref{eq:binary_B_A}: (a) with a cut connecting input and output
  and (b) without such a cut. For each value of $p_A$ and 
  $(\theta_{2B} - \theta_{2A} )/\pi$, characterizing the binary disorder, a single
  random realization was used, on a system of size $L_x=39$, $L_y=60$,
  calculated up to $t_\text{max}=2000$ timesteps.}
\label{fig:binary_transmission}
\end{figure}

\begin{figure}[h]
\subfigure{\includegraphics[width = \columnwidth]{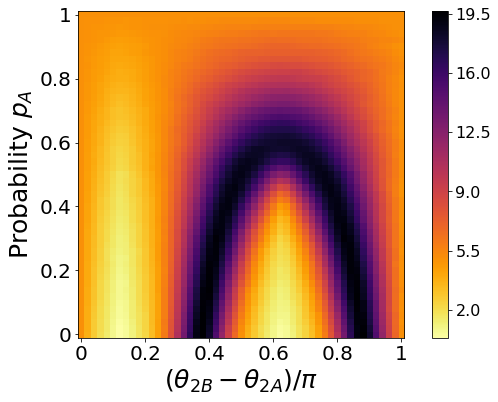}}
\caption{Root-mean-square size (square root of the variance) of the
  disorder-averaged position distribution after 170 timesteps, for the
  quantum walk with binary disorder. 100 disorder realizations were
  used, on a system of size $L_x=200 \times L_y=200$, with absorbing
  boundary conditions.}
\label{fig:binary_rms}
\end{figure}

In the main text we have observed delocalization of the system when
the rotation angles are randomized, i.e., $(\theta_1, \theta_2)$ are
position-dependent, picked from continuous random distributions.
Thus, each $(x,y)$ has (typically) a different value of $\theta_1$ and
$\theta_2$.
Now we ask what happens with \emph{binary disorder}, i.e., when the
angles $(\theta_1, \theta_2)$ are position-dependent, but picked from
the simplest discrete distribution: a probabilistic mixture (parameter
$p_A$) of two sets of values,
\begin{align}
  \theta_1(x,y) &= \theta_{1A};&
  \theta_2(x,y) &= \theta_{2A},\quad
  \text{prob.}~p_A~; \nonumber\\
  \theta_1(x,y) &= \theta_{1B};&
  \theta_2(x,y) &= \theta_{2B}, \quad
  \text{prob.}~(1-p_A), 
  \label{eq:binary_disorder_def}
\end{align}
while the other coin parameters $\alpha_{1,2}$, $\beta_{1,2}$ and
$\phi$ are uniformly random in range $[-\pi, \pi)$.
Set $A$ corresponds to a winding number of $\nu_A=+1$.

For a numerical look into binary disorder, we varied the parameter sets
$A$ and $B$ as follows. 
We fixed parameter set $A$, 
\begin{align} \label{eq:binary_theta_A}
  \theta_{1A} &= \frac{5\pi}{8};& \theta_{2A} &= - \frac{\pi}{8}.
\end{align}
The winding number of set $A$, as per
Eq.~\eqref{eq:clean_topo_number}---which holds for the case with or
without phase disorder---is $+1$. 
We varied the $\theta_1$ and $\theta_2$ parameters of set $B$
together, such that
\begin{align}
  \label{eq:binary_B_A}
\theta_{2B} - \theta_{2A} = \theta_{1A} - \theta_{1B}
\end{align}
is always respected.
Thus, when $\pi/4 <
\theta_{2B} < 3\pi/4$ , the winding number of set $B$ is $-1$,
otherwise it is $+1$.

We show the numerical results on diffusive/localized
behavior as functions of the difference of the $\theta_2$ parameters
and the probability $p_A$ of the set $A$ in
Figs.~\ref{fig:binary_transmission} and \ref{fig:binary_rms}.
Fig.~\ref{fig:binary_transmission} shows increased transmission at
the phase transition, accompanied by a change in the quantized
transmission in the case with a cut (of type $B$), as in
Fig.~\ref{fig:topo_trans}; Fig.~\ref{fig:binary_rms} shows the
increased spreading of the walk in the critical case.
When we have no values from set $A$, i.e., $p_A=0$,---bottom line of
plots---we see the expected signatures of the phase transition
between anomalous Floquet-Anderson localized phases.
Mixing in sites with coins in set $A$, i.e., increasing $p_A$, we find
that the size of the phase with the winding number $-1$ shrinks.
The critical value of $p_A$ needed for the winding number of $A$ to
dominate the mixture is largest when $B$ is in a flat-band limit,
$\theta_{1B}=0$ and $\theta_{2B} = \pi/2$.
This critical $p_A$ is $1/2$ when $\theta_{1B}=\theta_{1A}-\pi/2$ and
$\theta_{2B}=\theta_{2A}+\pi/2$; this is the case when the two sets
$A$ and $B$ describe disorder-averaged dynamics that map unto each
other under a mirror reflection.
We remark that for these numerical results we used disorder in $\phi,
\alpha, \beta$, both in the system, the cuts, and the leads, and the
$x$-$y$ basis instead of the rotated basis.

\bibliography{walkbib}

\begin{thebibliography}{49}%
\makeatletter
\providecommand \@ifxundefined [1]{%
 \@ifx{#1\undefined}
}%
\providecommand \@ifnum [1]{%
 \ifnum #1\expandafter \@firstoftwo
 \else \expandafter \@secondoftwo
 \fi
}%
\providecommand \@ifx [1]{%
 \ifx #1\expandafter \@firstoftwo
 \else \expandafter \@secondoftwo
 \fi
}%
\providecommand \natexlab [1]{#1}%
\providecommand \enquote  [1]{``#1''}%
\providecommand \bibnamefont  [1]{#1}%
\providecommand \bibfnamefont [1]{#1}%
\providecommand \citenamefont [1]{#1}%
\providecommand \href@noop [0]{\@secondoftwo}%
\providecommand \href [0]{\begingroup \@sanitize@url \@href}%
\providecommand \@href[1]{\@@startlink{#1}\@@href}%
\providecommand \@@href[1]{\endgroup#1\@@endlink}%
\providecommand \@sanitize@url [0]{\catcode `\\12\catcode `\$12\catcode
  `\&12\catcode `\#12\catcode `\^12\catcode `\_12\catcode `\%12\relax}%
\providecommand \@@startlink[1]{}%
\providecommand \@@endlink[0]{}%
\providecommand \url  [0]{\begingroup\@sanitize@url \@url }%
\providecommand \@url [1]{\endgroup\@href {#1}{\urlprefix }}%
\providecommand \urlprefix  [0]{URL }%
\providecommand \Eprint [0]{\href }%
\providecommand \doibase [0]{http://dx.doi.org/}%
\providecommand \selectlanguage [0]{\@gobble}%
\providecommand \bibinfo  [0]{\@secondoftwo}%
\providecommand \bibfield  [0]{\@secondoftwo}%
\providecommand \translation [1]{[#1]}%
\providecommand \BibitemOpen [0]{}%
\providecommand \bibitemStop [0]{}%
\providecommand \bibitemNoStop [0]{.\EOS\space}%
\providecommand \EOS [0]{\spacefactor3000\relax}%
\providecommand \BibitemShut  [1]{\csname bibitem#1\endcsname}%
\let\auto@bib@innerbib\@empty
%</preamble>
\bibitem [{\citenamefont {Kempe}(2003)}]{kempe2003quantum}%
  \BibitemOpen
  \bibfield  {author} {\bibinfo {author} {\bibfnamefont {Julia}\ \bibnamefont
  {Kempe}},\ }\bibfield  {title} {\enquote {\bibinfo {title} {Quantum random
  walks: an introductory overview},}\ }\href
  {https://www.tandfonline.com/doi/abs/10.1080/00107151031000110776} {\bibfield
   {journal} {\bibinfo  {journal} {Contemporary Physics}\ }\textbf {\bibinfo
  {volume} {44}},\ \bibinfo {pages} {307--327} (\bibinfo {year}
  {2003})}\BibitemShut {NoStop}%
\bibitem [{\citenamefont {Genske}\ \emph {et~al.}(2013)\citenamefont {Genske},
  \citenamefont {Alt}, \citenamefont {Steffen}, \citenamefont {Werner},
  \citenamefont {Werner}, \citenamefont {Meschede},\ and\ \citenamefont
  {Alberti}}]{genske2013electric}%
  \BibitemOpen
  \bibfield  {author} {\bibinfo {author} {\bibfnamefont {Maximilian}\
  \bibnamefont {Genske}}, \bibinfo {author} {\bibfnamefont {Wolfgang}\
  \bibnamefont {Alt}}, \bibinfo {author} {\bibfnamefont {Andreas}\ \bibnamefont
  {Steffen}}, \bibinfo {author} {\bibfnamefont {Albert~H}\ \bibnamefont
  {Werner}}, \bibinfo {author} {\bibfnamefont {Reinhard~F}\ \bibnamefont
  {Werner}}, \bibinfo {author} {\bibfnamefont {Dieter}\ \bibnamefont
  {Meschede}}, \ and\ \bibinfo {author} {\bibfnamefont {Andrea}\ \bibnamefont
  {Alberti}},\ }\bibfield  {title} {\enquote {\bibinfo {title} {Electric
  quantum walks with individual atoms},}\ }\href
  {https://journals.aps.org/prl/abstract/10.1103/PhysRevLett.110.190601}
  {\bibfield  {journal} {\bibinfo  {journal} {Phys. Rev. Lett.}\ }\textbf
  {\bibinfo {volume} {110}},\ \bibinfo {pages} {190601} (\bibinfo {year}
  {2013})}\BibitemShut {NoStop}%
\bibitem [{\citenamefont {Ambainis}\ \emph {et~al.}(2020)\citenamefont
  {Ambainis}, \citenamefont {Gily{\'e}n}, \citenamefont {Jeffery},\ and\
  \citenamefont {Kokainis}}]{ambainis2020quadratic}%
  \BibitemOpen
  \bibfield  {author} {\bibinfo {author} {\bibfnamefont {Andris}\ \bibnamefont
  {Ambainis}}, \bibinfo {author} {\bibfnamefont {Andr{\'a}s}\ \bibnamefont
  {Gily{\'e}n}}, \bibinfo {author} {\bibfnamefont {Stacey}\ \bibnamefont
  {Jeffery}}, \ and\ \bibinfo {author} {\bibfnamefont {Martins}\ \bibnamefont
  {Kokainis}},\ }\bibfield  {title} {\enquote {\bibinfo {title} {Quadratic
  speedup for finding marked vertices by quantum walks},}\ }in\ \href@noop {}
  {\emph {\bibinfo {booktitle} {Proceedings of the 52nd Annual ACM SIGACT
  Symposium on Theory of Computing}}}\ (\bibinfo {year} {2020})\ pp.\ \bibinfo
  {pages} {412--424}\BibitemShut {NoStop}%
\bibitem [{\citenamefont {Nagaoka}(1985)}]{nagaoka1985theory}%
  \BibitemOpen
  \bibfield  {author} {\bibinfo {author} {\bibfnamefont {Yosuke}\ \bibnamefont
  {Nagaoka}},\ }\bibfield  {title} {\enquote {\bibinfo {title} {{Theory of
  {A}nderson Localization: — A Historical Survey —}},}\ }\href {\doibase
  10.1143/PTPS.84.1} {\bibfield  {journal} {\bibinfo  {journal} {Progress of
  Theoretical Physics Supplement}\ }\textbf {\bibinfo {volume} {84}},\ \bibinfo
  {pages} {1--15} (\bibinfo {year} {1985})}\BibitemShut {NoStop}%
\bibitem [{\citenamefont {Sepehrinia}\ and\ \citenamefont
  {Sheikhan}(2011)}]{sepehrinia2011numerical}%
  \BibitemOpen
  \bibfield  {author} {\bibinfo {author} {\bibfnamefont {Reza}\ \bibnamefont
  {Sepehrinia}}\ and\ \bibinfo {author} {\bibfnamefont {Ameneh}\ \bibnamefont
  {Sheikhan}},\ }\bibfield  {title} {\enquote {\bibinfo {title} {Numerical
  simulation of {A}nderson localization},}\ }\href
  {https://ieeexplore.ieee.org/document/5756278} {\bibfield  {journal}
  {\bibinfo  {journal} {Computing in Science \& Engineering}\ }\textbf
  {\bibinfo {volume} {13}},\ \bibinfo {pages} {74--83} (\bibinfo {year}
  {2011})}\BibitemShut {NoStop}%
\bibitem [{\citenamefont {Joye}(2012)}]{joye2012dynamical}%
  \BibitemOpen
  \bibfield  {author} {\bibinfo {author} {\bibfnamefont {Alain}\ \bibnamefont
  {Joye}},\ }\bibfield  {title} {\enquote {\bibinfo {title} {Dynamical
  localization for d-dimensional random quantum walks},}\ }\href
  {https://link.springer.com/article/10.1007/s11128-012-0406-7} {\bibfield
  {journal} {\bibinfo  {journal} {Quantum Information Processing}\ }\textbf
  {\bibinfo {volume} {11}},\ \bibinfo {pages} {1251--1269} (\bibinfo {year}
  {2012})}\BibitemShut {NoStop}%
\bibitem [{\citenamefont {Vakulchyk}\ \emph {et~al.}(2017)\citenamefont
  {Vakulchyk}, \citenamefont {Fistul}, \citenamefont {Qin},\ and\ \citenamefont
  {Flach}}]{vakulchyk2017anderson}%
  \BibitemOpen
  \bibfield  {author} {\bibinfo {author} {\bibfnamefont {I.}~\bibnamefont
  {Vakulchyk}}, \bibinfo {author} {\bibfnamefont {M.~V.}\ \bibnamefont
  {Fistul}}, \bibinfo {author} {\bibfnamefont {P.}~\bibnamefont {Qin}}, \ and\
  \bibinfo {author} {\bibfnamefont {S.}~\bibnamefont {Flach}},\ }\bibfield
  {title} {\enquote {\bibinfo {title} {Anderson localization in generalized
  discrete-time quantum walks},}\ }\href {\doibase 10.1103/PhysRevB.96.144204}
  {\bibfield  {journal} {\bibinfo  {journal} {Phys. Rev. B}\ }\textbf {\bibinfo
  {volume} {96}},\ \bibinfo {pages} {144204} (\bibinfo {year}
  {2017})}\BibitemShut {NoStop}%
\bibitem [{\citenamefont {Schreiber}\ \emph {et~al.}(2011)\citenamefont
  {Schreiber}, \citenamefont {Cassemiro}, \citenamefont {Poto{\v{c}}ek},
  \citenamefont {G{\'a}bris}, \citenamefont {Jex},\ and\ \citenamefont
  {Silberhorn}}]{schreiber2011decoherence}%
  \BibitemOpen
  \bibfield  {author} {\bibinfo {author} {\bibfnamefont {A}~\bibnamefont
  {Schreiber}}, \bibinfo {author} {\bibfnamefont {KN}~\bibnamefont
  {Cassemiro}}, \bibinfo {author} {\bibfnamefont {V}~\bibnamefont
  {Poto{\v{c}}ek}}, \bibinfo {author} {\bibfnamefont {A}~\bibnamefont
  {G{\'a}bris}}, \bibinfo {author} {\bibfnamefont {I}~\bibnamefont {Jex}}, \
  and\ \bibinfo {author} {\bibfnamefont {Ch}~\bibnamefont {Silberhorn}},\
  }\bibfield  {title} {\enquote {\bibinfo {title} {Decoherence and disorder in
  quantum walks: from ballistic spread to localization},}\ }\href
  {https://journals.aps.org/prl/abstract/10.1103/PhysRevLett.106.180403}
  {\bibfield  {journal} {\bibinfo  {journal} {Phys. Rev. Lett.}\ }\textbf
  {\bibinfo {volume} {106}},\ \bibinfo {pages} {180403} (\bibinfo {year}
  {2011})}\BibitemShut {NoStop}%
\bibitem [{\citenamefont {Pankov}\ \emph {et~al.}(2019)\citenamefont {Pankov},
  \citenamefont {Vatnik}, \citenamefont {Churkin},\ and\ \citenamefont
  {Derevyanko}}]{pankov2019anderson}%
  \BibitemOpen
  \bibfield  {author} {\bibinfo {author} {\bibfnamefont {Artem~V}\ \bibnamefont
  {Pankov}}, \bibinfo {author} {\bibfnamefont {Ilya~D}\ \bibnamefont {Vatnik}},
  \bibinfo {author} {\bibfnamefont {Dmitry~V}\ \bibnamefont {Churkin}}, \ and\
  \bibinfo {author} {\bibfnamefont {Stanislav~A}\ \bibnamefont {Derevyanko}},\
  }\bibfield  {title} {\enquote {\bibinfo {title} {Anderson localization in
  synthetic photonic lattice with random coupling},}\ }\href
  {https://www.osapublishing.org/oe/abstract.cfm?uri=oe-27-4-4424} {\bibfield
  {journal} {\bibinfo  {journal} {Optics express}\ }\textbf {\bibinfo {volume}
  {27}},\ \bibinfo {pages} {4424--4434} (\bibinfo {year} {2019})}\BibitemShut
  {NoStop}%
\bibitem [{\citenamefont {D{\"u}r}\ \emph {et~al.}(2002)\citenamefont
  {D{\"u}r}, \citenamefont {Raussendorf}, \citenamefont {Kendon},\ and\
  \citenamefont {Briegel}}]{dur2002quantum}%
  \BibitemOpen
  \bibfield  {author} {\bibinfo {author} {\bibfnamefont {Wolfgang}\
  \bibnamefont {D{\"u}r}}, \bibinfo {author} {\bibfnamefont {Robert}\
  \bibnamefont {Raussendorf}}, \bibinfo {author} {\bibfnamefont {Vivien~M}\
  \bibnamefont {Kendon}}, \ and\ \bibinfo {author} {\bibfnamefont {H-J}\
  \bibnamefont {Briegel}},\ }\bibfield  {title} {\enquote {\bibinfo {title}
  {Quantum walks in optical lattices},}\ }\href
  {https://journals.aps.org/pra/abstract/10.1103/PhysRevA.66.052319} {\bibfield
   {journal} {\bibinfo  {journal} {Phys. Rev. A}\ }\textbf {\bibinfo {volume}
  {66}},\ \bibinfo {pages} {052319} (\bibinfo {year} {2002})}\BibitemShut
  {NoStop}%
\bibitem [{\citenamefont {Koll{\'a}r}\ \emph {et~al.}(2015)\citenamefont
  {Koll{\'a}r}, \citenamefont {Kiss},\ and\ \citenamefont
  {Jex}}]{kollar2015strongly}%
  \BibitemOpen
  \bibfield  {author} {\bibinfo {author} {\bibfnamefont {B{\'a}lint}\
  \bibnamefont {Koll{\'a}r}}, \bibinfo {author} {\bibfnamefont {Tam{\'a}s}\
  \bibnamefont {Kiss}}, \ and\ \bibinfo {author} {\bibfnamefont {Igor}\
  \bibnamefont {Jex}},\ }\bibfield  {title} {\enquote {\bibinfo {title}
  {Strongly trapped two-dimensional quantum walks},}\ }\href
  {https://journals.aps.org/pra/abstract/10.1103/PhysRevA.91.022308} {\bibfield
   {journal} {\bibinfo  {journal} {Phys. Rev. A}\ }\textbf {\bibinfo {volume}
  {91}},\ \bibinfo {pages} {022308} (\bibinfo {year} {2015})}\BibitemShut
  {NoStop}%
\bibitem [{\citenamefont {Machida}\ and\ \citenamefont
  {Chandrashekar}(2015)}]{machida2015localization}%
  \BibitemOpen
  \bibfield  {author} {\bibinfo {author} {\bibfnamefont {Takuya}\ \bibnamefont
  {Machida}}\ and\ \bibinfo {author} {\bibfnamefont {CM}~\bibnamefont
  {Chandrashekar}},\ }\bibfield  {title} {\enquote {\bibinfo {title}
  {Localization and limit laws of a three-state alternate quantum walk on a
  two-dimensional lattice},}\ }\href
  {https://journals.aps.org/pra/abstract/10.1103/PhysRevA.92.062307} {\bibfield
   {journal} {\bibinfo  {journal} {Phys. Rev. A}\ }\textbf {\bibinfo {volume}
  {92}},\ \bibinfo {pages} {062307} (\bibinfo {year} {2015})}\BibitemShut
  {NoStop}%
\bibitem [{\citenamefont {Koll{\'a}r}\ \emph {et~al.}(2020)\citenamefont
  {Koll{\'a}r}, \citenamefont {Gily{\'e}n}, \citenamefont
  {Tk{\'a}{\v{c}}ov{\'a}}, \citenamefont {Kiss}, \citenamefont {Jex},\ and\
  \citenamefont {{\v{S}}tefa{\v{n}}{\'a}k}}]{kollar2020complete}%
  \BibitemOpen
  \bibfield  {author} {\bibinfo {author} {\bibfnamefont {B}~\bibnamefont
  {Koll{\'a}r}}, \bibinfo {author} {\bibfnamefont {A}~\bibnamefont
  {Gily{\'e}n}}, \bibinfo {author} {\bibfnamefont {I}~\bibnamefont
  {Tk{\'a}{\v{c}}ov{\'a}}}, \bibinfo {author} {\bibfnamefont {T}~\bibnamefont
  {Kiss}}, \bibinfo {author} {\bibfnamefont {I}~\bibnamefont {Jex}}, \ and\
  \bibinfo {author} {\bibfnamefont {M}~\bibnamefont
  {{\v{S}}tefa{\v{n}}{\'a}k}},\ }\bibfield  {title} {\enquote {\bibinfo {title}
  {Complete classification of trapping coins for quantum walks on the
  two-dimensional square lattice},}\ }\href
  {https://journals.aps.org/pra/abstract/10.1103/PhysRevA.102.012207}
  {\bibfield  {journal} {\bibinfo  {journal} {Phys. Rev. A}\ }\textbf {\bibinfo
  {volume} {102}},\ \bibinfo {pages} {012207} (\bibinfo {year}
  {2020})}\BibitemShut {NoStop}%
\bibitem [{\citenamefont {Hasan}\ and\ \citenamefont
  {Kane}(2010)}]{hasan2010colloquium}%
  \BibitemOpen
  \bibfield  {author} {\bibinfo {author} {\bibfnamefont {M~Zahid}\ \bibnamefont
  {Hasan}}\ and\ \bibinfo {author} {\bibfnamefont {Charles~L}\ \bibnamefont
  {Kane}},\ }\bibfield  {title} {\enquote {\bibinfo {title} {Colloquium:
  topological insulators},}\ }\href
  {https://journals.aps.org/rmp/abstract/10.1103/RevModPhys.82.3045} {\bibfield
   {journal} {\bibinfo  {journal} {Rev. {M}od. {P}hys.}\ }\textbf {\bibinfo
  {volume} {82}},\ \bibinfo {pages} {3045} (\bibinfo {year}
  {2010})}\BibitemShut {NoStop}%
\bibitem [{\citenamefont {Kitagawa}\ \emph {et~al.}(2010)\citenamefont
  {Kitagawa}, \citenamefont {Rudner}, \citenamefont {Berg},\ and\ \citenamefont
  {Demler}}]{kitagawa2010exploring}%
  \BibitemOpen
  \bibfield  {author} {\bibinfo {author} {\bibfnamefont {Takuya}\ \bibnamefont
  {Kitagawa}}, \bibinfo {author} {\bibfnamefont {Mark~S.}\ \bibnamefont
  {Rudner}}, \bibinfo {author} {\bibfnamefont {Erez}\ \bibnamefont {Berg}}, \
  and\ \bibinfo {author} {\bibfnamefont {Eugene}\ \bibnamefont {Demler}},\
  }\bibfield  {title} {\enquote {\bibinfo {title} {Exploring topological phases
  with quantum walks},}\ }\href {\doibase 10.1103/PhysRevA.82.033429}
  {\bibfield  {journal} {\bibinfo  {journal} {Phys. Rev. A}\ }\textbf {\bibinfo
  {volume} {82}},\ \bibinfo {pages} {033429} (\bibinfo {year}
  {2010})}\BibitemShut {NoStop}%
\bibitem [{\citenamefont {Tarasinski}\ \emph {et~al.}(2014)\citenamefont
  {Tarasinski}, \citenamefont {Asb\'oth},\ and\ \citenamefont
  {Dahlhaus}}]{tarasinski2014scattering}%
  \BibitemOpen
  \bibfield  {author} {\bibinfo {author} {\bibfnamefont {B.}~\bibnamefont
  {Tarasinski}}, \bibinfo {author} {\bibfnamefont {J.~K.}\ \bibnamefont
  {Asb\'oth}}, \ and\ \bibinfo {author} {\bibfnamefont {J.~P.}\ \bibnamefont
  {Dahlhaus}},\ }\bibfield  {title} {\enquote {\bibinfo {title} {Scattering
  theory of topological phases in discrete-time quantum walks},}\ }\href
  {\doibase 10.1103/PhysRevA.89.042327} {\bibfield  {journal} {\bibinfo
  {journal} {Phys. Rev. A}\ }\textbf {\bibinfo {volume} {89}},\ \bibinfo
  {pages} {042327} (\bibinfo {year} {2014})}\BibitemShut {NoStop}%
\bibitem [{\citenamefont {Cedzich}\ \emph {et~al.}(2018)\citenamefont
  {Cedzich}, \citenamefont {Geib}, \citenamefont {Gr{\"u}nbaum}, \citenamefont
  {Stahl}, \citenamefont {Vel{\'a}zquez}, \citenamefont {Werner},\ and\
  \citenamefont {Werner}}]{cedzich2018topological}%
  \BibitemOpen
  \bibfield  {author} {\bibinfo {author} {\bibfnamefont {C}~\bibnamefont
  {Cedzich}}, \bibinfo {author} {\bibfnamefont {T}~\bibnamefont {Geib}},
  \bibinfo {author} {\bibfnamefont {FA}~\bibnamefont {Gr{\"u}nbaum}}, \bibinfo
  {author} {\bibfnamefont {C}~\bibnamefont {Stahl}}, \bibinfo {author}
  {\bibfnamefont {L}~\bibnamefont {Vel{\'a}zquez}}, \bibinfo {author}
  {\bibfnamefont {AH}~\bibnamefont {Werner}}, \ and\ \bibinfo {author}
  {\bibfnamefont {RF}~\bibnamefont {Werner}},\ }\bibfield  {title} {\enquote
  {\bibinfo {title} {The topological classification of one-dimensional
  symmetric quantum walks},}\ }in\ \href
  {https://link.springer.com/article/10.1007/s00023-017-0630-x} {\emph
  {\bibinfo {booktitle} {Annales Henri Poincar{\'e}}}},\ Vol.~\bibinfo {volume}
  {19}\ (\bibinfo {organization} {Springer},\ \bibinfo {year} {2018})\ pp.\
  \bibinfo {pages} {325--383}\BibitemShut {NoStop}%
\bibitem [{\citenamefont {Rudner}\ \emph {et~al.}(2013)\citenamefont {Rudner},
  \citenamefont {Lindner}, \citenamefont {Berg},\ and\ \citenamefont
  {Levin}}]{rudner2013anomalous}%
  \BibitemOpen
  \bibfield  {author} {\bibinfo {author} {\bibfnamefont {Mark~S.}\ \bibnamefont
  {Rudner}}, \bibinfo {author} {\bibfnamefont {Netanel~H.}\ \bibnamefont
  {Lindner}}, \bibinfo {author} {\bibfnamefont {Erez}\ \bibnamefont {Berg}}, \
  and\ \bibinfo {author} {\bibfnamefont {Michael}\ \bibnamefont {Levin}},\
  }\bibfield  {title} {\enquote {\bibinfo {title} {Anomalous edge states and
  the bulk-edge correspondence for periodically driven two-dimensional
  systems},}\ }\href {\doibase 10.1103/PhysRevX.3.031005} {\bibfield  {journal}
  {\bibinfo  {journal} {Phys. Rev. X}\ }\textbf {\bibinfo {volume} {3}},\
  \bibinfo {pages} {031005} (\bibinfo {year} {2013})}\BibitemShut {NoStop}%
\bibitem [{\citenamefont {Asb{\'o}th}\ and\ \citenamefont
  {Edge}(2015)}]{asboth2015edge}%
  \BibitemOpen
  \bibfield  {author} {\bibinfo {author} {\bibfnamefont {J{\'a}nos~K.}\
  \bibnamefont {Asb{\'o}th}}\ and\ \bibinfo {author} {\bibfnamefont
  {Jonathan~M.}\ \bibnamefont {Edge}},\ }\bibfield  {title} {\enquote {\bibinfo
  {title} {Edge-state-enhanced transport in a two-dimensional quantum walk},}\
  }\href {\doibase 10.1103/PhysRevA.91.022324} {\bibfield  {journal} {\bibinfo
  {journal} {Phys. Rev. A}\ }\textbf {\bibinfo {volume} {91}},\ \bibinfo
  {pages} {022324} (\bibinfo {year} {2015})}\BibitemShut {NoStop}%
\bibitem [{\citenamefont {Edge}\ and\ \citenamefont
  {Asb{\'o}th}(2015)}]{edge2015localization}%
  \BibitemOpen
  \bibfield  {author} {\bibinfo {author} {\bibfnamefont {Jonathan~M.}\
  \bibnamefont {Edge}}\ and\ \bibinfo {author} {\bibfnamefont {J{\'a}nos~K.}\
  \bibnamefont {Asb{\'o}th}},\ }\bibfield  {title} {\enquote {\bibinfo {title}
  {Localization, delocalization, and topological transitions in disordered
  two-dimensional quantum walks},}\ }\href {\doibase
  10.1103/PhysRevB.91.104202} {\bibfield  {journal} {\bibinfo  {journal} {Phys.
  Rev. B}\ }\textbf {\bibinfo {volume} {91}},\ \bibinfo {pages} {104202}
  (\bibinfo {year} {2015})}\BibitemShut {NoStop}%
\bibitem [{\citenamefont {Titum}\ \emph {et~al.}(2016)\citenamefont {Titum},
  \citenamefont {Berg}, \citenamefont {Rudner}, \citenamefont {Refael},\ and\
  \citenamefont {Lindner}}]{titum2016anomalous}%
  \BibitemOpen
  \bibfield  {author} {\bibinfo {author} {\bibfnamefont {Paraj}\ \bibnamefont
  {Titum}}, \bibinfo {author} {\bibfnamefont {Erez}\ \bibnamefont {Berg}},
  \bibinfo {author} {\bibfnamefont {Mark~S.}\ \bibnamefont {Rudner}}, \bibinfo
  {author} {\bibfnamefont {Gil}\ \bibnamefont {Refael}}, \ and\ \bibinfo
  {author} {\bibfnamefont {Netanel~H.}\ \bibnamefont {Lindner}},\ }\bibfield
  {title} {\enquote {\bibinfo {title} {Anomalous {F}loquet-{A}nderson insulator
  as a nonadiabatic quantized charge pump},}\ }\href {\doibase
  10.1103/PhysRevX.6.021013} {\bibfield  {journal} {\bibinfo  {journal} {Phys.
  Rev. X}\ }\textbf {\bibinfo {volume} {6}},\ \bibinfo {pages} {021013}
  (\bibinfo {year} {2016})}\BibitemShut {NoStop}%
\bibitem [{\citenamefont {Zeng}\ and\ \citenamefont
  {Yong}(2017)}]{zeng2017discrete}%
  \BibitemOpen
  \bibfield  {author} {\bibinfo {author} {\bibfnamefont {Meng}\ \bibnamefont
  {Zeng}}\ and\ \bibinfo {author} {\bibfnamefont {Ee~Hou}\ \bibnamefont
  {Yong}},\ }\bibfield  {title} {\enquote {\bibinfo {title} {Discrete-time
  quantum walk with phase disorder: localization and entanglement entropy},}\
  }\href {https://www.nature.com/articles/s41598-017-12077-0} {\bibfield
  {journal} {\bibinfo  {journal} {Scientific Reports}\ }\textbf {\bibinfo
  {volume} {7}},\ \bibinfo {pages} {1--9} (\bibinfo {year} {2017})}\BibitemShut
  {NoStop}%
\bibitem [{\citenamefont {Mendes}\ \emph {et~al.}(2019)\citenamefont {Mendes},
  \citenamefont {Almeida}, \citenamefont {Lyra},\ and\ \citenamefont
  {de~Moura}}]{mendes2019localization}%
  \BibitemOpen
  \bibfield  {author} {\bibinfo {author} {\bibfnamefont {CVC}\ \bibnamefont
  {Mendes}}, \bibinfo {author} {\bibfnamefont {GMA}\ \bibnamefont {Almeida}},
  \bibinfo {author} {\bibfnamefont {ML}~\bibnamefont {Lyra}}, \ and\ \bibinfo
  {author} {\bibfnamefont {FABF}\ \bibnamefont {de~Moura}},\ }\bibfield
  {title} {\enquote {\bibinfo {title} {Localization-delocalization transition
  in discrete-time quantum walks with long-range correlated disorder},}\ }\href
  {https://journals.aps.org/pre/abstract/10.1103/PhysRevE.99.022117} {\bibfield
   {journal} {\bibinfo  {journal} {Phys. Rev. E}\ }\textbf {\bibinfo {volume}
  {99}},\ \bibinfo {pages} {022117} (\bibinfo {year} {2019})}\BibitemShut
  {NoStop}%
\bibitem [{\citenamefont {Obuse}\ and\ \citenamefont
  {Kawakami}(2011)}]{obuse2011topological}%
  \BibitemOpen
  \bibfield  {author} {\bibinfo {author} {\bibfnamefont {Hideaki}\ \bibnamefont
  {Obuse}}\ and\ \bibinfo {author} {\bibfnamefont {Norio}\ \bibnamefont
  {Kawakami}},\ }\bibfield  {title} {\enquote {\bibinfo {title} {Topological
  phases and delocalization of quantum walks in random environments},}\ }\href
  {https://journals.aps.org/prb/abstract/10.1103/PhysRevB.84.195139} {\bibfield
   {journal} {\bibinfo  {journal} {Phys. Rev. B}\ }\textbf {\bibinfo {volume}
  {84}},\ \bibinfo {pages} {195139} (\bibinfo {year} {2011})}\BibitemShut
  {NoStop}%
\bibitem [{\citenamefont {Zhao}\ and\ \citenamefont
  {Gong}(2015)}]{zhao2015disordered}%
  \BibitemOpen
  \bibfield  {author} {\bibinfo {author} {\bibfnamefont {Qifang}\ \bibnamefont
  {Zhao}}\ and\ \bibinfo {author} {\bibfnamefont {Jiangbin}\ \bibnamefont
  {Gong}},\ }\bibfield  {title} {\enquote {\bibinfo {title} {From disordered
  quantum walk to physics of off-diagonal disorder},}\ }\href
  {https://journals.aps.org/prb/abstract/10.1103/PhysRevB.92.214205} {\bibfield
   {journal} {\bibinfo  {journal} {Physical Review B}\ }\textbf {\bibinfo
  {volume} {92}},\ \bibinfo {pages} {214205} (\bibinfo {year}
  {2015})}\BibitemShut {NoStop}%
\bibitem [{\citenamefont {Rakovszky}\ and\ \citenamefont
  {Asb{\'o}th}(2015)}]{rakovszky2015localization}%
  \BibitemOpen
  \bibfield  {author} {\bibinfo {author} {\bibfnamefont {Tibor}\ \bibnamefont
  {Rakovszky}}\ and\ \bibinfo {author} {\bibfnamefont {J{\'a}nos~K.}\
  \bibnamefont {Asb{\'o}th}},\ }\bibfield  {title} {\enquote {\bibinfo {title}
  {Localization, delocalization, and topological phase transitions in the
  one-dimensional split-step quantum walk},}\ }\href {\doibase
  10.1103/PhysRevA.92.052311} {\bibfield  {journal} {\bibinfo  {journal} {Phys.
  Rev. A}\ }\textbf {\bibinfo {volume} {92}},\ \bibinfo {pages} {052311}
  (\bibinfo {year} {2015})}\BibitemShut {NoStop}%
\bibitem [{\citenamefont {Kitagawa}(2012)}]{kitagawa2012topological}%
  \BibitemOpen
  \bibfield  {author} {\bibinfo {author} {\bibfnamefont {Takuya}\ \bibnamefont
  {Kitagawa}},\ }\bibfield  {title} {\enquote {\bibinfo {title} {Topological
  phenomena in quantum walks: elementary introduction to the physics of
  topological phases},}\ }\href
  {https://link.springer.com/article/10.1007/s11128-012-0425-4} {\bibfield
  {journal} {\bibinfo  {journal} {Quantum Information Processing}\ }\textbf
  {\bibinfo {volume} {11}},\ \bibinfo {pages} {1107--1148} (\bibinfo {year}
  {2012})}\BibitemShut {NoStop}%
\bibitem [{\citenamefont {Yal{\c{c}}{\i}nkaya}\ and\ \citenamefont
  {Gedik}(2015)}]{yalccinkaya2015two}%
  \BibitemOpen
  \bibfield  {author} {\bibinfo {author} {\bibfnamefont {{\.{I}}.}~\bibnamefont
  {Yal{\c{c}}{\i}nkaya}}\ and\ \bibinfo {author} {\bibfnamefont
  {Z.}~\bibnamefont {Gedik}},\ }\bibfield  {title} {\enquote {\bibinfo {title}
  {Two-dimensional quantum walk under artificial magnetic field},}\ }\href
  {\doibase 10.1103/PhysRevA.92.042324} {\bibfield  {journal} {\bibinfo
  {journal} {Phys. Rev. A}\ }\textbf {\bibinfo {volume} {92}},\ \bibinfo
  {pages} {042324} (\bibinfo {year} {2015})}\BibitemShut {NoStop}%
\bibitem [{\citenamefont {Arnault}\ and\ \citenamefont
  {Debbasch}(2016)}]{arnault2016quantum}%
  \BibitemOpen
  \bibfield  {author} {\bibinfo {author} {\bibfnamefont {Pablo}\ \bibnamefont
  {Arnault}}\ and\ \bibinfo {author} {\bibfnamefont {Fabrice}\ \bibnamefont
  {Debbasch}},\ }\bibfield  {title} {\enquote {\bibinfo {title} {Quantum walks
  and discrete gauge theories},}\ }\href
  {https://journals.aps.org/pra/abstract/10.1103/PhysRevA.93.052301} {\bibfield
   {journal} {\bibinfo  {journal} {Phys. Rev. A}\ }\textbf {\bibinfo {volume}
  {93}},\ \bibinfo {pages} {052301} (\bibinfo {year} {2016})}\BibitemShut
  {NoStop}%
\bibitem [{\citenamefont {Sajid}\ \emph {et~al.}(2019)\citenamefont {Sajid},
  \citenamefont {Asb\'oth}, \citenamefont {Meschede}, \citenamefont {Werner},\
  and\ \citenamefont {Alberti}}]{sajid2018creating}%
  \BibitemOpen
  \bibfield  {author} {\bibinfo {author} {\bibfnamefont {Muhammad}\
  \bibnamefont {Sajid}}, \bibinfo {author} {\bibfnamefont {J\'anos~K.}\
  \bibnamefont {Asb\'oth}}, \bibinfo {author} {\bibfnamefont {Dieter}\
  \bibnamefont {Meschede}}, \bibinfo {author} {\bibfnamefont {Reinhard~F.}\
  \bibnamefont {Werner}}, \ and\ \bibinfo {author} {\bibfnamefont {Andrea}\
  \bibnamefont {Alberti}},\ }\bibfield  {title} {\enquote {\bibinfo {title}
  {Creating anomalous {F}loquet {C}hern insulators with magnetic quantum
  walks},}\ }\href {\doibase 10.1103/PhysRevB.99.214303} {\bibfield  {journal}
  {\bibinfo  {journal} {Phys. Rev. B}\ }\textbf {\bibinfo {volume} {99}},\
  \bibinfo {pages} {214303} (\bibinfo {year} {2019})}\BibitemShut {NoStop}%
\bibitem [{\citenamefont {Cedzich}\ \emph {et~al.}(2019)\citenamefont
  {Cedzich}, \citenamefont {Geib}, \citenamefont {Werner},\ and\ \citenamefont
  {Werner}}]{cedzich2019quantum}%
  \BibitemOpen
  \bibfield  {author} {\bibinfo {author} {\bibfnamefont {C.}~\bibnamefont
  {Cedzich}}, \bibinfo {author} {\bibfnamefont {T.}~\bibnamefont {Geib}},
  \bibinfo {author} {\bibfnamefont {A.~H.}\ \bibnamefont {Werner}}, \ and\
  \bibinfo {author} {\bibfnamefont {R.~F.}\ \bibnamefont {Werner}},\ }\bibfield
   {title} {\enquote {\bibinfo {title} {Quantum walks in external gauge
  fields},}\ }\href {\doibase 10.1063/1.5054894} {\bibfield  {journal}
  {\bibinfo  {journal} {Journal of Mathematical Physics}\ }\textbf {\bibinfo
  {volume} {60}},\ \bibinfo {pages} {012107} (\bibinfo {year} {2019})},\
  \Eprint {http://arxiv.org/abs/https://doi.org/10.1063/1.5054894}
  {https://doi.org/10.1063/1.5054894} \BibitemShut {NoStop}%
\bibitem [{\citenamefont {Mallick}(2018)}]{Mallick:2019hvd}%
  \BibitemOpen
  \bibfield  {author} {\bibinfo {author} {\bibfnamefont {Arindam}\ \bibnamefont
  {Mallick}},\ }\emph {\bibinfo {title} {{Quantum Simulation of Neutrino
  Oscillation and Dirac Particle Dynamics in Curved Space-time}}},\ \href@noop
  {} {Ph.D. thesis},\ \bibinfo  {school} {IMSc, Chennai} (\bibinfo {year}
  {2018}),\ \Eprint {http://arxiv.org/abs/1901.04014} {arXiv:1901.04014
  [quant-ph]} \BibitemShut {NoStop}%
%%CITATION = ARXIV:1901.04014;%%
\bibitem [{\citenamefont {Asb{\'o}th}\ and\ \citenamefont
  {Obuse}(2013)}]{asboth2013bulk}%
  \BibitemOpen
  \bibfield  {author} {\bibinfo {author} {\bibfnamefont {J{\'a}nos~K.}\
  \bibnamefont {Asb{\'o}th}}\ and\ \bibinfo {author} {\bibfnamefont {Hideaki}\
  \bibnamefont {Obuse}},\ }\bibfield  {title} {\enquote {\bibinfo {title}
  {Bulk-boundary correspondence for chiral symmetric quantum walks},}\ }\href
  {\doibase 10.1103/PhysRevB.88.121406} {\bibfield  {journal} {\bibinfo
  {journal} {Phys. Rev. B}\ }\textbf {\bibinfo {volume} {88}},\ \bibinfo
  {pages} {121406} (\bibinfo {year} {2013})}\BibitemShut {NoStop}%
\bibitem [{\citenamefont {Asb\'oth}\ and\ \citenamefont
  {Alberti}(2017)}]{asboth2017spectral}%
  \BibitemOpen
  \bibfield  {author} {\bibinfo {author} {\bibfnamefont {J\'anos~K.}\
  \bibnamefont {Asb\'oth}}\ and\ \bibinfo {author} {\bibfnamefont {Andrea}\
  \bibnamefont {Alberti}},\ }\bibfield  {title} {\enquote {\bibinfo {title}
  {Spectral flow and global topology of the {H}ofstadter butterfly},}\ }\href
  {\doibase 10.1103/PhysRevLett.118.216801} {\bibfield  {journal} {\bibinfo
  {journal} {Phys. Rev. Lett.}\ }\textbf {\bibinfo {volume} {118}},\ \bibinfo
  {pages} {216801} (\bibinfo {year} {2017})}\BibitemShut {NoStop}%
\bibitem [{\citenamefont {Fulga}\ and\ \citenamefont
  {Maksymenko}(2016)}]{fulga2016scattering}%
  \BibitemOpen
  \bibfield  {author} {\bibinfo {author} {\bibfnamefont {I.~C.}\ \bibnamefont
  {Fulga}}\ and\ \bibinfo {author} {\bibfnamefont {M.}~\bibnamefont
  {Maksymenko}},\ }\bibfield  {title} {\enquote {\bibinfo {title} {Scattering
  matrix invariants of {F}loquet topological insulators},}\ }\href {\doibase
  10.1103/PhysRevB.93.075405} {\bibfield  {journal} {\bibinfo  {journal} {Phys.
  Rev. B}\ }\textbf {\bibinfo {volume} {93}},\ \bibinfo {pages} {075405}
  (\bibinfo {year} {2016})}\BibitemShut {NoStop}%
\bibitem [{\citenamefont {Rodr{\'\i}guez-Mena}\ and\ \citenamefont
  {Torres}(2019)}]{rodriguez2019topological}%
  \BibitemOpen
  \bibfield  {author} {\bibinfo {author} {\bibfnamefont {Esteban~A}\
  \bibnamefont {Rodr{\'\i}guez-Mena}}\ and\ \bibinfo {author} {\bibfnamefont
  {LEF~Foa}\ \bibnamefont {Torres}},\ }\bibfield  {title} {\enquote {\bibinfo
  {title} {Topological signatures in quantum transport in anomalous
  {F}loquet-{A}nderson insulators},}\ }\href
  {https://journals.aps.org/prb/abstract/10.1103/PhysRevB.100.195429}
  {\bibfield  {journal} {\bibinfo  {journal} {Phys. Rev. B}\ }\textbf {\bibinfo
  {volume} {100}},\ \bibinfo {pages} {195429} (\bibinfo {year}
  {2019})}\BibitemShut {NoStop}%
\bibitem [{\citenamefont {Liu}\ \emph {et~al.}(2020)\citenamefont {Liu},
  \citenamefont {Fulga},\ and\ \citenamefont {Asb{\'o}th}}]{liu2020anomalous}%
  \BibitemOpen
  \bibfield  {author} {\bibinfo {author} {\bibfnamefont {Hui}\ \bibnamefont
  {Liu}}, \bibinfo {author} {\bibfnamefont {Ion~Cosma}\ \bibnamefont {Fulga}},
  \ and\ \bibinfo {author} {\bibfnamefont {J{\'a}nos~K}\ \bibnamefont
  {Asb{\'o}th}},\ }\bibfield  {title} {\enquote {\bibinfo {title} {Anomalous
  levitation and annihilation in floquet topological insulators},}\ }\href
  {https://journals.aps.org/prresearch/abstract/10.1103/PhysRevResearch.2.022048}
  {\bibfield  {journal} {\bibinfo  {journal} {Phys. Rev. Research}\ }\textbf
  {\bibinfo {volume} {2}},\ \bibinfo {pages} {022048} (\bibinfo {year}
  {2020})}\BibitemShut {NoStop}%
\bibitem [{\citenamefont {Barkhofen}\ \emph {et~al.}(2017)\citenamefont
  {Barkhofen}, \citenamefont {Nitsche}, \citenamefont {Elster}, \citenamefont
  {Lorz}, \citenamefont {G\'abris}, \citenamefont {Jex},\ and\ \citenamefont
  {Silberhorn}}]{barkhofen2017measuring}%
  \BibitemOpen
  \bibfield  {author} {\bibinfo {author} {\bibfnamefont {Sonja}\ \bibnamefont
  {Barkhofen}}, \bibinfo {author} {\bibfnamefont {Thomas}\ \bibnamefont
  {Nitsche}}, \bibinfo {author} {\bibfnamefont {Fabian}\ \bibnamefont
  {Elster}}, \bibinfo {author} {\bibfnamefont {Lennart}\ \bibnamefont {Lorz}},
  \bibinfo {author} {\bibfnamefont {Aur\'el}\ \bibnamefont {G\'abris}},
  \bibinfo {author} {\bibfnamefont {Igor}\ \bibnamefont {Jex}}, \ and\ \bibinfo
  {author} {\bibfnamefont {Christine}\ \bibnamefont {Silberhorn}},\ }\bibfield
  {title} {\enquote {\bibinfo {title} {Measuring topological invariants in
  disordered discrete-time quantum walks},}\ }\href {\doibase
  10.1103/PhysRevA.96.033846} {\bibfield  {journal} {\bibinfo  {journal} {Phys.
  Rev. A}\ }\textbf {\bibinfo {volume} {96}},\ \bibinfo {pages} {033846}
  (\bibinfo {year} {2017})}\BibitemShut {NoStop}%
\bibitem [{\citenamefont {Nazarov}\ \emph {et~al.}(2009)\citenamefont
  {Nazarov}, \citenamefont {Nazarov},\ and\ \citenamefont
  {Blanter}}]{nazarov2009quantum}%
  \BibitemOpen
  \bibfield  {author} {\bibinfo {author} {\bibfnamefont {Yuli~V}\ \bibnamefont
  {Nazarov}}, \bibinfo {author} {\bibfnamefont {Yuli}\ \bibnamefont {Nazarov}},
  \ and\ \bibinfo {author} {\bibfnamefont {Yaroslav~M}\ \bibnamefont
  {Blanter}},\ }\href@noop {} {\emph {\bibinfo {title} {Quantum transport:
  introduction to nanoscience}}}\ (\bibinfo  {publisher} {Cambridge University
  Press},\ \bibinfo {year} {2009})\BibitemShut {NoStop}%
\bibitem [{\citenamefont {Asb{\'o}th}(2012)}]{asboth2012symmetries}%
  \BibitemOpen
  \bibfield  {author} {\bibinfo {author} {\bibfnamefont {J{\'a}nos~K}\
  \bibnamefont {Asb{\'o}th}},\ }\bibfield  {title} {\enquote {\bibinfo {title}
  {Symmetries, topological phases, and bound states in the one-dimensional
  quantum walk},}\ }\href
  {https://journals.aps.org/prb/abstract/10.1103/PhysRevB.86.195414} {\bibfield
   {journal} {\bibinfo  {journal} {Phys. Rev. B}\ }\textbf {\bibinfo {volume}
  {86}},\ \bibinfo {pages} {195414} (\bibinfo {year} {2012})}\BibitemShut
  {NoStop}%
\bibitem [{\citenamefont {Lehoucq}\ \emph {et~al.}(1998)\citenamefont
  {Lehoucq}, \citenamefont {Sorensen},\ and\ \citenamefont
  {Yang}}]{lehoucq1998arpack}%
  \BibitemOpen
  \bibfield  {author} {\bibinfo {author} {\bibfnamefont {Richard~B}\
  \bibnamefont {Lehoucq}}, \bibinfo {author} {\bibfnamefont {Danny~C}\
  \bibnamefont {Sorensen}}, \ and\ \bibinfo {author} {\bibfnamefont {Chao}\
  \bibnamefont {Yang}},\ }\href@noop {} {\emph {\bibinfo {title} {ARPACK users'
  guide: solution of large-scale eigenvalue problems with implicitly restarted
  Arnoldi methods}}},\ Vol.~\bibinfo {volume} {6}\ (\bibinfo  {publisher}
  {Siam},\ \bibinfo {year} {1998})\BibitemShut {NoStop}%
\bibitem [{\citenamefont {Obuse}\ and\ \citenamefont
  {Yakubo}(2005)}]{obuse2005critical}%
  \BibitemOpen
  \bibfield  {author} {\bibinfo {author} {\bibfnamefont {H}~\bibnamefont
  {Obuse}}\ and\ \bibinfo {author} {\bibfnamefont {K}~\bibnamefont {Yakubo}},\
  }\bibfield  {title} {\enquote {\bibinfo {title} {Critical level statistics
  and anomalously localized states at the anderson transition},}\ }\href
  {https://journals.aps.org/prb/abstract/10.1103/PhysRevB.71.035102} {\bibfield
   {journal} {\bibinfo  {journal} {Phys. Rev. B}\ }\textbf {\bibinfo {volume}
  {71}},\ \bibinfo {pages} {035102} (\bibinfo {year} {2005})}\BibitemShut
  {NoStop}%
\bibitem [{\citenamefont {Mezzadri}(2006)}]{mezzadri2006generate}%
  \BibitemOpen
  \bibfield  {author} {\bibinfo {author} {\bibfnamefont {Francesco}\
  \bibnamefont {Mezzadri}},\ }\bibfield  {title} {\enquote {\bibinfo {title}
  {How to generate random matrices from the classical compact groups},}\
  }\href@noop {} {\bibfield  {journal} {\bibinfo  {journal} {arXiv preprint
  math-ph/0609050}\ } (\bibinfo {year} {2006})}\BibitemShut {NoStop}%
\bibitem [{\citenamefont {Ozols}(2009)}]{ozols2009generate}%
  \BibitemOpen
  \bibfield  {author} {\bibinfo {author} {\bibfnamefont {Maris}\ \bibnamefont
  {Ozols}},\ }\bibfield  {title} {\enquote {\bibinfo {title} {How to generate a
  random unitary matrix},}\ }\href
  {http://home.lu.lv/~sd20008/papers/essays.html} {\bibfield  {journal}
  {\bibinfo  {journal} {unpublished essay on http://home.lu.lv/sd20008}\ }
  (\bibinfo {year} {2009})}\BibitemShut {NoStop}%
\bibitem [{\citenamefont {Kim}\ \emph {et~al.}(2020)\citenamefont {Kim},
  \citenamefont {Bagrets}, \citenamefont {Micklitz},\ and\ \citenamefont
  {Altland}}]{woo2019quantum}%
  \BibitemOpen
  \bibfield  {author} {\bibinfo {author} {\bibfnamefont {Kun~Woo}\ \bibnamefont
  {Kim}}, \bibinfo {author} {\bibfnamefont {Dmitry}\ \bibnamefont {Bagrets}},
  \bibinfo {author} {\bibfnamefont {Tobias}\ \bibnamefont {Micklitz}}, \ and\
  \bibinfo {author} {\bibfnamefont {Alexander}\ \bibnamefont {Altland}},\
  }\bibfield  {title} {\enquote {\bibinfo {title} {Quantum hall criticality in
  floquet topological insulators},}\ }\href
  {https://journals.aps.org/prb/abstract/10.1103/PhysRevB.101.165401}
  {\bibfield  {journal} {\bibinfo  {journal} {Physical Review B}\ }\textbf
  {\bibinfo {volume} {101}},\ \bibinfo {pages} {165401} (\bibinfo {year}
  {2020})}\BibitemShut {NoStop}%
\bibitem [{\citenamefont {Huckestein}(1995)}]{huckestein1995scaling}%
  \BibitemOpen
  \bibfield  {author} {\bibinfo {author} {\bibfnamefont {Bodo}\ \bibnamefont
  {Huckestein}},\ }\bibfield  {title} {\enquote {\bibinfo {title} {Scaling
  theory of the integer quantum hall effect},}\ }\href
  {https://journals.aps.org/rmp/abstract/10.1103/RevModPhys.67.357} {\bibfield
  {journal} {\bibinfo  {journal} {Rev. Mod. Phys.}\ }\textbf {\bibinfo {volume}
  {67}},\ \bibinfo {pages} {357} (\bibinfo {year} {1995})}\BibitemShut
  {NoStop}%
\bibitem [{\citenamefont {Evers}\ \emph {et~al.}(2008)\citenamefont {Evers},
  \citenamefont {Mildenberger},\ and\ \citenamefont
  {Mirlin}}]{evers2008multifractality}%
  \BibitemOpen
  \bibfield  {author} {\bibinfo {author} {\bibfnamefont {F}~\bibnamefont
  {Evers}}, \bibinfo {author} {\bibfnamefont {A}~\bibnamefont {Mildenberger}},
  \ and\ \bibinfo {author} {\bibfnamefont {AD}~\bibnamefont {Mirlin}},\
  }\bibfield  {title} {\enquote {\bibinfo {title} {Multifractality at the
  quantum hall transition: Beyond the parabolic paradigm},}\ }\href
  {https://journals.aps.org/prl/abstract/10.1103/PhysRevLett.101.116803}
  {\bibfield  {journal} {\bibinfo  {journal} {Phys. Rev. Lett.}\ }\textbf
  {\bibinfo {volume} {101}},\ \bibinfo {pages} {116803} (\bibinfo {year}
  {2008})}\BibitemShut {NoStop}%
\bibitem [{\citenamefont {Ahlbrecht}\ \emph {et~al.}(2011)\citenamefont
  {Ahlbrecht}, \citenamefont {Scholz},\ and\ \citenamefont
  {Werner}}]{ahlbrecht2011disordered}%
  \BibitemOpen
  \bibfield  {author} {\bibinfo {author} {\bibfnamefont {Andre}\ \bibnamefont
  {Ahlbrecht}}, \bibinfo {author} {\bibfnamefont {Volkher~B}\ \bibnamefont
  {Scholz}}, \ and\ \bibinfo {author} {\bibfnamefont {Albert~H}\ \bibnamefont
  {Werner}},\ }\bibfield  {title} {\enquote {\bibinfo {title} {Disordered
  quantum walks in one lattice dimension},}\ }\href
  {https://aip.scitation.org/doi/10.1063/1.3643768} {\bibfield  {journal}
  {\bibinfo  {journal} {Journal of Mathematical Physics}\ }\textbf {\bibinfo
  {volume} {52}},\ \bibinfo {pages} {102201} (\bibinfo {year}
  {2011})}\BibitemShut {NoStop}%
\bibitem [{\citenamefont {Delplace}\ \emph {et~al.}(2017)\citenamefont
  {Delplace}, \citenamefont {Fruchart},\ and\ \citenamefont
  {Tauber}}]{delplace2017phase}%
  \BibitemOpen
  \bibfield  {author} {\bibinfo {author} {\bibfnamefont {Pierre}\ \bibnamefont
  {Delplace}}, \bibinfo {author} {\bibfnamefont {Michel}\ \bibnamefont
  {Fruchart}}, \ and\ \bibinfo {author} {\bibfnamefont {Cl{\'e}ment}\
  \bibnamefont {Tauber}},\ }\bibfield  {title} {\enquote {\bibinfo {title}
  {Phase rotation symmetry and the topology of oriented scattering networks},}\
  }\href {https://journals.aps.org/prb/abstract/10.1103/PhysRevB.95.205413}
  {\bibfield  {journal} {\bibinfo  {journal} {Phys. Rev. B}\ }\textbf {\bibinfo
  {volume} {95}},\ \bibinfo {pages} {205413} (\bibinfo {year}
  {2017})}\BibitemShut {NoStop}%
\end{thebibliography}%

\end{document}